\documentclass[12pt]{iopart}
\usepackage{graphicx}
\usepackage{cite}
\usepackage{comment}
\usepackage{url}
\usepackage{color,soul}
\usepackage{xcolor}
\usepackage{hyperref}

\expandafter\let\csname equation*\endcsname\relax
\expandafter\let\csname endequation*\endcsname\relax
\usepackage{amsmath}
\usepackage{amsfonts}

\begin{document}

%\title[Do Graph Neural Networks Dream of Landau Damping?]{Do Graph Neural Networks Dream of Landau Damping? Insights from Kinetic Simulations of a Plasma Sheet Model}
%\title{Learning the Physical Dynamics of a Fundamental Plasma Sheet Model with Graph Neural Networks}
%\title{Learning the Dynamics of a 1D Plasma Sheet Model with Graph Neural Networks}
%\title{Simulating a One-Dimensional Plasma Model with Graph Neural Networks}
\title[Learning the dynamics of a one-dimensional plasma model with GNNs]{Learning the dynamics of a one-dimensional plasma model with graph neural networks} 

\author{Diogo D Carvalho$^1$, Diogo R Ferreira$^2$ and Luís O Silva$^1$}
\address{$^1$ GoLP/Instituto de Plasmas e Fus\~{a}o Nuclear, Instituto Superior T\'{e}cnico, Universidade de Lisboa, 1049-001 Lisbon, Portugal}
\address{$^2$ Instituto de Plasmas e Fus\~{a}o Nuclear, Instituto Superior T\'{e}cnico, Universidade de Lisboa, 1049-001 Lisbon, Portugal}

\ead{diogo.d.carvalho@tecnico.ulisboa.pt}

\begin{abstract} % around 200 words

We explore the possibility of fully replacing a plasma physics kinetic simulator with a graph neural network-based simulator. We focus on this class of surrogate models given the similarity between their message-passing update mechanism and the traditional physics solver update, and the possibility of enforcing known physical priors into the graph construction and update. We show that our model learns the kinetic plasma dynamics of the one-dimensional plasma model, a predecessor of contemporary kinetic plasma simulation codes, and recovers a wide range of well-known kinetic plasma processes, including plasma thermalization, electrostatic fluctuations about thermal equilibrium, and the drag on a fast sheet and Landau damping. We compare the performance against the original plasma model in terms of run-time, conservation laws, and temporal evolution of key physical quantities. The limitations of the model are presented and possible directions for higher-dimensional surrogate models for kinetic plasmas are discussed.

\end{abstract}


\section{Introduction}
\label{sec:intro}

Simulating the kinetic behavior of a plasma~\cite{chen1984introduction} is a complex and computationally demanding task. Fully relativistic massively-parallelized Particle-in-Cell (PIC) codes are commonly used to model these phenomena and have been shown to correctly recover and predict plasma collective behavior~\cite{dawson1983particle, birdsall2004plasma, fonseca2002osiris}. 

To obtain computational speed-ups, there have been recent attempts to combine existing PIC codes with machine learning surrogate models. These efforts include approaches to accelerate \cite{kube2021machine} or fully replace\cite{nicolini2019model, aguilar2021deep} the field solver block, reduce the computational burden associated with the particle push and grid-particle/particle-grid interpolation~\cite{nayak2023accelerating, hesthaven2024adaptive},
and the integration of surrogate models into advanced physics extensions~\cite{badiali2022machine}.
In parallel, PIC simulations and machine learning algorithms have also been used to train fast surrogate models for plasma accelerator setups~\cite{djordjevic2021modeling, bethke2021invertible, schmitz2023modeling, sandberg2024synthesizing}, to learn closures for fluid simulations~\cite{joglekar2023machine}, to model hybrid plasma representations~\cite{wu2022learning}, and to recover reduced plasma models~\cite{alves2022data}.  However, obtaining a significant computational gain, while enforcing known physics constraints and reproducing the kinetic effects for a broad range of scenarios, is still an open research question.

Developments in machine learning introduced several physics-inspired surrogate models as an alternative to standard differential equation solvers \cite{chen2018neural, li2020fourier, karniadakis2021physics, brandstetter2022message} and $n$-body or mesh-based simulators ~\cite{sanchez2020learning, pfaff2020learning, zhong2021benchmarking, satorras2021n, brandstetter2022geometric, prantl2022guaranteed, lam2022graphcast}.

From the broad set of available surrogate models, one class of algorithms that can be of particular interest for kinetic plasma simulations are graph neural network-based approaches~\cite{scarselli2008graph, battaglia2018relational, bronstein2021geometric}, because of their capability to model both particle-particle~\cite{sanchez2020learning} and particle-mesh interactions~\cite{pfaff2020learning}, as well as the possibility of enforcing known invariances or symmetries into the network architecture~\cite{bronstein2021geometric, satorras2021n, brandstetter2022geometric}. These approaches have been successfully applied to fluid~\cite{sanchez2020learning, pfaff2020learning, lam2022graphcast}, rigid body~\cite{pfaff2020learning, allen2023learning}, and charged particle dynamics~\cite{cranmer2020discovering, satorras2021n, brandstetter2022geometric}. However, to the best of our knowledge, their applicability to model kinetic plasma simulations is still to be demonstrated. Graph-based representations and graph theory techniques have been explored in other branches of plasma physics, including low-temperature plasmas and plasma chemistry, but mostly for visualizing and reducing chemical reaction mechanisms\mbox{~\cite{holmes2021graph,sakai2022complexity,venturi2023uncertainty}}.

In this work we aim to model the predecessor of the PIC loop, the one-dimensional electrostatic sheet plasma model introduced by Dawson~\cite{dawson1962one, dawson1970electrostatic}. This is an ideal initial testbench since it provides a simpler scenario, in terms of the problem structure and possible computational gains, while capturing a wide range of kinetic plasma physics phenomena that go beyond ``collisionless'' physics, including Coulomb collisions, and collisional thermalization~\cite{dawson1962one, dawson1962some, dawson1964thermal, dawson1968some, dawson1970electrostatic}. Moreover, recent studies in the fundamental statistical physics processes in plasmas have been using the sheet model and/or direct extensions~\cite{gravier2023collision}. We show how to leverage previous work on graph neural network-based simulators by Sanchez-Gonzalez~et~al.~\cite{sanchez2020learning} for kinetic plasma simulations and for the one-dimensional sheet model by introducing domain knowledge into the graph construction and simulator update mechanisms which enforce the desired symmetries and invariances. We discuss the advantages and disadvantages of using our surrogate model when compared to the standard physics simulator in terms of accuracy, run-time, energy conservation, and generalization capabilities. In particular, we will show that when trained on data generated at high temporal resolution data (high-fidelity), the model is able to learn an improved algorithm to resolve sheet crossings (equivalent to resolving Coulomb collisions) for lower temporal resolutions when compared with the traditional physics solver. Futhermore, while the crossing correction algorithm implemented by the traditional solver is a serial routine, the model is also capable of reducing the overall run-time, since its operations are mostly parallelized. Finally, based on our findings, we comment on the expected impact of graph neural network-based simulators for the multi-dimensional PIC scenario.

\section{Electrostatic sheet model}
\label{sec:sm}

\begin{figure}[htb]
    \centering
    \includegraphics[scale=0.7]{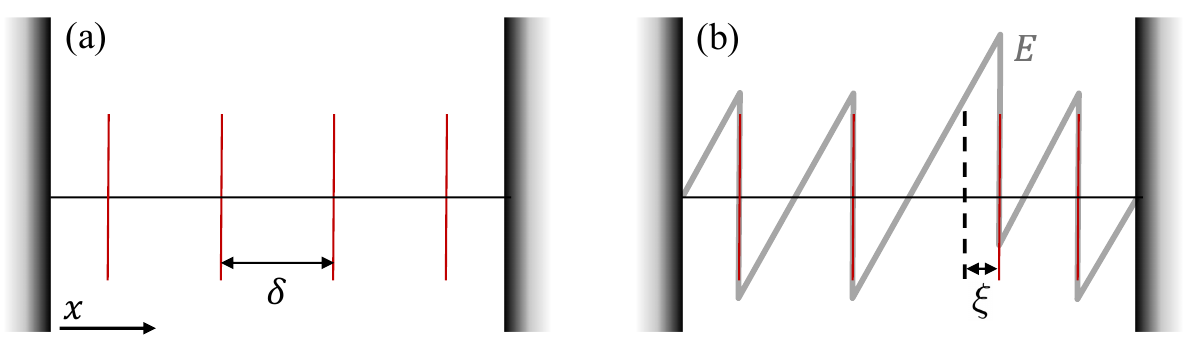}
    \caption{Schematic of the 1D single species electrostatic sheet model. (a)~In equilibrium, the negatively charged sheets (red) are equally spaced inside the box; (b)~If one sheet is displaced from its equilibrium position, the average electric field on the sheet is not zero, due to the charge imbalance. Adapted from Dawson~\cite{dawson1962one}.}
    \label{fig:sheet_model}
\end{figure}

The single-species one-dimensional electrostatic sheet model introduced by Dawson~\cite{dawson1962one, dawson1970electrostatic} represents a plasma as a group of identical negatively charged sheets, moving freely over a uniformly neutralizing positive background (see~\fref{fig:sheet_model}). In a one-dimensional system, this model is exact and describes, within classical physics, the dynamics of a non-relativistic electron plasma. The sheets can be regarded as being composed by electrons, while the background is composed by heavier, immovable ions.

For a system of $N$ sheets inside a simulation box of length $L$ with a background ion number density $n_0$, if the elementary charge is $-e$ for electrons and $+e$ for ions, then the total background charge $Q\!=\!e n_0 L$ must be balanced by $N$ sheets with a total charge of $-Q$. Therefore, each sheet has a charge of $-e n_0 \delta$, with $\delta\!=\!L/N$, and a mass of $m_e n_0 \delta$, where $m_e$ is the mass of the electron.

When the system is in equilibrium (\fref{fig:sheet_model}(a)) the sheets are at rest and equally spaced by $\delta$. In this scenario, the electric field can be represented by a sawtooth function. To understand why, one can apply Gauss's law on a closed surface around a sheet. If the surface enlarges, more background charge is enclosed, which means that the electric field grows linearly along the $x$-axis by a factor of $4\pi e n_0$ (in c.g.s.~units). On the other hand, when the Gaussian surface is large enough to include the next sheet, the enclosed charge drops by $-e n_0 \delta$, and the electric field jumps by $-4\pi e n_0 \delta$.

At their equilibrium positions, the net electric force on each sheet is zero. However, if a sheet moves a certain distance $\xi$ from its equilibrium position (\fref{fig:sheet_model}(b)), it experiences an electric field $E = 4\pi e n_0 \xi$ and a corresponding electric force $F = -e n_0 \delta E$. Since the sheet mass is $m_e n_0 \delta$, the Newtonian equation of motion can be written as:
\begin{equation}
\ddot{\xi} = - \omega_p^2 \xi
\label{eq:sheet_motion}
\end{equation}
where $\omega_p = \sqrt{4 \pi n_0 e^2 / m_e}$ is the plasma frequency. This result implies that, for small displacements, each sheet behaves as an independent harmonic oscillator.

\begin{figure}[t]
    \centering
    \includegraphics[width=\textwidth]{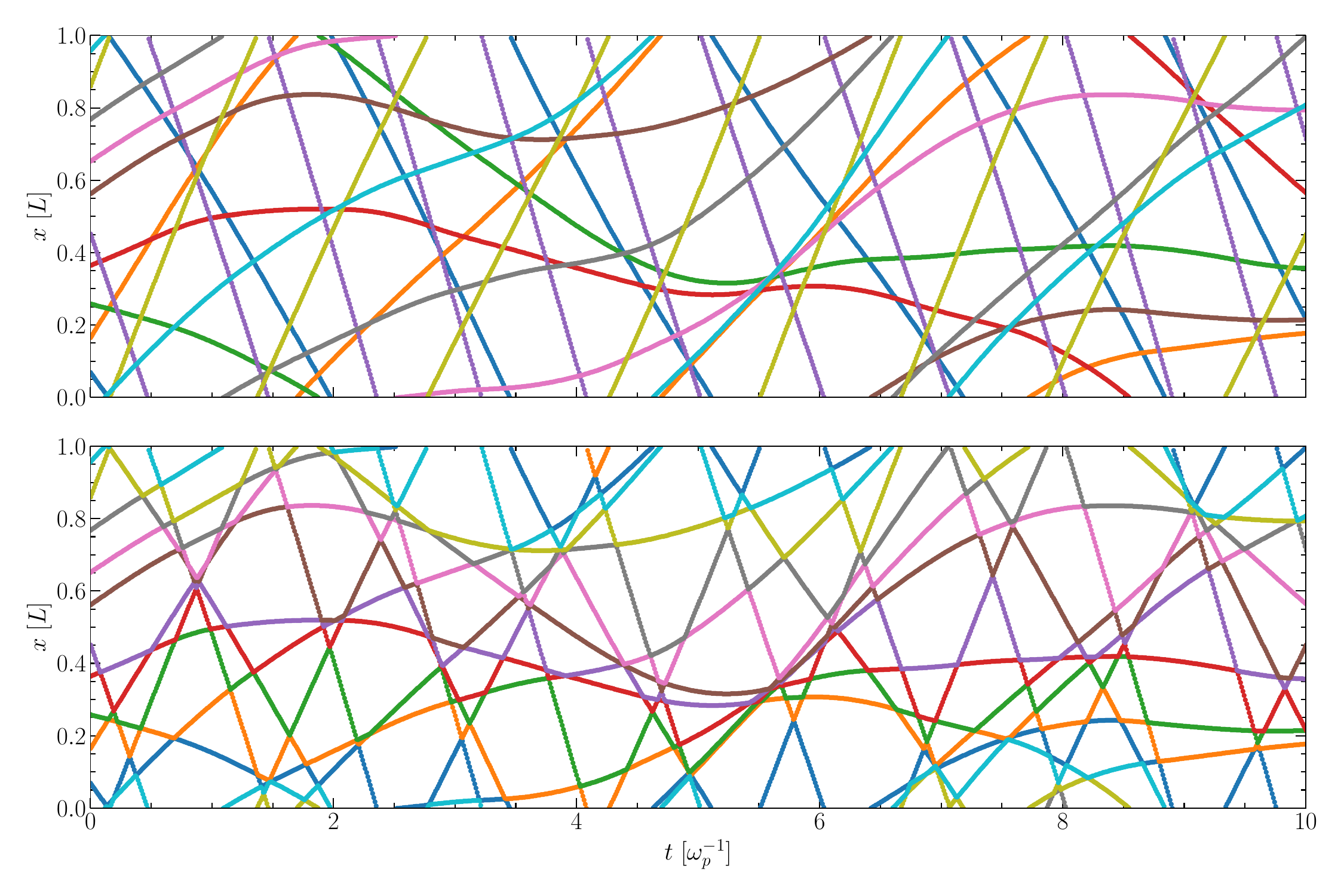}
    \caption{Comparison of charged sheet trajectories when considering sheet interactions as crossings (top) \textit{versus} binary collisions (bottom). The system consists of 10 sheets (represented by different colors) moving on a periodic box of length $L$. Initial velocities were randomly chosen. We will learn to model the dynamics of the first case (crossings) as this is considerably easier, mainly due to the smoothness of the sheet trajectories. More details on the difficulties that arise when attempting to learn collisional dynamics are provided in~\ref{app:collisions}.}
    \label{fig:crossings_vs_collisions}
\end{figure}

For larger displacements, it is possible that consecutive sheets cross each other, corresponding to a one-dimensional Coulomb collision, meaning their equilibrium positions switch. This interaction can also be modeled as an elastic collision, i.e. one can simply switch the velocities of the sheets at the instant of crossing (instead of their equilibrium positions) as it results from the conservation of energy and momentum. An illustration of the difference in the resulting individual trajectories is provided in \fref{fig:crossings_vs_collisions}. 

To simulate this system, two computational directions can be used~\cite{dawson1970electrostatic}: a synchronous method, and an asynchronous method. The synchronous method first updates the sheet dynamics according to~\eqref{eq:sheet_motion} considering a fixed $\Delta t$. It then detects crossings by testing the condition $x_i^{t+1} > x_{j}^{t+1}$ for $j > i$, and proceeds to use an iterative method to estimate the crossing times and correct the motion of the corresponding sheets (more details are provided in~\ref{app:sheetmodel}). This method does not correctly resolve crossings involving more than two sheets in a single time step, which leads to an overall energy loss in the system if the time step is too large compared with the inverse of the typical collision frequency. For this reason, for higher sheet velocities it is necessary to use smaller simulation time steps (the collision frequency increases with increasing thermal velocity).

The asynchronous method advances the simulation until the next crossing. The next crossing time can be predicted analytically from (\ref{eq:sheet_motion}) by solving for $x_i(t) = x_j(t)$ with respect to $t$ for all pairs of adjacent sheets. This algorithm guarantees energy conservation but implies additional computational effort since a sorted table containing all expected crossing times between neighboring sheets must be pre-computed and updated after each crossing is resolved.

Since the Graph Network Simulator (GNS) is a synchronous model, we use the synchronous sheet model algorithm (SM, illustrated in~\fref{fig:simulators}) for both data generation and testing purposes to allow for accuracy comparisons at equivalent simulation time steps. Additionally, we introduce a synchronous Modified Sheet Model algorithm (MSM) which only checks for crossings with at most the $n$-th neighboring sheets. This allows us to compare the GNS with an algorithm that only has access to the equivalent amount of neighboring sheets when correcting for crossings. For completion, we also implement the asynchronous version of the sheet model algorithm (ASM) for run-time comparisons. More details regarding the different implementations are provided in~\ref{app:sheetmodel}.

\section{Graph Network Simulator}
\label{sec:gns}

\begin{figure}[t]
    \centering
    \includegraphics[width=\columnwidth]{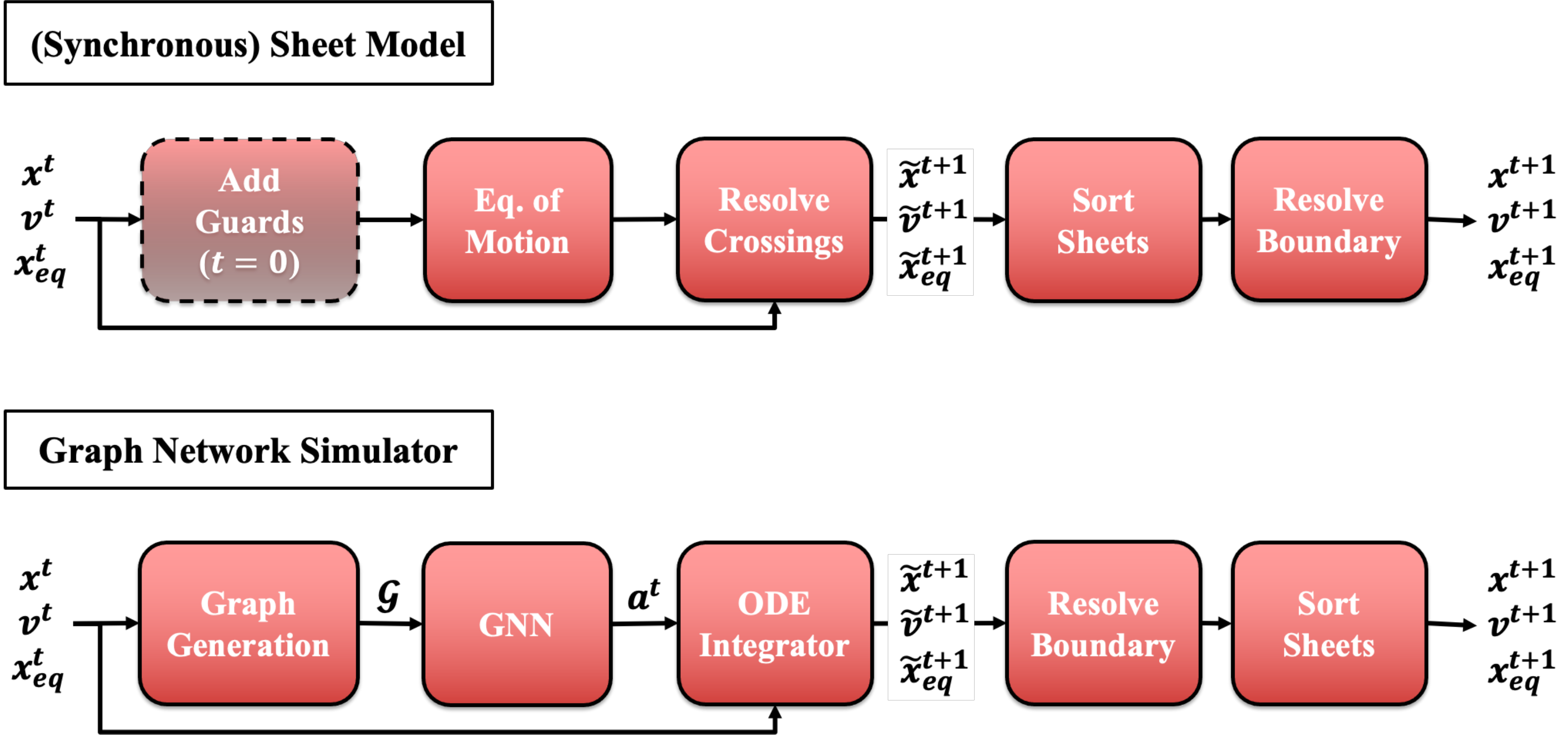}
    \caption{Schematic representations of the synchronous electrostatic sheet model algorithm~\cite{ dawson1970electrostatic} and the proposed Graph Network Simulator.}
    \label{fig:simulators}
\end{figure}

The GNS architecture used here is inspired by the work of Sanchez-Gonzalez et~al.~\cite{sanchez2020learning}, while taking into consideration the specifics of the electrostatic sheet model. The main building blocks are presented in~\fref{fig:simulators}.

Based on the sheet positions $\bi{x}^t$, velocities $\bi{v}^t$, equilibrium positions $\bi{x}_{eq}^t$, and boundary conditions, we generate a graph representation $\mathcal{G}$ of the plasma. A Graph Neural Network (GNN) predicts each individual sheet acceleration $\bi{a}^t$ which will be used to update the positions $\bi{x}^t$ and velocities $\bi{v}^t$. We enforce the boundary conditions by re-injecting sheets that crossed the boundary, sorting the sheets by their position, and updating their equilibrium positions. This process can be repeated to generate longer simulation rollouts.

Our GNS does not treat sheet interaction as binary collisions. Instead, we learn to predict the changes in velocity as sheets move through one another. This choice makes it significantly easier for the network to learn the dynamics of the system and also reduces the graph and model complexity. We provide this comparison, as well as the structural changes required to the simulator when considering crossings as collisions, in~\ref{app:collisions}.

\subsection{Graph representation}
\label{sec:graph_representation}

\begin{figure}[t]
    \centering
    \includegraphics[width=0.8\columnwidth]{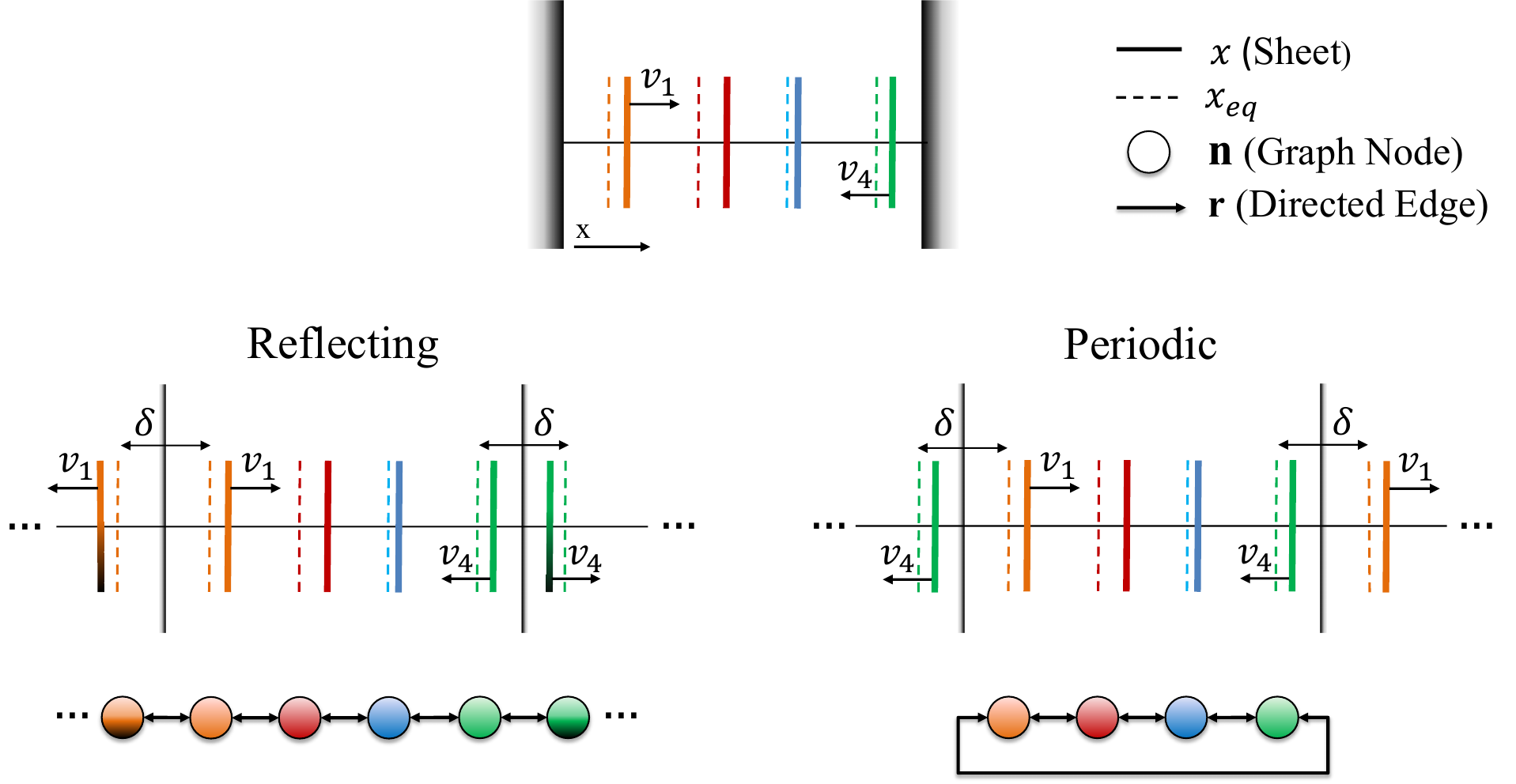}
    \caption{Graph representation of a four-sheet system for different boundary conditions. Each graph node corresponds to a physical sheet. Directed edges connect neighboring nodes and represent the possible interaction between sheets. Computationally, nodes are represented by a vector and edges by a scalar according to equation~\eqref{eq:graph_vectors}. For the sheet model, boundaries are modeled using \textit{guard sheets}~\cite{dawson1970electrostatic}. For reflecting boundaries they represent mirrored versions of the sheet(s) closer to the boundary, for periodic boundaries they represent equal versions of the sheet(s) closer to the opposite wall. For the GNS, we adopt a similar representation for reflecting boundaries by adding guard nodes representing mirrored versions of the sheets closer to the boundary. The number of guard nodes is set to the number of message-passing steps of the GNN. For periodic boundaries, we instead add directed edges between the first and last node.}
    \label{fig:sheet_graph}
\end{figure}

The plasma is represented as a graph $\mathcal{G}$ with a set of nodes $\{\mathbf{n}_i\}$ representing the negatively charged sheets, and a set of directed edges~$\{r_{ij}\}$ which connect neighboring sheets (\fref{fig:sheet_graph}). In our case, we opted to connect only the first neighbors (additional comments on the trade-off between higher connectivity and a deeper GNN are provided throughout the text). Each node $\mathbf{n}_i$ is represented by a vector containing the information relative to the corresponding negatively charged sheet, while each edge $r_{ij}$ contains the relative displacement of sheet $i$ in relation to sheet $j$. They are defined as:
\begin{equation}
\eqalign{
    \mathbf{n}^t_i &= \left[\xi^t_i, v^t_i\right] \cr 
    r^t_{ij} &= x^t_j - x^t_i 
    \label{eq:graph_vectors}
}
\end{equation}
where $\xi_i^t$ corresponds to the displacement of the $i$-th sheet from its equilibrium position $\left(x^t_i - x^t_{eq_i}\right)$, and $v^t_i$ the finite difference velocity $\left(x_i^t - x_i^{t-1}\right)/\Delta t$. To allow the model to generalize to different box sizes and number of sheets, we normalize all distances and velocities by the intersheet distance at equilibrium~$\delta$. This transformation makes the inputs of the network invariant to the system size (box length and number of sheets). This is also the reason why we include in the node vector the displacement from equilibrium instead of the sheet position inside the box, which additionally enforces an invariance to the sheet rank.

When considering reflecting boundary conditions, extra ``guard'' nodes representing mirrored versions of the sheets closer to the respective boundary are added to the graph~(\fref{fig:sheet_graph}, mirrored versions follow $\xi^{m}_i = -\xi_i$, $v^{m}_i = -v_i$, for the left boundary $x^{m}_{eq_i} = x_{eq_i} - i\cdot \delta$, for the right boundary $x^{m}_{eq_i} = x_{eq_i} + (N + 1 - i)\cdot \delta$)). 
Ideally, the number of mirror sheets should be large enough so that no boundary sheet crosses all mirror sheets in a single time step (in our case we set it to the number of message-passing steps, which we cover in Section~\ref{sec:gnn}). On the other hand, for periodic boundaries, the graph becomes cyclic by introducing edges between the first and last nodes and considering the distance between the corresponding sheets to be equal to the distance through the walls of the simulation box ($r_{1N} = -r_{N1} = x_N - x_1 - L$). In subsequent sections, we show that representing boundaries in this way allows the GNN model to be applied to different boundary conditions than the ones it was trained on, since it learns solely interactions between sheets (not interactions with the wall). However, this comes at the cost of an extra hard-coded step, which re-injects sheets that crossed the boundary (procedure to be explained in more detail in later sections).

We experimented with several other possible representations of the system, (e.g. different boundary representations, node and edge parameters) but these variants either produced worse results (in terms of accuracy or generalization capabilities) or introduced extra complexity and memory requirements that did not result in meaningful accuracy and/or speed-up improvements for the tested scenarios. Connecting $n$-nearest neighbors allowed us to achieve similar accuracies for shallower GNNs (since information propagates faster through the graph). However, it introduced extra memory requirements. It is therefore a trade-off that should be taken into account for future test scenarios.

\subsection{Graph Neural Network architecture}
\label{sec:gnn}

\begin{figure}[t]
    \centering
    \includegraphics[width=0.8\columnwidth]{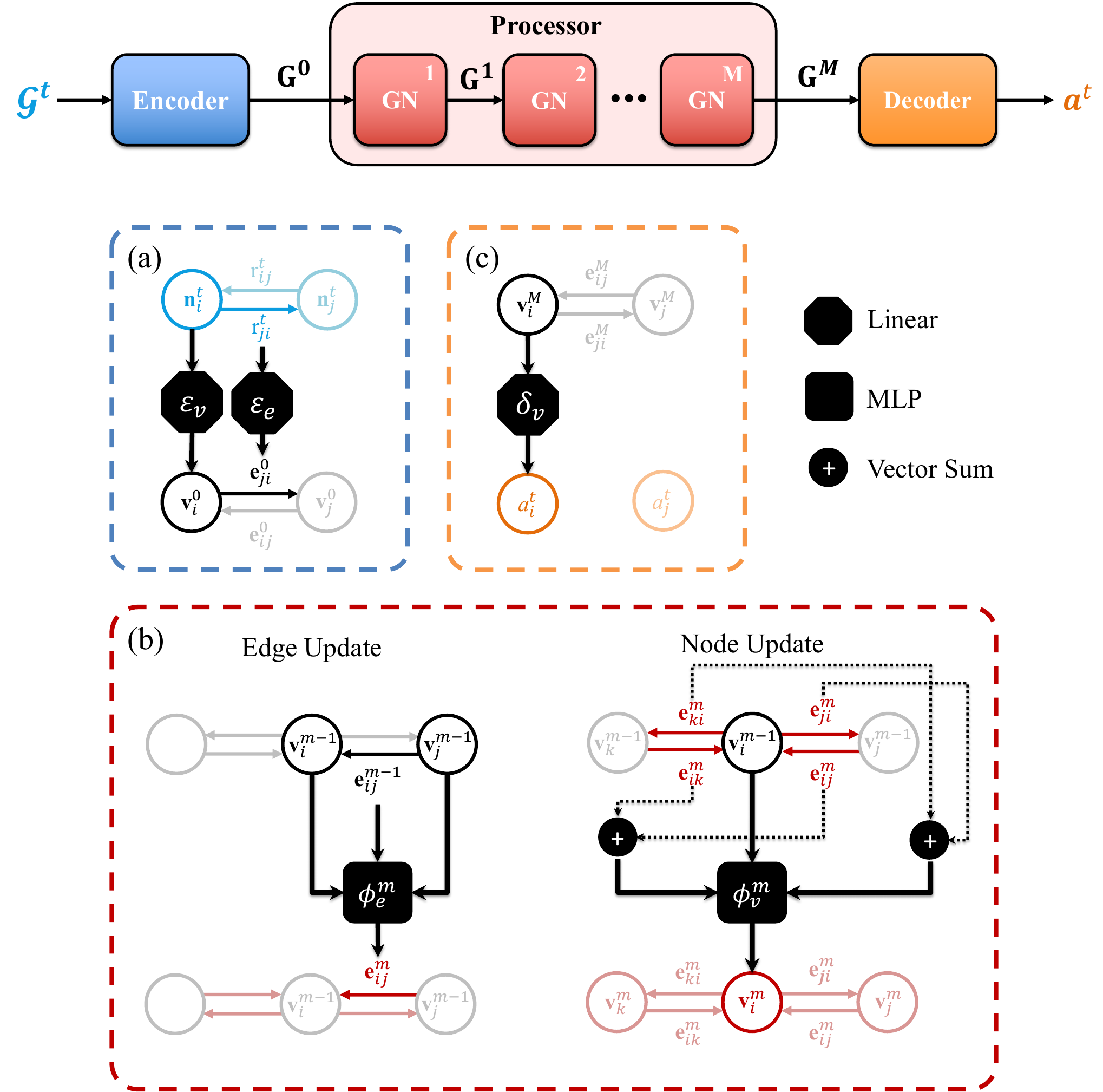}
    \caption{Schematic of the GNN encoder-processor-decoder architecture.  The processor block is composed of $M$ independent Graph Network (GN) blocks similar to~\cite{battaglia2018relational}. (a) The encoder block converts the input graph $\mathcal{G}^t$ node/edges ($\mathbf{n}_i^t$, $r_{ij}^t$) into a higher-dimension latent space graph representation $\mathbf{G}^0$ with node/edge vectors ($\mathbf{v}_i^0$, $\mathbf{e}_{ij}^0$). (b) Each $m$-th GN block updates the latent edge and node vectors of $\mathbf{G}^{m-1}\rightarrow\mathbf{G}^m$. The update functions $\phi^m_{e,v}$ are parameterized by a Multi-Layer Perceptron (MLP, in our case a 2-layer dense neural network). MLP weights are different for each GN block. (c) The decoder block computes the sheet acceleration $a_i^t$ from the corresponding final latent node representation $\mathbf{v}_i^M$. The linear transformations in the encoder/decoder block and the MLP weights from the GN blocks are learned during training.}
    \label{fig:gnn_architecture}
\end{figure}

The GNN module used (\fref{fig:gnn_architecture}) follows an encoder-processor-decoder architecture as introduced by Battaglia et~al.~\cite{battaglia2018relational}. This architecture allows us to model the interactions between sheets by performing a series of message-passing steps between the connected nodes in the graph (representing the sheets) and therefore predict each individual sheet acceleration for the corresponding time step (which includes the contribution from the crossings and its oscillatory motion). The main building blocks are formalized as follows.

\vspace*{+10pt}
\noindent\textbf{Encoder}: The encoder (\fref{fig:gnn_architecture}a) transforms the graph $\mathcal{G}^t$ nodes $\mathbf{n}_i \in \mathbb{R}^2$ and edge $r_{ij} \in \mathbb{R}$ vectors into a (higher-dimensional) latent space graph representation $\mathbf{G}^0$, whose latent nodes $\mathbf{v}_i \in \mathbb{R}^{\mathrm{L}}$ and latent edges $\mathbf{e}_{ij}\in \mathbb{R}^{\mathrm{L}}$, are given by:
\begin{equation}
\eqalign{
    \mathbf{v}_i = \varepsilon_v\left(\mathbf{n}_{i}\right) \cr
    \mathbf{e}_{ij} = \varepsilon_e\left(r_{ij}\right)
}
\end{equation}
where $\varepsilon_v$ and $\varepsilon_e$ are learnable functions and L is the latent space size. \\

\textbf{Processor}: The processor consists of a series of $M$ Graph Network (GN) blocks (\fref{fig:gnn_architecture}b) adapted from~\cite{battaglia2018relational}. (These blocks were modified to include the sent edge information in the node update function, which is equivalent to the GN implementation available in Jraph~\cite{jraph2020github}.) Each $m$-th GN block updates the latent graph outputted by the previous layer $\mathrm{GN}^m(\mathbf{G}^{m-1})\rightarrow \mathbf{G}^{m}$ according to:
\begin{equation}
\eqalign{
    \mathbf{e}_{ij}^{m} & = \phi^{m}_e\left(\mathbf{e}^{m-1}_{ij}, \mathbf{v}^{m-1}_i, \mathbf{v}^{m-1}_j\right) \cr 
    \overline{\mathbf{e}}_{r_i}^{m} &= \sum_{j \in \mathcal{N}(i)}  \mathbf{e}^{m}_{ij} \cr
    \overline{\mathbf{e}}_{s_i}^{m} &= \sum_{j \in \mathcal{N}(i)}  \mathbf{e}^{m}_{ji} \cr
    \mathbf{v}_{i}^{m} &= \phi_{v}^{m}\left(\overline{\mathbf{e}}^{m}_{r_i}, \overline{\mathbf{e}}^{m}_{s_i}, \mathbf{v}^{m-1}_i\right) \label{eq:gn}
}
\end{equation}
where the superscript denotes the block number, $\mathcal{N}(i)$ the set of nodes connected to $i$, and $\phi^m_v$, $\phi^m_e$ are learnable functions. The value of $M$ is set depending on the training time step and the maximum velocity of the sheets present in the training simulations. Ideally, $M$ should be larger than the maximum number of neighboring sheets that a given sheet crosses in any particular time step, since a graph node will at most receive/send information from/to the $M^{\text{th}}$ neighbor. This condition can be relaxed if the $N$-nearest neighboring nodes are directly connected (with the cost of additional memory requirements), and we have observed similar performance in our tested scenarios for models with an equivalent $M\!\times\!N$ factor. However, for higher crossing frequencies than the ones presented in this work, it might be preferable to increase $N$ to avoid possible information bottlenecks~\cite{topping2022understanding}.\\

\noindent\textbf{Decoder}: The decoder block (\fref{fig:gnn_architecture}c) transforms the node vectors of the last latent graph $\mathbf{G}^M$ into a scalar:
\begin{equation}
    y_i = \delta_v\left(\mathbf{v}_i^M\right) \equiv a_i^t
\end{equation}
where $\delta_v$ is a learnable function. In our case, the output $y_i$ is a single real value that corresponds to an estimate of the individual finite difference sheet acceleration $a^t_i = (v^{t+1} - v^{t})/\Delta t = (x^{t+1} - 2x^{t} - x^{t-1})/\Delta t$ normalized to the intersheet spacing $\delta$. \\

We parameterize the encoder and decoder functions ($\varepsilon_e$, $\varepsilon_v$, $\delta_v$) as (learnable) linear transformations. As for the processor functions ($\phi^m_e$, $\phi^m_v$), they are given by two-layer dense neural networks (one for each GN block) following: Input $\rightarrow$ \{LinearLayer  $\rightarrow$ ReLU $\rightarrow$ LinearLayer\} $\rightarrow$ Output. In every block (encoder, processor, and decoder), we use a latent space of size~$128$ (all hidden and output layers have this dimension). A summary of the hyper-parameter tuning experiments that led to these final values is provided in~\ref{app:parameter_scans}.

Although this GNN architecture, by design, does not enforce equivariance with respect to reflections over an equilibrium position (a symmetry present in the sheet model, i.e.\ if the simulation box is flipped the absolute value of the predicted accelerations should simply switch signs), we observed that the network was nonetheless capable of correctly approximating this symmetry within the training data range. In fact, we developed an alternative architecture that enforced this equivariance but did not observe relevant gains concerning the required number of training simulations nor improved rollout accuracy or energy conservation capabilities (in fact we observed a deterioration for considerable out-of-training data distribution values). Similarly, not using the sent messages for the node update mechanism led to poorer energy conservation for (considerably) out-of-training data distribution values. More details about these comparisons and the equivariant architecture are provided in~\ref{app:equivariantgnn}.

\subsection{Position and velocity update}

The predicted accelerations $\bi{a}^t$ provided by the GNN decoder are used to update the sheets dynamics. For this purpose, we use a semi-implicit first-order Euler integration scheme as follows:
\begin{equation}
\eqalign{\tilde{v}_i^{t+1} &=  v_i^t + a_i^t \Delta t  \cr
    \tilde{x}_i^{t+1} &= x_i^t + \tilde{v}_i^{t+1} \Delta t \ } 
\end{equation}
which corresponds to the ``ODE Integrator'' block in \fref{fig:simulators}.

After this update, we resolve the boundary crossings. When considering reflecting boundary conditions, we flip the position and velocity of the sheets that went beyond the simulation box. No change is applied to their equilibrium positions. When considering periodic boundaries, we instead re-insert the sheets through the opposite boundary without changing their velocities. Additionally, the equilibrium positions are updated to take into consideration sheets that crossed the boundaries (additional information is provided in \ref{app:boundary_crossings}).

Finally, we sort the sheets by their position inside the box. This step is required to correctly attribute equilibrium positions and ensure the necessary relative ordering for graph construction. 

\section{Implementation}
\label{sec:implementation}

For reference, we implemented the synchronous version of the original electrostatic sheet model~\cite{dawson1962one, dawson1970electrostatic} in Python, using NumPy~\cite{harris2020array}. This code is used to generate all ground truth training and test data at a high temporal resolution.  The modified synchronous algorithm and the asynchronous algorithm were implemented in a similar fashion and are only used for testing purposes. The GNS was also implemented in Python using JAX~\cite{jax2018github}, Jraph~\cite{jraph2020github}, Haiku~\cite{haiku2020github}, and Optax~\cite{deepmind2020jax}.

Additionally, from here onwards we will adopt a system of units similar to Dawson~\cite{dawson1962one}. Time will be shown in units of the plasma period $\omega_p^{-1}$ (with $\omega_p$ as defined right after~\eqref{eq:sheet_motion}), distances will be presented in units of the intersheet spacing in equilibrium $\delta$, resulting in velocities in units of $\delta\!\cdot\!\omega_p$ and accelerations in units of $\delta \!\cdot\! \omega_p^2$. 
Note that in the adopted units the Debye length $\lambda_D$ is equivalent to the thermal velocity since, by definition, $v_{th} = \lambda_D\omega_p$~\cite{chen1984introduction}, and the length of the simulation box $L$ is equivalent to the number of sheets since $L = N_{sheets}\delta$.

\subsection{Generating the ground truth data}
\label{sec:training}

Using the electrostatic sheet model, we generate 10,000 simulations of systems consisting of 10 sheets moving inside a periodic box.
We use only 10 sheets for training to emphasize the capabilities of the simulator to generalize to significantly larger system sizes (over several orders of magnitude) at test time. It would be possible to train with a larger system size, and this would allow using a smaller training set. On the other hand, the opposite idea of reducing the system size even further is not advisable, since this could limit the capability of the GNS to learn crossings involving multiple sheets (to be explained in more detail in later sections).

All training simulations are run for a duration of $t_{max}=10~\omega_p^{-1}$ using a time step of $\Delta t_{sim} = 10^{-4}~\omega_p^{-1}$. The initial displacements from the equilibrium positions and velocities of the sheets are randomly sampled from uniform distributions. The maximum initial displacement equals $\xi_{max}^0=0.2~\delta$ and the maximum initial velocity is $v_{max}^0=10~\delta\!\cdot\!\omega_p$. In addition, we ensured that the total energy of the system did not vary more than a predefined threshold $\left(\Delta\varepsilon/\varepsilon_0 = 10^{-6}\right)$ during the full simulation by discarding simulations that did not fulfill this criterion. This guarantees that all crossings are well resolved by the sheet model. We have also tested training models using synchronous data generated from asynchronous simulations (with energy variations of $\Delta\epsilon/\epsilon_0\ < 10^{-10}$) but did not observe performance differences.

\subsection{Data preprocessing and augmentation}
\label{sec:training2}

Before training the GNN models, we apply the following preprocessing steps. First, we downsample the data to the desired training time step (e.g. $\Delta t_{train}=10^{-2}~\omega_p^{-1}$). To take advantage of the system symmetries, we proceed to augment the training dataset by mirroring the simulations along the $x$- and time-axis (the latter is not equivalent to simply changing the sign of the velocities since we are using the finite difference velocity). We proceed to generate pairs of input graphs and output target accelerations, where each input graph corresponds to a full simulation rollout (and corresponding augmented versions). More details on the impact of the training dataset size and data augmentation are provided in~\ref{app:dataset_size}.

\subsection{Training}
\label{sec:training3}

To train the models, we hold out 100 simulations for validation purposes. We proceed to minimize the mean squared error between the predicted and target accelerations using the Adam optimizer. We use an exponential learning rate scheduler, similarly to Sanchez-Gonzalez et~al.~\cite{sanchez2020learning}, for which $\alpha(j) =  \alpha_{final} + (\alpha_{start} - \alpha_{final}) \cdot 0.1^{j\cdot10^6}$, where $j$ represents the gradient update step, and the initial and final learning rates are given by $\alpha_{start}=10^{-4}$ and $\alpha_{final}=10^{-6}$. We set the batch size to~$1$ (one graph corresponds to a full simulation) and compute the validation loss on the full validation set once a full cycle over the training dataset is completed. The training procedure is then run to a maximum of $1 \times 10^6$ gradient updates for $\Delta t_{train} = 10^{-1}~\omega_p^{-1}$, and $1.5 \times 10^6$ gradient updates for $\Delta t_{train} = 10^{-2}~\omega_p^{-1}$. The final weights of the model are those obtained for the smallest recorded validation loss. The full training procedure lasts approximately $4$~hours for $\Delta t_{train}=10^{-1}~\omega_p^{-1}$ and $M=5$, and 1~day for $\Delta t_{train}=10^{-2}~\omega_p^{-1}$ and $M=3$, on a single Nvidia Titan X GPU. For each value of $\Delta t_{train}$ we train 5 equivalent models using different random seeds in order to assess the dependence of performance on weight initialization.

\begin{figure}[tb]
    \centering
    \includegraphics[width=0.95\columnwidth]{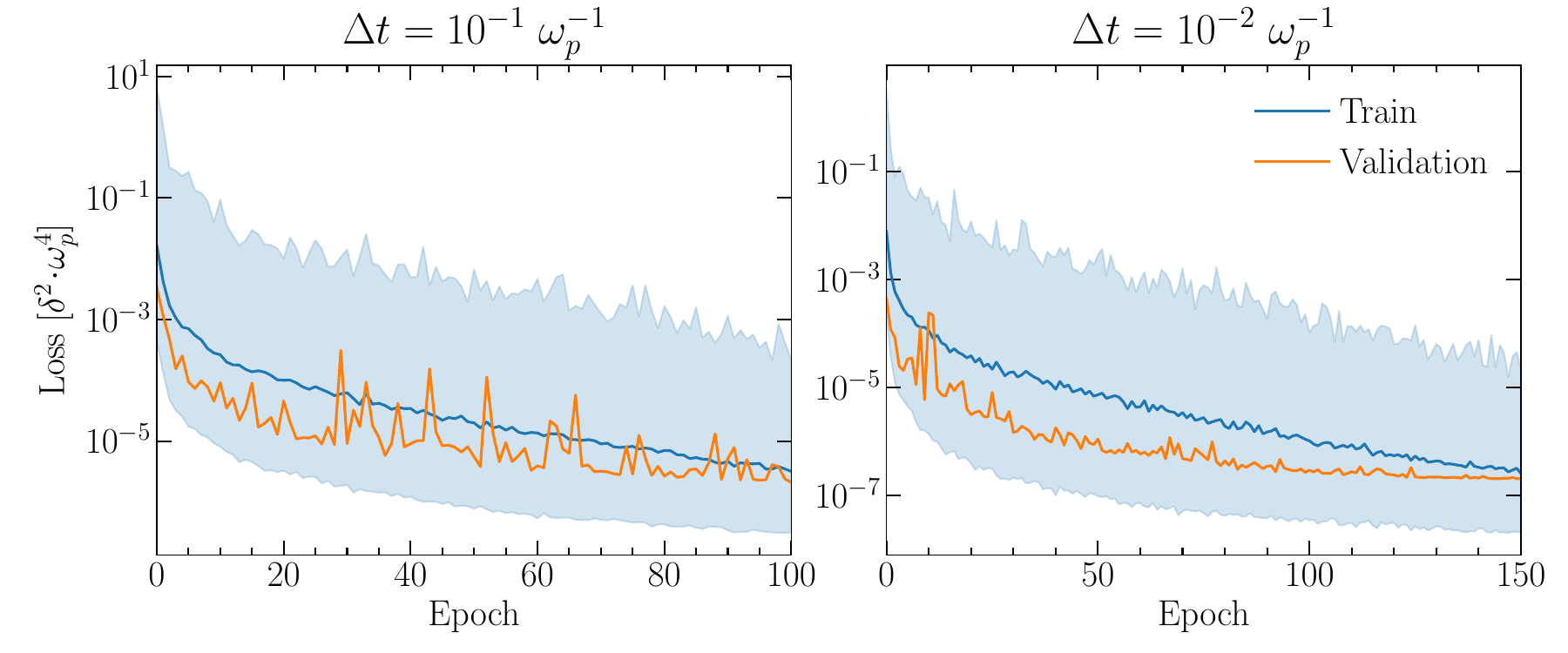}
    \caption{Examples of the training and validation loss evolution during training. For each epoch, $9900$ gradient updates are performed (1 per training simulation). The shaded area represents the minimum and maximum train loss across the batches for the corresponding epoch, mean value is presented in full line. Validation loss is computed at the end of the epoch over 100 held-out simulations. The results show that the training loss, on average, decreases monotonically and that the validation loss plateaus close to the defined epoch limit. No relevant gains were observed for longer training runs.}
    \label{fig:train_curves_ex}
\end{figure}

Examples of the obtained training loss curves are depicted in \fref{fig:train_curves_ex}. We observe a similar behavior across other models, with the validation loss stabilizing close to the last epoch. The learning rate scheduler is particularly helpful in the later stage of training to reduce the training loss oscillations across batches.

\section{Model benchmark}
\label{sec:benchmark}

In this section, we assess the capability of the GNS to predict individual sheet trajectories. We showcase the generalization capabilities already hinted at in Section~\ref{sec:graph_representation} by evaluating the model accuracy for systems of different sizes and boundary conditions. Additionally, we compare its energy conservation capabilities and run-time against the different sheet model algorithms and discuss the identified GNS limitations.

\subsection{Trajectory prediction error}
\label{sec:trajectoryprediction}

In order to benchmark the rollout accuracy and generalization capabilities of the GNS, we evaluate its accuracy on multiple test sets consisting of systems with different numbers of sheets and boundary conditions. Each test set contains 100 simulations with a similar duration, temporal resolution, maximum initial displacement and velocity as the ones present in the training set. Our evaluation metrics are the rollout mean absolute error (MAE) and the earth mover's distance (EMD)~\cite{villani2009optimal} between the predicted and the ground truth sheet trajectories, calculated, for each time step, as
\begin{equation}
    \eqalign{
    \mathrm{MAE} =&\ \frac{1}{N}\sum_i^{N} \left|x_i^{GNS} - x_i^{True}\right|\cr
    \mathrm{EMD} =&\ \min \frac{1}{N} \sum_i^{N}\sum_j^{N} \left|x_i^{GNS} - x_j^{True}\right| \cdot c_{ij}, \cr &\ \mathrm{subject \ to} \ \sum_i c_{ij} = 1 \ , \ \sum_j c_{ij} = 1 \ , \ \mathrm{and} \
    c_{ij} \in \{0,1\}
    }
\end{equation}
and then averaged over the full simulation rollout. In the case of periodic boundaries, we modify both metrics to consider the absolute distance to be the minimum of the distances through the box or through the walls.

\begin{figure}[tb]
    \centering
    \includegraphics[width=\columnwidth]{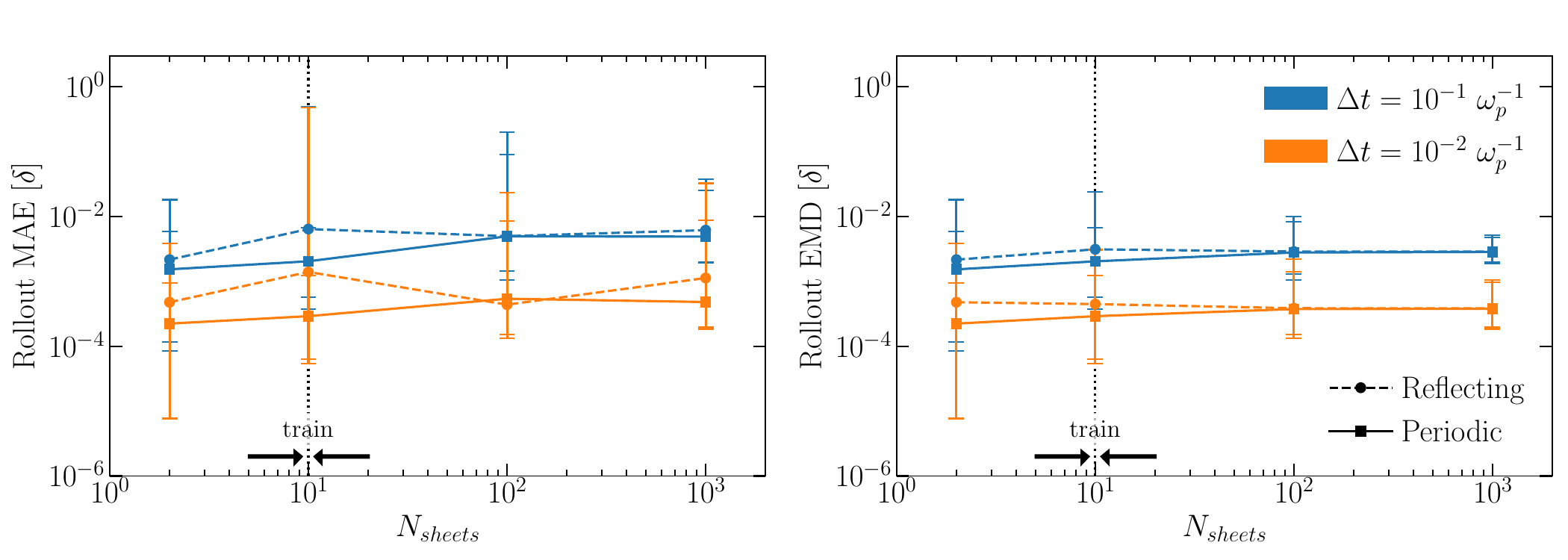}
    \caption{Rollout error metrics for the GNS in the test set simulations. For each value of $\Delta t$ we compute the metrics for 5 equivalent GNNs trained using different random seeds. The presented mean values are computed by averaging over sheets, time steps, simulations, and GNN models (for a detailed comparison between different models see~\ref{app:seeds}). The error bars represent the minimum and maximum rollout error achieved for the corresponding set  of test simulations across all models. The results demonstrate that even though the training data contains solely  systems consisting of 10 sheets moving over a periodic box, the GNS is capable of generalizing to smaller/larger system sizes and different boundary conditions. Furthermore, the reported errors are considerably small.}
    \label{fig:benchmark_rollout}
\end{figure}

\begin{figure}[htb]
    \centering
    Periodic Boundary - $\Delta t_{GNS} = 10^{-2}~\omega_p^{-1}$ 
    \includegraphics[width=\columnwidth]{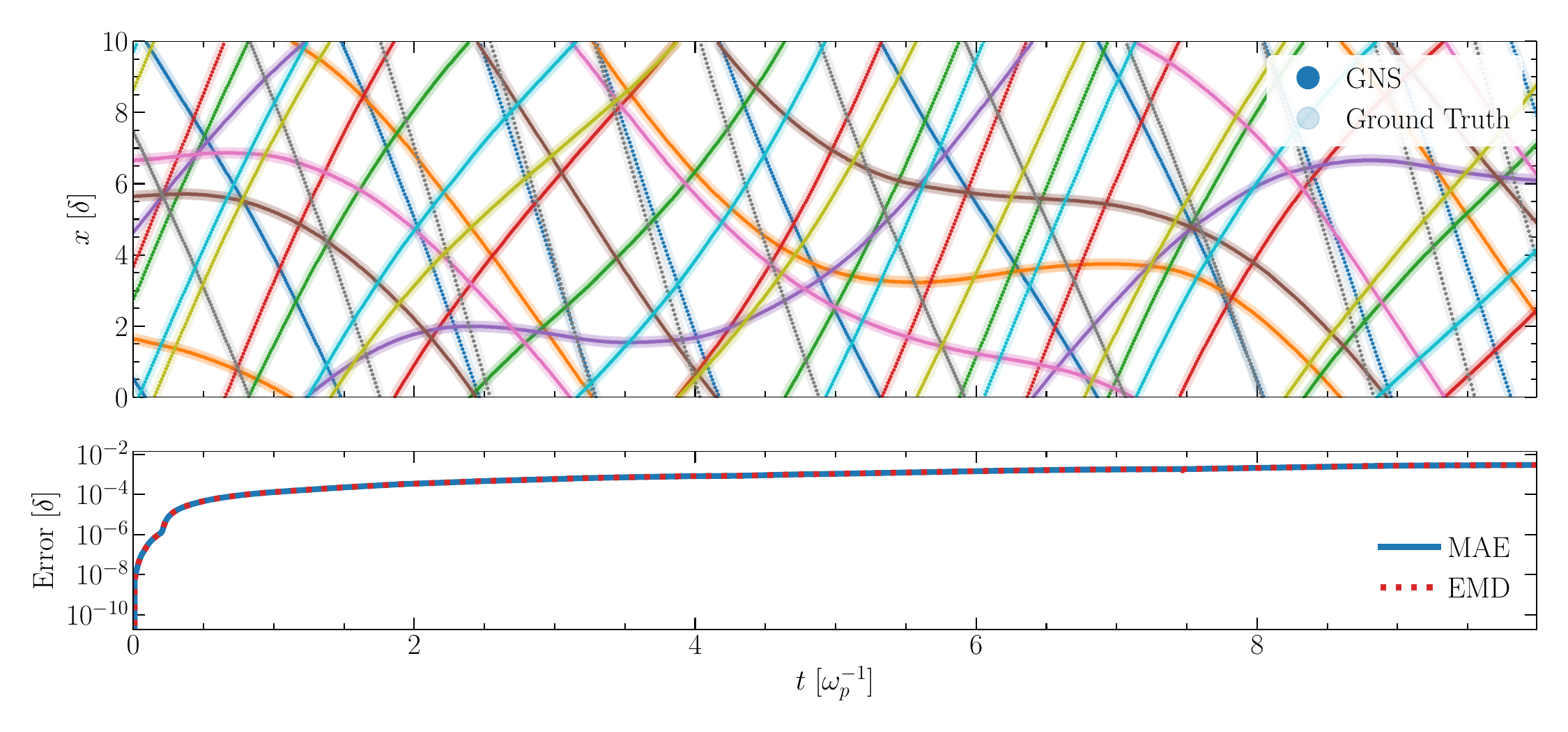} \\
    Reflecting Boundary - $\Delta t_{GNS} = 10^{-1}~\omega_p^{-1}$ 
    \includegraphics[width=\columnwidth]{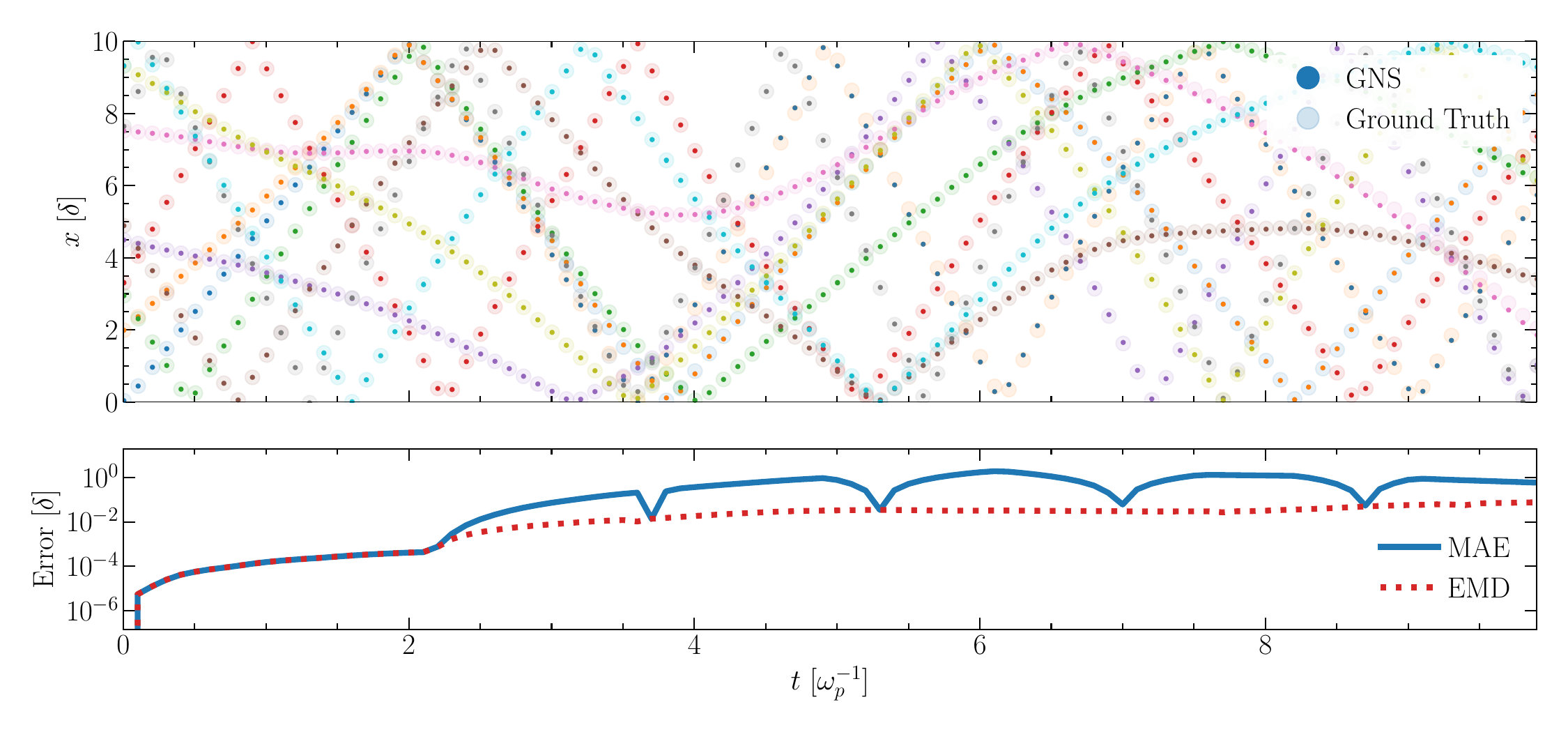} 
    \caption{Example of simulation rollouts observed for test simulations of 10-sheet systems. These examples correspond to the worst-performing rollouts (largest EMD across all models) for the indicated simulator time step and boundary conditions. The predicted and ground truth trajectories and the MAE/EMD evolution are shown (per time step average over sheets). In both cases, the ground truth trajectories (obtained with the Sheet Model using $\Delta t = 10^{-4}~\omega_p^{-1}$) are downsampled to the same simulation time step as the GNS. We plot the ground truth trajectories with a larger marker size in order to be possible to distinguish them with respect to the prediction. It is clear that the GNS is capable of correctly modeling the sheet trajectories for longer rollouts, even though it was solely trained on single-step prediction.}
    \label{fig:example_sim}
\end{figure}

The results presented in \fref{fig:benchmark_rollout} allow us to draw some conclusions.
We observe the rollout errors obtained are considerably small (note that they are presented in units of the intersheet spacing $\delta$) demonstrating that, despite training solely on single-step acceleration prediction, we achieve a stable rollout accuracy. To provide additional insight into the small scale of the errors, we showcase in~\fref{fig:example_sim} the worst test simulation rollouts (highest rollout EMD across all models) for different time steps and boundary conditions. 

We also observe the simulator accuracy is independent of the number of sheets and boundary conditions, without additional re-training/fine-tuning. These invariances, already hinted at in Section~\ref{sec:graph_representation}, are a direct consequence of the chosen graph representation of the system.

Finally, \fref{fig:benchmark_rollout} illustrates the importance of using both the MAE and the EMD as complementary evaluation metrics. The larger intervals associated with the MAE are produced by a small set of simulations where, due to the accumulation of small prediction errors, two sheets switch their expected relative order during a tangential crossing (i.e. when they are moving in the same direction with very similar velocities). This results in a permutation of their predicted trajectories with respect to their ground truth, leading to larger MAE values (see for example \fref{fig:example_sim} for reflecting boundary, at $t=2~\omega_p^{-1}$ the orange and dark blue trajectories permute after reflection). The reason why the error intervals decrease significantly for the EMD case is because this metric is invariant to permutations of sheet trajectories. This invariance (which the MAE does not provide) is an important property for our case study since a simple permutation of sheet trajectories does not change the distribution function of the system (i.e.~the systems are equivalent). Therefore, the EMD provides a better assessment of the accuracy of the simulator to model the collective plasma dynamics.

We observed that, overall, equivalent models trained with different random seeds converge to very similar rollout errors (detailed comparisons provided in~\ref{app:seeds}). The only exception was one of the models trained for $\Delta t=10^{-2}~\omega_p^{-1}$ which revealed a slightly worse rollout performance. We attribute this larger error to its worse single-step prediction capabilities given the validation loss at train time was approximately double that of equivalent models.

\subsection{Energy conservation}
\label{sec:energy_conservation}

In order to check for energy conservation, we run simulations using two types of initial velocity distributions: thermal -- velocities sampled from a normal distribution with standard deviation equal to $v_{th}$; and oscillation -- sheets share the same initial velocity $v_0$ (no crossings should occur). For both initial conditions, we perform a scan over the initial thermal/oscillating velocity (one simulation per value) and sheets are always initialized at their equilibrium positions. All simulations consider a system of $10^3$ sheets moving over a periodic box for a total of $t_{max}=5 \times 2\pi~\omega_p^{-1}$.

While for the sheet model the energy decreases monotonically, we observed this was not the case for the GNS (energy might increase, decrease, or oscillate, examples are provided in~\ref{app:energy_conservation}). Furthermore, since the GNS uses the finite difference velocities instead of the instantaneous velocities, there is an oscillation in the energy associated with the plasma period (the period of the energy oscillation is equal to half a plasma period) which is clearly dominant for lower thermal velocities. This oscillation is merely an artifact of the finite different velocities used to compute the energy of the system. 

To allow for a fair comparison between the sheet model and GNS we then compute the total energy variation as follows: we skip the first plasma oscillation, apply a moving average with a window size of $\Delta t =  2\pi~\omega_p^{-1}$ to the remaining datapoints and retrieve the maximum deviation from the initial energy of the system (these steps are further justified in~\ref{app:energy_conservation}). For each time step, the energy of the system $\epsilon$ is computed according to:
\begin{equation}
    \epsilon = \frac{1}{2} m_e\sum_i^N \left(v_i^2 + \omega_p^2 \xi_i^2 \right).
\end{equation}

\begin{figure}[t]
    \centering
    \includegraphics[width=\columnwidth]{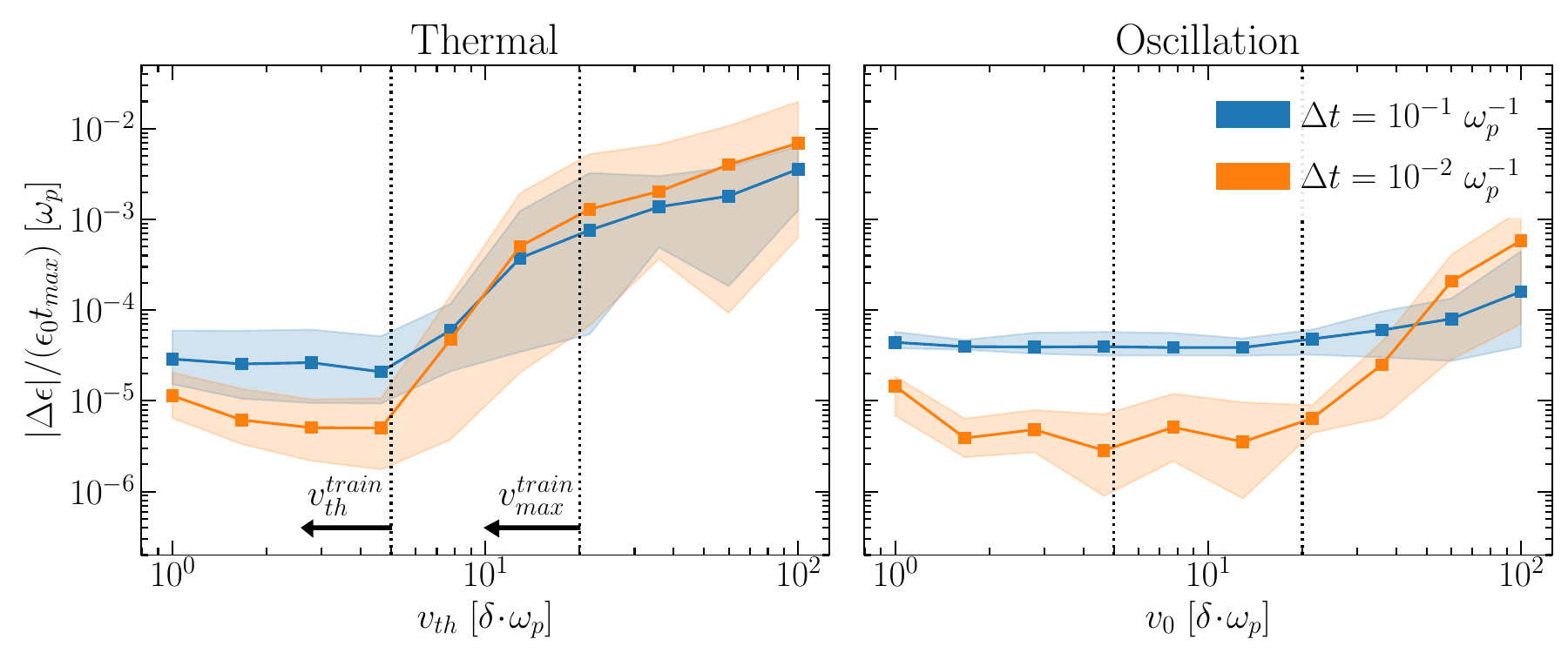}
    \caption{Energy variation rate of the GNS for simulations of $10^3$ sheets moving over a periodic box for $t_{max}=5\times 2\pi~\omega_p^{-1}$. Initial sheet velocities are either sampled from a normal distribution (thermal) or all equal to $v_0$ (oscillation). We run, per trained $\Delta t$, the exact same simulations using 5 equivalent GNN models (trained with different random seeds). The mean values across GNN models are presented in full line, and the min/max values are represented by the shaded region (for a detailed comparison between different seeds, see~\ref{app:seeds}). The GNS performance is stable within the training region ($v_{th} \leq v_{th}^{train}$) but starts to significantly degrade for larger thermal velocities.}
    \label{fig:energy_conservation}
\end{figure}

\begin{figure}[t]
    \centering
    \includegraphics[width=\columnwidth]{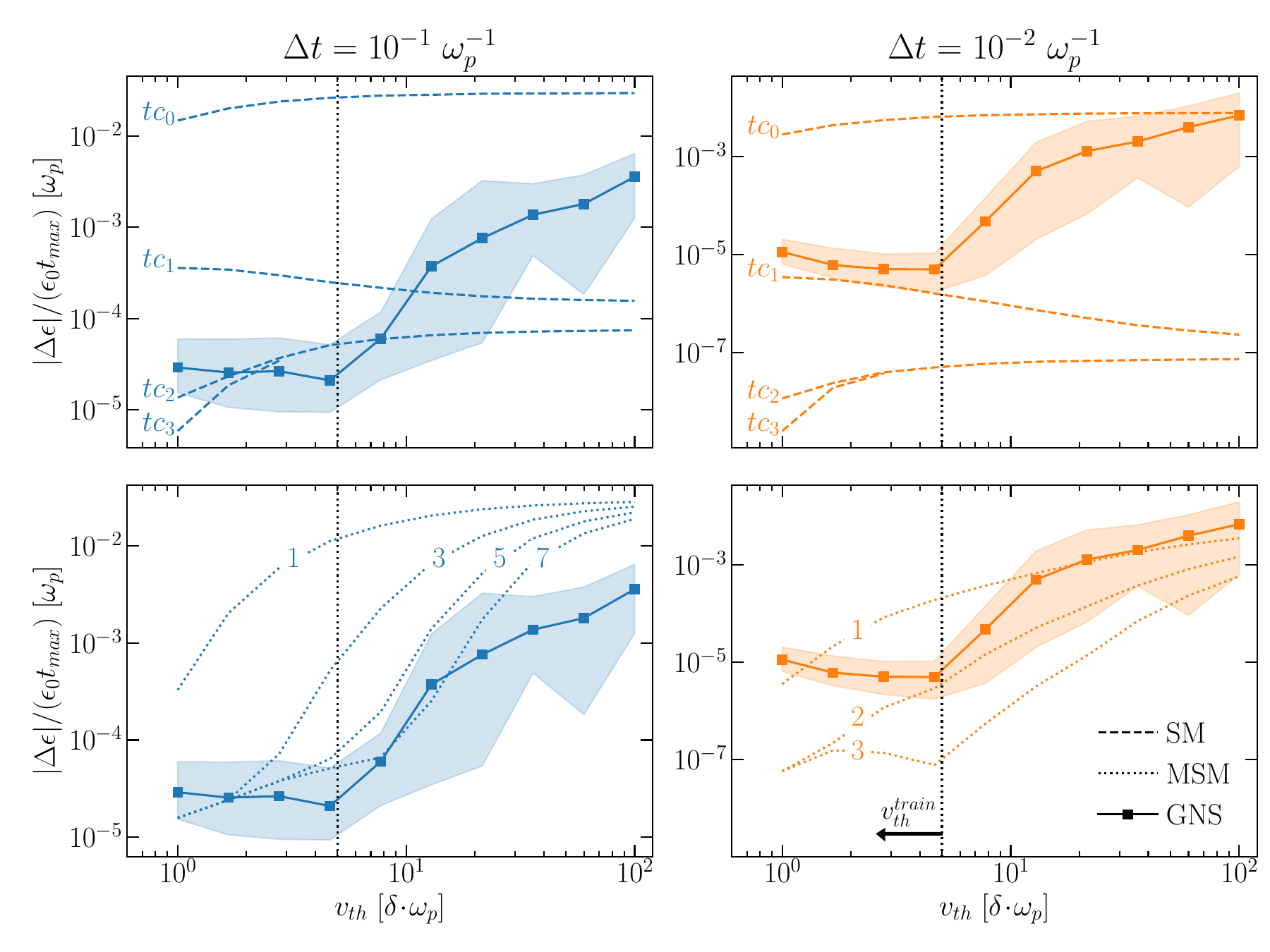}
    \caption{Comparison of the energy variation rate for the GNS,  the synchronous Sheet Model (SM), and the Modified Sheet Model (MSM). We use the same setup as in~\fref{fig:energy_conservation} (thermal initialization, $10^3$ sheets, periodic box, $t_{max}=5\times 2\pi~\omega_p^{-1}$). For the SM and MSM, we run a single simulation per set of initial conditions and simulation time step $\Delta t$. For the SM, $tc_k$ indicates the order of the correction for the crossing time ($k=0$ is no correction). For the MSM, we use $k=2$, and the numbers indicate the maximum neighbor checked for crossings. No significant gains in energy conservation are observed when comparing the GNS against the SM which uses higher order corrections ($tc_{2/3}$). The MSM exhibits a behavior similar to the GNS (steep increase in error at certain $v_{th}$), reinforcing the conclusion that the performance degradation of the GNS for $v_{t} > v_{th}^{train}$ is mainly due to the chosen graph connectivity and number of message-passing steps ($M=5$ for $\Delta t=10^{-1}~\omega_p^{-1}$ and $M=3$ for $\Delta t = 10^{-2} \omega_p^{-1}$).}
    \label{fig:energy_conservation_comparison}
\end{figure}

The final results for the GNS, containing scans performed for different initial velocity distributions, are presented in~\fref{fig:energy_conservation}. Comparisons between the GNS and the synchronous sheet model (original and modified version) are presented in \fref{fig:energy_conservation_comparison} for initial thermal conditions. No comparisons are provided for initial oscillatory conditions since the sheet model conserves perfectly the energy in this scenario. Similarly, no comparisons are provided with the asynchronous model since by definition it conserves perfectly the energy of the system.

It is observed that the accuracy of the GNS models remains approximately constant for the thermal velocities for which it was trained ($v_{th} \leq v_{th}^{train}$ in~\fref{fig:energy_conservation}). However, for out-of-distribution scenarios the energy variation starts increasing, more noticeably for $\Delta t=10^{-2}~\omega_p^{-1}$. This behavior is not visible for the oscillatory initial conditions, where we observe a very stable energy loss rate across all models for $v_{0} \leq v_{max}^{train}$. Therefore, the lower performance for higher $v_{th}$ should be attributed to a failure in correctly modeling all crossing events in such conditions. 

When comparing the GNS to the sheet model within the training data distribution, we observe once again two distinct behaviors for the different temporal resolutions (\fref{fig:energy_conservation_comparison}). For $\Delta t=10^{-1}~\omega_p^{-1}$ the GNS is on par with the sheet model, and actually improves (on average) upon the sheet model energy conservation for $v_{th}\simeq v_{th}^{train}$. This happens since the GNS learns a better correction algorithm for crossings involving $n>2$ sheets (refer to \ref{app:sheetmodel} on why the synchronous sheet model does not correctly model higher order crossings). On the other hand, for $\Delta t=10^{-2}~\omega_p^{-1}$, the GNS energy variation is worse than the SM using solely the first order correction (although the value is still considerably small). We believe the limiting factor here is simply the training error, which we could not further reduce (validation loss values are already $\approx 10^{-8}$).

A performance degradation is expected for the GNS outside its training data region since: (a) the number of message-passing steps associated with each model ($M=3$ for $\Delta t = 10^{-2}~\omega_p^{-1}$, $M=5$ for $\Delta t=10^{-1}~\omega_p^{-1}$) limits the GNS capability to correctly resolve crossings involving a larger number of sheets (which are more likely to occur at larger values of $v_{th}$; the same behavior is observed for the modified sheet model in \fref{fig:energy_conservation_comparison}); and (b) the GNN is only trained on crossings involving $n \leq n_{max}$, where $n_{max}$ is the maximum number of sheets involved in any crossing within the training data. However, this does not explain why outside the training data the performance of the GNS at different time resolutions is equivalent. We attribute this behavior to two main effects. 

Firstly, there is one particular GNN model (for $\Delta t =10^{-2}\omega_p^{-1})$ that consistently performed worse across all metrics (validation loss, rollout accuracy, and energy conservation) which biased significantly the average value of the energy variation rate. Removing this model from the average calculation changes considerably the behavior for $v_{th}^{train} < v_{th} < v^{max}_{train}$ (more details in~\ref{app:seeds}). 

Secondly, for the higher temporal resolution scenario, the majority of the training samples do not include any crossings ($>90\%$, more details in~\ref{app:energy_conservation}) or crossings involving more than 2 sheets ($\simeq 99.9\%$). This can bias the training procedure to significantly reduce the prediction error for events that do not involve crossings to the detriment of the (smaller) subset that contains crossings (which becomes problematic at test time for scenarios where crossings dominate the overall dynamics). Additional support for this claim is the fact that the higher temporal resolution models seem to be ``overfitting'' the purely oscillatory dynamics within the training data range since they fail to generalize to larger oscillation amplitudes (steep increase in energy variation for the oscillating initial conditions for $v_0 > v_{max}^{train}$). Furthermore, while for $\Delta t = 10^{-1}~\omega_p^{-1}$ the GNS performs significantly better than the ``equivalent'' MSM (GNS uses $M=5$), for $\Delta t = 10^{-2}~\omega_p^{-1}$ the GNS can not obtain similar results (GNS uses $M=3$).

The impact of the aforementioned effects could be investigated in future work (and mitigated if necessary) by using alternative data sampling strategies. However, the main focus should be the improvement of the performance for lower temporal resolutions and higher crossing rates since this is where larger computational gains are expected (accompanied by improved energy conservation).

It would also be important to study why some models are capable of achieving better energy conservation capabilities at larger $v_{th}$ (which might indicate that they learn a more robust crossing resolution algorithm), how to consistently achieve this level of performance (e.g. by using a regularizing penalty~\cite{cranmer2020discovering}), and if penalizing for rollout accuracy at train time could improve stability~\cite{prantl2022guaranteed, lam2022graphcast}. Additional evaluation metrics/tests should also be devised since neither the validation loss nor the current rollout accuracy tests seem to be good predictors for improved energy conservation capabilities. These new tests could include, for instance, measurements of rollout accuracy for significantly longer simulations and higher thermal velocities.

\subsection{Run-time}
\label{sec:runtime}

\begin{figure}[t]
    \centering
    \includegraphics[width=\columnwidth]{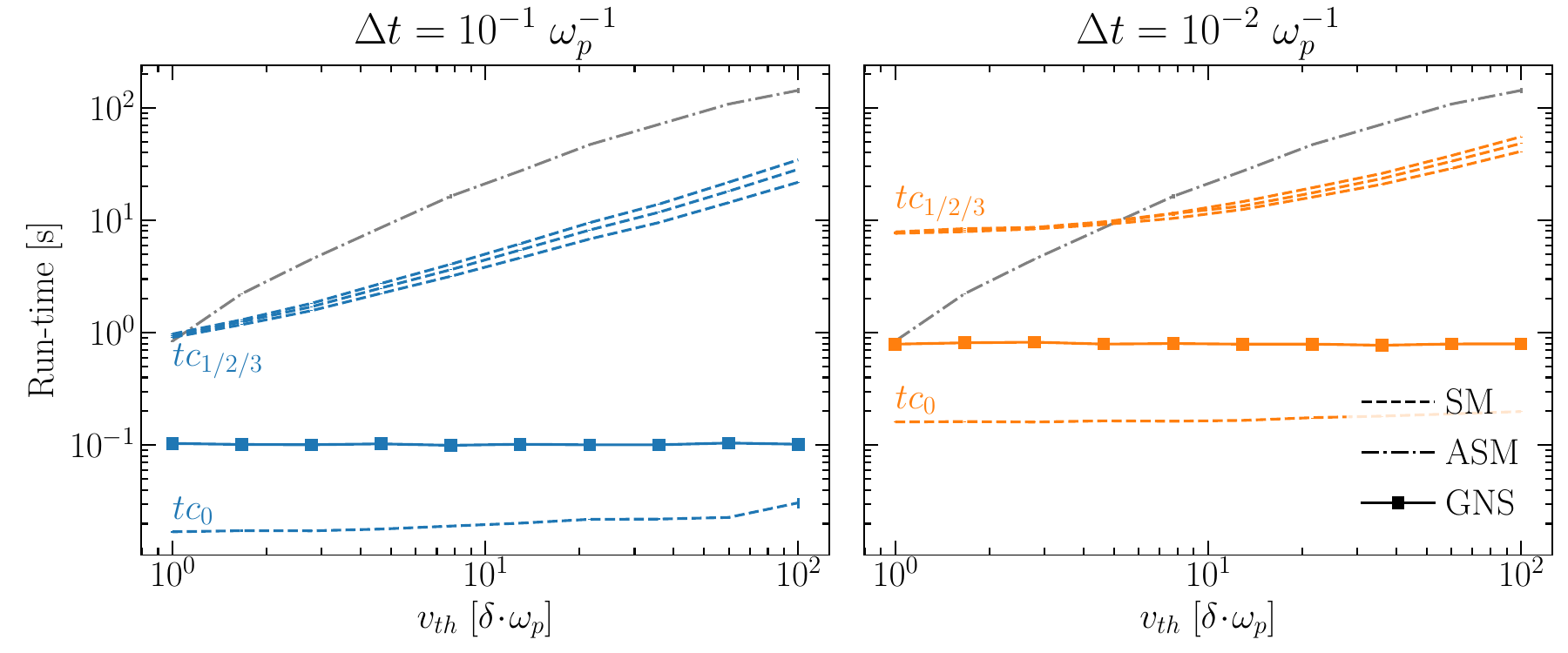}
    \caption{Run-time of the synchronous (SM) and asynchronous (ASM) sheet model \textit{vs.} the GNS for systems of $10^3$ sheets moving on a periodic box over the period of a single plasma oscillation ($t_{max}=2\pi~\omega_p^{-1}$). Values shown are averages over 5 simulations. For a fairer comparison among different time resolutions, the rollout data is only saved to a buffer every $\Delta t=10^{-1} ~\omega_p^{-1}$. The run-time of the modified sheet model (MSM), which is not shown, is equivalent to that of the SM. The just-in-time compilation time for the GNS is not included since it is a fixed cost that does not change for longer simulations (where it is considerably diluted). It amounts to $t_{JIT} = 1.2$~s for $\Delta t_{GNS} = 10^{-2}~\omega_p^{-1}$ and $t_{JIT} = 2.3$~s for $\Delta t_{GNS} = 10^{-1}~\omega_p^{-1}$. The results demonstrate that the (A)SM run-time increases with $v_{th}$ (higher number of crossings) while the GNS run-time is constant. By comparing these results with those of~\fref{fig:energy_conservation_comparison} it is observed that the GNS, at the lower temporal resolution, is faster than the SM at equivalent energy variation rates ($v_{th} \leq v_{th}^{train}$).}
    \label{fig:run_time}
\end{figure}

Using the same setup for thermal initial conditions, we analyzed the run-time of the different sheet model algorithms \textit{versus} the GNS.  The results obtained are presented in \fref{fig:run_time}. The sheet model run-time increases with the crossing frequency (for all versions), while the GNS does not. This happens because, with higher crossing frequencies, the sheet model verifies a higher number of neighbors for each sheet, while the GNS maintains the same amount of message-passing steps and graph structure. Furthermore, most of the operations of the sheet model algorithm are sequential, while for the GNS, most operations occur in parallel.

We observe that the GNS is one order of magnitude faster than the synchronous sheet model algorithm for $\Delta t =10^{-1}~\omega_p^{-1}$ at similar energy conservation rates (i.e. within the training regime).
However, it is important to highlight that the different models are implemented with different packages (NumPy \textit{vs} JAX) and running on different hardware (CPU \textit{vs} GPU) which influences their respective run-time. Furthermore, we make no claims that our implementations are optimal, meaning they could both benefit from further speed-ups. 

Additionally, if one is capable of training similarly accurate GNS models for larger simulation steps and crossing rates, without significantly increasing the number of message-passing steps (by instead increasing the graph connectivity statically and/or dynamically without triggering recompilation), relevant computational gains could be obtained. On the other hand, if one is capable of modifying the synchronous sheet model crossing correction routines in a way that they become parallelizable, the observed gap could be significantly mitigated. Furthermore, while the asynchronous model can not be parallelized and presents a higher run-time than both the synchronous algorithm (for large $\Delta t$) and the GNS, it provides significantly better energy conservation capabilities (limited by rounding errors) which might be preferable depending on the application.

\subsection{Limitations}
\label{sec:limitations}

The main limitations that we identified for the GNS are the requirement to use a fixed simulation step (equal to the training simulation step) and the performance degradation on out-of-training data distribution scenarios (as demonstrated in \fref{fig:energy_conservation}). 

The first constraint arises because the network has to learn to predict sheet crossings, which implicitly forces it to know what is the simulation step. To train a single model for different simulation steps, it would be necessary to provide $\Delta t$ as an input to the network (in the GNS the time step only appears explicitly in the ODE integrator). Alternatively, as we have shown, different models can be trained, one per $\Delta t$. As long as enough training data is provided and the model architecture is scaled accordingly (e.g.~by increasing the number of message passing steps for larger $\Delta t$ or connecting the $n^{\text{th}}$-closest neighbors) there is no limitation on the time step which can be used to train the model. In contrast, the crossing correction algorithm  in the SM requires that $\Delta t \leq \pi / 2~\omega_p^{-1}$ (more details in~\ref{app:sheetmodel}).

Regarding the second constraint, high-fidelity simulations for larger values of $v_{th}$ can be generated in order to fine-tune or retrain models for a broader dynamic range (while scaling the graph connectivity accordingly). These training simulations should be produced for larger system sizes, to ensure that sheets cross multiple neighboring sheets (and not their periodic versions) during the time step to be used for the GNS. Only by doing this can the GNS learn to correctly model the dynamics of sheets that cross a large number of neighbors. It is also important to highlight that, if provided with a representative set of crossings, the GNS learns a better crossing correction routine than the synchronous sheet model which is limited to the equivalent number of neighbors (GNS \textit{vs.} MSM in \fref{fig:energy_conservation_comparison}).

\section{Recovering known kinetic plasma processes}
\label{sec:plasma_processes}

In order to provide stronger evidence of the GNS generalization capabilities, we showcase a broad range of known kinetic plasma processes that the simulator is able to recover. These examples, present in both the original sheet model benchmarks~\cite{dawson1962one, dawson1962some, dawson1964thermal, dawson1968some, dawson1970electrostatic} and other kinetic codes benchmarks~\cite{dawson1983particle, opher2001nuclear, birdsall2004plasma, hesthaven2024adaptive, gravier2023collision}, aim to demonstrate the capability of the GNS to simulate collective behavior in accordance with known kinetic theory. An important point to stress is that the surrogate simulator was not explicitly trained to reproduce these effects. The GNN only learned to (correctly) model single-step updates over a reduced system size (10 sheets). However, when we apply it to larger system sizes and longer time durations, we observe the emergence of the expected kinetic collective plasma dynamics.

The results presented hereafter are produced using the GNN trained for $\Delta t = 10^{-1}~\omega_p^{-1}$ which showcased the best energy conservation capabilities (Model $\#4$ in \ref{app:seeds}). The same collective plasma dynamics are recovered for the equivalent GNN models trained using different random seeds, and those trained using the larger time step $\Delta t = 10^{-2}~\omega_p^{-1}$.

\subsection{Plasma thermalization}

In~\cite{dawson1962one, dawson1964thermal}, Dawson demonstrated that, independently of the initial velocity distribution of the sheets, it is expected that over time the system will move towards thermal equilibrium, and that this happens due to crossings/collisions involving more than 2 sheets~\cite{dawson1964thermal} (cf.~Section~\ref{sec:sm} for a discussion on the physics of $n=2$ crossing/elastic collisions and why they do not modify the distribution function). The distribution function of the sheet velocities is expected to converge to a normal distribution whose standard deviation corresponds to the thermal velocity of the plasma.

\begin{figure}[htb]
    \centering
    \includegraphics[width=\columnwidth]{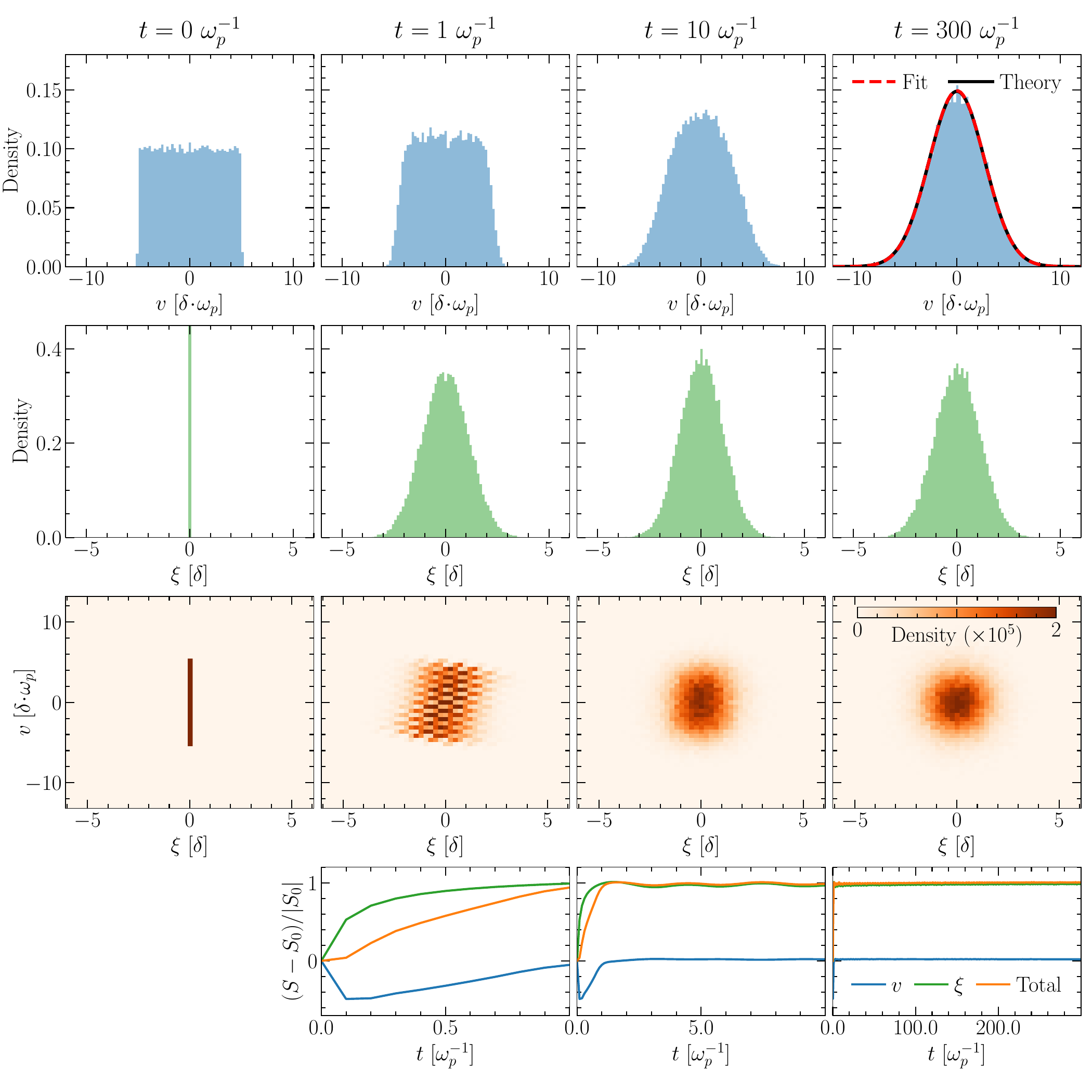}
    \caption{Evolution of the density functions for the velocity and displacement from the equilibrium position, and the entropy ($S$) of the system. Histograms represent the density function ensemble averaged over 50 simulations at the corresponding time. The entropy variation of the different phase-space components ($\xi$, $v$) is obtained using a diagnostic similar to the one implemented by Liang~et~al.~\cite{liang2019decomposition}. For the calculation of the distribution functions, we discretized the ($\xi$, $v$) phase-space for the range $\xi \in \left[-6.5, 6.5\right]\,\delta$ and $v \in \left[-12.6, 12.6\right]\,\delta\!\cdot\!\omega_p$ using 51 bins along each axis. These results demonstrate that the GNS is capable of correctly modeling the process of plasma thermalization from a non-equilibrium state, with the thermal velocity of the system in equilibrium $v_{th} = 2.671\,\delta\cdot\omega_p$ (measured by fitting a Gaussian to the final distribution)  in excellent agreement with the theoretical prediction $v_{th} = 2.679\,\delta\!\cdot\!\omega_p$.}
    \label{fig:thermalization_multisim}
\end{figure}

We demonstrate this behavior by performing $50$ simulations of systems consisting of $10^3$ sheets with initial velocities randomly sampled from a uniform distribution ($v\in[-5, 5]\,\delta\!\cdot\!\omega_p$). We provide snapshots of the evolution of the distribution function (averaged over simulations) in \fref{fig:thermalization_multisim}. 

It is clear that the system does indeed thermalize, and that the measured thermal velocity $v_{th} = 2.671\,\delta\!\cdot\!\omega_p$ is in accordance with the expected value $v_{th} = 2.679\,\delta\!\cdot\!\omega_p$. The latter is computed according to $v_{th}^2 = 1/3 \ v_{max}^2 r_{kin}$~\cite{dawson1970electrostatic}, where $v_{max}$ corresponds to the initial uniform distribution maximum value, and $r_{kin}$ represents the ratio of the available kinetic energy with respect to the total energy of the system (estimated by averaging over time steps and simulations), since a percentage of the initial kinetic energy is deposited in the fields.

Additionally, using a diagnostic similar to the one introduced by Liang~et~al.~\cite{liang2019decomposition}, \fref{fig:thermalization_multisim} demonstrates that there is a steep increase in the entropy ($S$) of the system until $t\approx1.25~\omega_p^{-1}$. This increase is associated with the establishment of correlations between sheets as crossings start to occur, and the length of this time interval is actually independent of the initial velocity range~\cite{dawson1964thermal}. 

\subsection{Debye shielding}

Another fundamental property of plasmas is their quasi-neutrality~\cite{chen1984introduction}, i.e.~on a macroscopic scale the overall charge density of positive and negative particles will cancel out. However, within local regions of characteristic length $\lambda_D = v_{th} / \omega_p$ (referred to as the Debye length) the local electric fields generated by a charged particle will not be fully screened by the oppositely charged particles. We expect to observe the same behavior for the sheet model. More precisely, the density of sheets at a certain distance from a test position is expected to follow~\cite{dawson1962one}:
\begin{equation}
    n(x) = n_0 \left(1 - \frac{\delta}{2\lambda_D} e^{-|x|/\lambda_D}\right).
\end{equation}

To test the GNS, we initialize systems of $10^4$ sheets following different initial thermal distributions ($v_{th} = [1.5, \ 2.5, \ 5.0]\,\delta\!\cdot\!\omega_p$). The simulations are run for $t_{max} = 80\,\tau_{relax}$, where $\tau_{relax} = \sqrt{2 \pi} \lambda_D /\delta \ \omega_P^{-1}$ is an estimate of the relaxation time
of the system~\cite{dawson1962one}, i.e.~the time it takes for the system to forget its current state. 

\begin{figure}[tb]
    \centering
    \includegraphics[width=0.6\columnwidth]{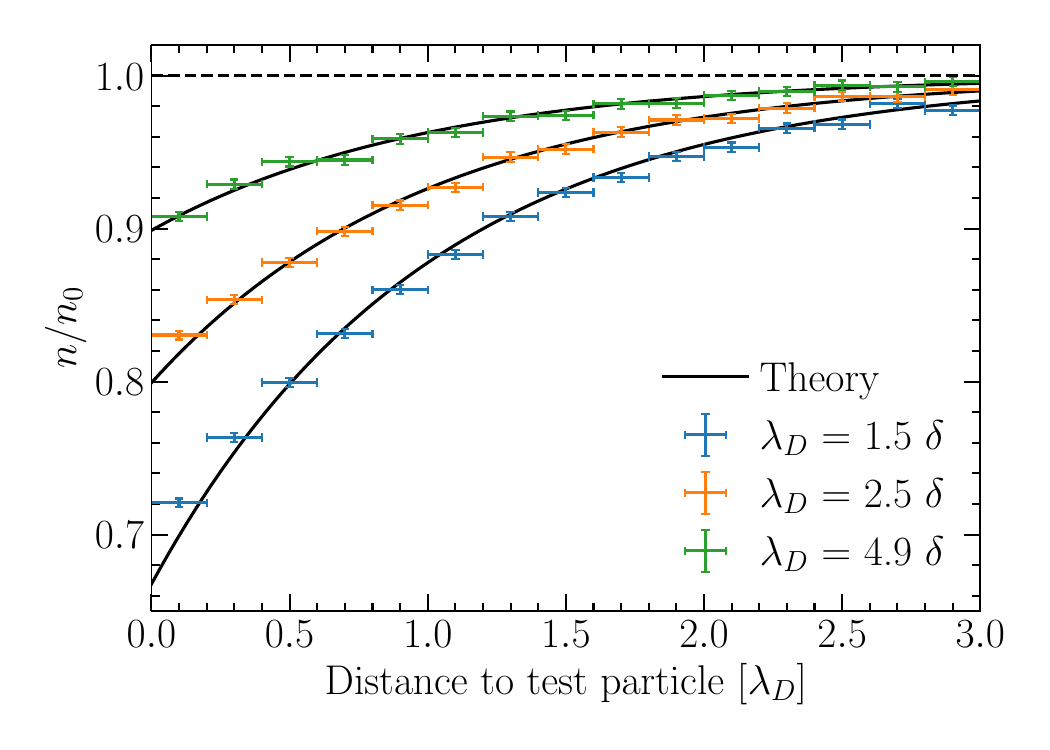}
    \caption{Example of Debye shielding for systems with different Debye lengths. The GNS can correctly recover the expected density profiles for all the tested scenarios.}
    \label{fig:debye_shielding}
\end{figure}

To compute the sheet density profiles shown in \fref{fig:debye_shielding}, we follow a similar procedure as Dawson~\cite{dawson1962one}. We choose a set of equally spaced test sheets and, for each of them, measure the number of neighboring sheets within a pre-defined range of increasing distances. In our case, we compute the number of sheets within a distance $d \in \ ]0.2i, 0.2(i+1)]\,\lambda_D$ up to $3\,\lambda_D$ ($i=15$). We repeat this procedure for every \mbox{$(3\lambda_D/\delta)$-th} sheet, over multiple independent time steps ($t_j=j\cdot \tau_{relax}$ for $j > 0$). The counts are then averaged over the number of test sheets, and time steps. 
It is clear from the results presented in \fref{fig:debye_shielding} that the expected behavior is recovered.

\subsection{Electrostatic fluctuations}

Although the plasma is in thermal equilibrium, there are constant exchanges of energy between the sheets and the (electrostatic) waves propagating inside the plasma. This leads to the appearance of electrostatic fluctuations, with an average power spectrum that follows~\cite{dawson1983particle
}:
\begin{equation}
    \frac{\langle E^2(k)\rangle}{8\pi}\ = \frac{k_BT}{2L\left(1 + k^2\lambda_D^2\right)}
\end{equation}
where $k$ represents the wave vector, $k_B$ the Boltzmann constant, $T$ the plasma temperature ($k_B T = m\,v_{th}^2$), and $\langle \cdot \rangle$ the time average. 

\begin{figure}[tb]
    \centering
    \includegraphics[width=0.6\columnwidth]{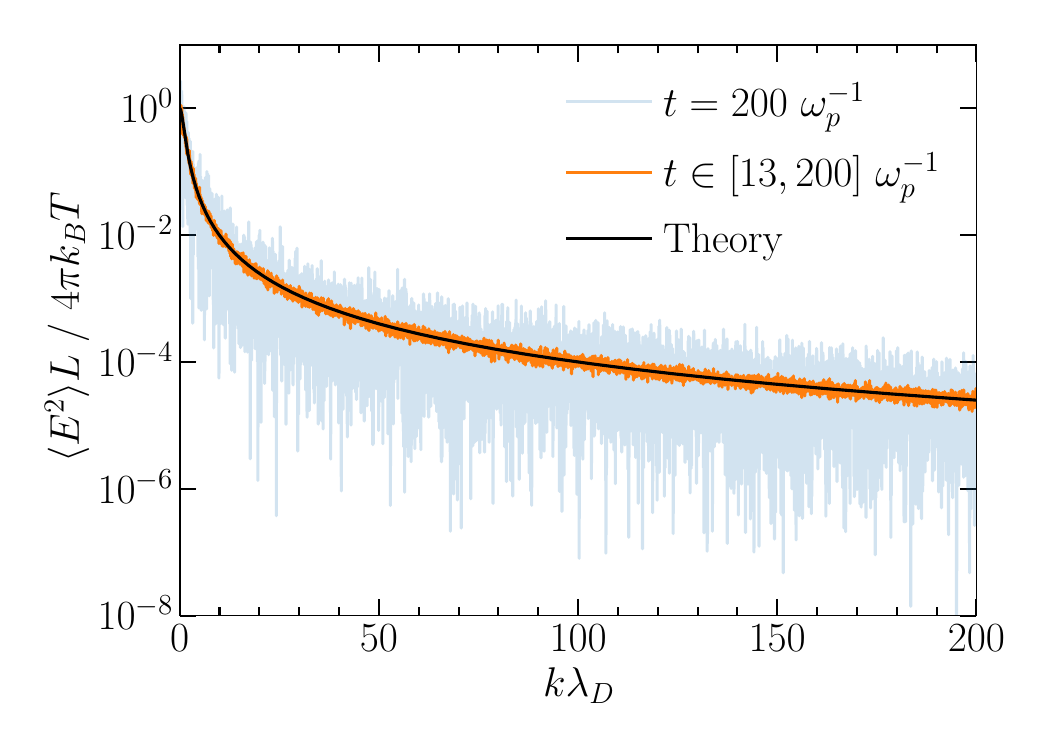}
    \caption{Electric field power spectrum for a system in thermal equilibrium. The power spectrum for the last time step and the temporal average (computed over relaxation periods $\Delta t = \tau_{relax} \approx 13 \ \omega_p^{-1}$) are shown. The time-averaged power spectrum retrieved from the GNS simulation matches the theoretical curve, thus demonstrating that it correctly models the electrostatic fluctuations around thermal equilibrium.}
    \label{fig:e_field}
\end{figure}

In \fref{fig:e_field} we recover this spectrum for a system of $10^3$ sheets with $\lambda_D = 5\,\delta$. We make use of the ergodic theorem~\cite{huang2008statistical} to compute the statistical average of the power spectrum by averaging over independent time steps (separated by $\Delta t = \tau_{relax}$).

\subsection{Dispersion relation}

The electrostatic waves propagating inside the plasma are expected to obey a particular dispersion relation, i.e.~for a given angular wavenumber $k$ only a certain wave angular frequency $\omega$ is allowed. The ratio between the two quantities defines the wave phase velocity $v_{ph} = \omega/k$. For electrostatic waves, also known as Langmuir waves, the dispersion relation is given by~\cite{chen1984introduction}:
\begin{equation}
    1 = \frac{\omega_p^2}{k^2} \int^\infty_{-\infty} \frac{\partial \hat{f}_0/\partial v}{v - (\omega/k)} dv
\label{eq:dispersion_relation}
\end{equation}
where $\hat{f}_0$ corresponds to the distribution function in velocity space. The solution will have both a real and imaginary part. The real part corresponds to the wave angular frequency, while the imaginary part corresponds to the inverse of the wave damping time (a phenomenon known as Landau damping which will be explored in~Section~\ref{sec:landau_damping}).

\begin{figure}[tb]
    \centering
    \includegraphics[width=0.7\columnwidth]{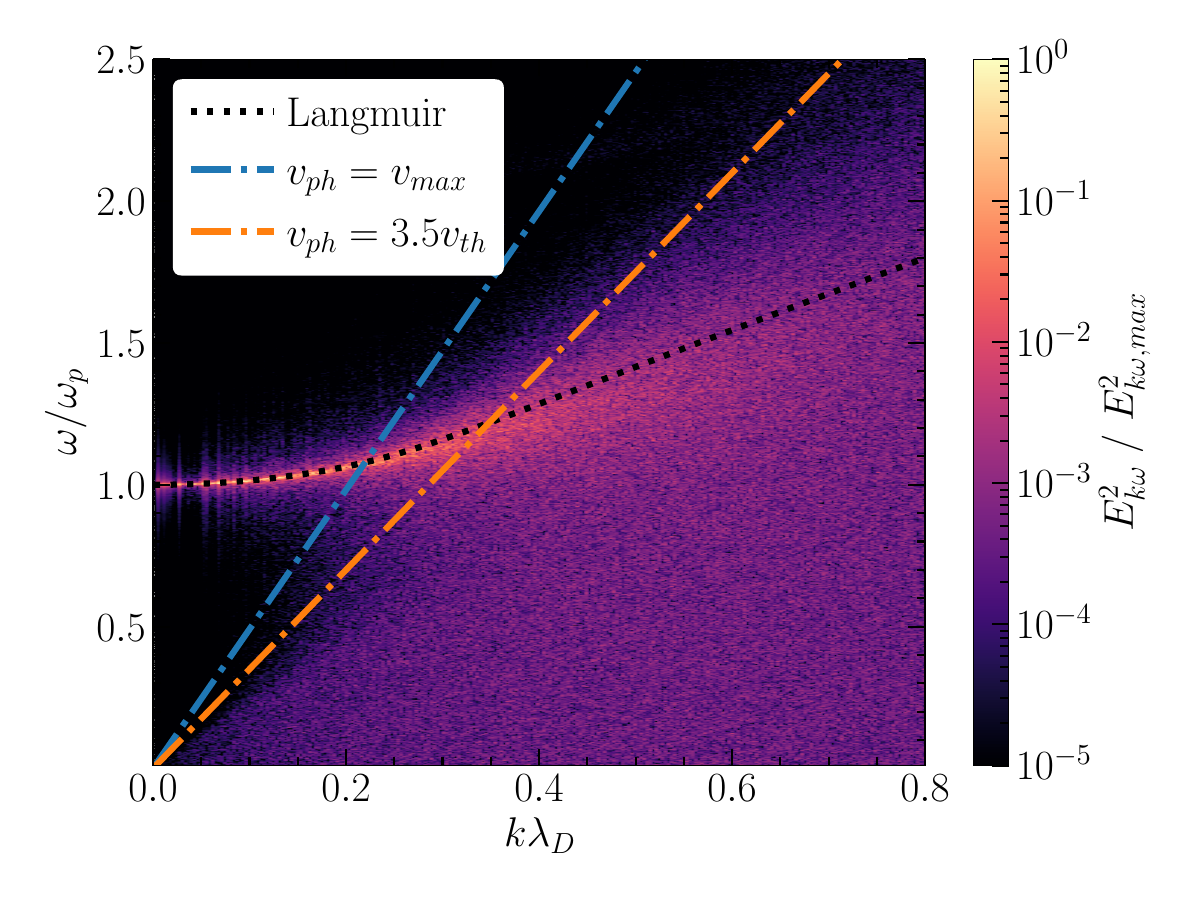}
    \caption{Dispersion relation obtained for the electrostatic waves propagating inside a thermal plasma. The GNS recovers the expected Langmuir waves dispersion relation and also reveals the presence of the non-collective ballistic (free-streaming) highly damped modes associated with single particle effects (phase velocity $v_{ph}$ corresponds to the sheet velocity). The ballistic modes can only exist for $v_{ph} \leq v_{max}$ (maximum sheet velocity). In this figure, they are mostly visible until $v_{ph} \approx 3.5v_{th}$ which encompasses $99.95\%$ of the sheet velocities across all time steps.}
    \label{fig:dispersion_relation}
\end{figure}

To demonstrate that the GNS recovers the expected dispersion relation we perform a simulation of a thermal plasma with $\lambda_D = 5~\delta$ and $N_{sheets}=10^5$ until $t_{max} = 2\times10^3~\omega_p^{-1}$. For each time step, we compute the electric field inside the box using a resolution of $\Delta x = 1~\delta$ and then perform a 2D-FFT along the time and $x$-axis. The obtained power spectrum is presented in \fref{fig:dispersion_relation}.
The signal corresponding to the Langmuir waves is clearly visible and in agreement with the numerical solution of the dispersion relation~\cite{jackson1960longitudinal}. Additionally, it is also visible the presence of non-collective ballistic (free-streaming) modes~\cite{krall1973principles, grismayer2011time} associated with the individual particle motion ($\omega=kv_{sheet}$ where $v_{sheet}$ is the velocity of the sheet).

\subsection{Drag on a fast sheet}

A fast sheet ($v_{sheet} \gg v_{th}$) moving through the plasma is expected to feel a constant drag given by~\cite{dawson1962one}: 
\begin{equation}
    \frac{dv}{dt} = -\frac{\omega_p^2 \delta}{2} .
\end{equation}

This drag is independent of the velocity of the sheet and is caused by the excitation of a electrostatic wake on the rear of the fast sheet, i.e. the sheet transfers energy to the electrostatic wake.

\begin{figure}[tb]
    \centering
    \includegraphics[width=0.6\columnwidth]{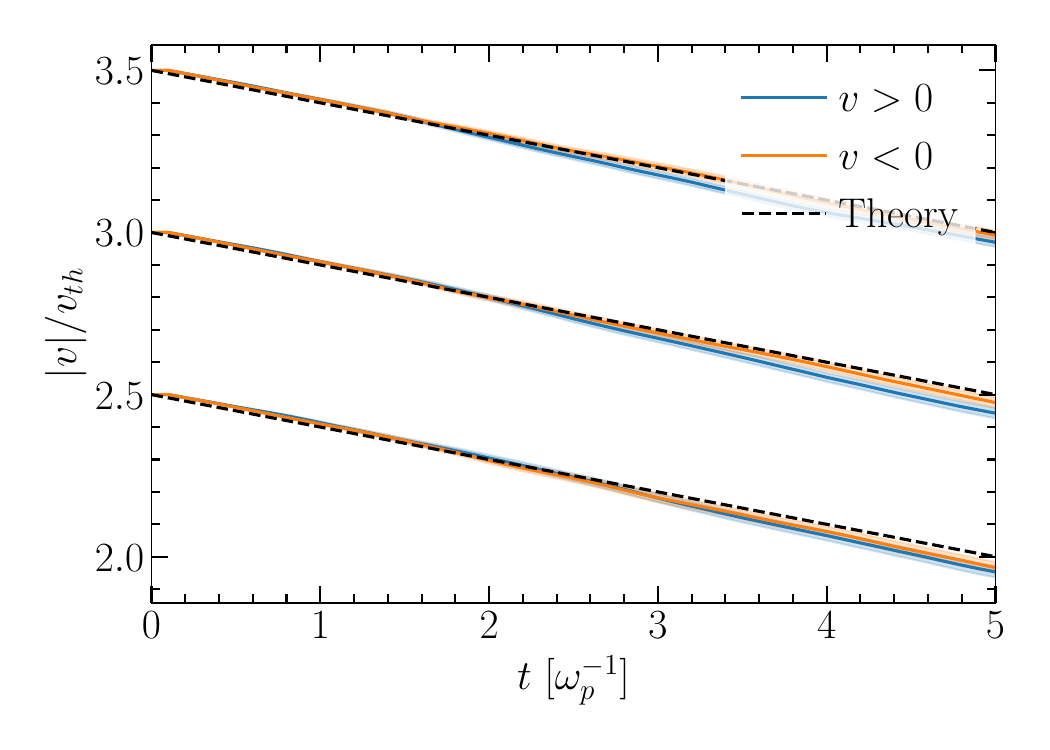}
    \caption{Average drag on fast sheets of different initial velocities ($v_0 \gg v_{th} = 5\,\delta\cdot\omega_p$). The GNS recovers the expected drag felt by the fast sheets independently of their initial velocity and propagation direction.}
    \label{fig:fast_sheet}

\end{figure}

In \fref{fig:fast_sheet} we demonstrate this behavior. The results were obtained by performing simulations of periodic systems of 100 sheets with $\lambda_D = 5\,\delta$ ($v_{th} = 5\,\delta\!\cdot\!\omega_p$) over a period of $t_{max} = 5~\omega_p^{-1}$. For each simulation, we set the initial velocity of the first sheet to $v_0=\pm\,\alpha\,v_{th}$ and track its evolution over time. We then average over simulations, 1000 for each initial sheet velocity (accounting for the sign change).

\subsection{Landau damping}
\label{sec:landau_damping}

While fast sheets are able to excite an electrostatic wake in their rear, the resulting electrostatic wake can also accelerate sheets moving slightly slower than its phase velocity~\cite{chen1984introduction, dawson1962one}.
Electrostatic modes are therefore self-consistently generated by particles moving close to its phase velocity, while being damped by particles moving slightly slower. However, since a plasma in thermal equilibrium follows a Maxwellian distribution in velocity space, there exist on average more particles moving faster than the wave, than those moving slower. Therefore, on average, the modes will be damped. More specifically, for a given mode $m$, with a wavelength $\lambda_m = 2L/m$ and wave vector $k_m = 2\pi/\lambda_m$ we can compute its wave frequency and damping time by finding numerically the roots of the dispersion relation~\eqref{eq:dispersion_relation}. This damping mechanism is known as Landau damping~\cite{chen1984introduction, grismayer2011time} and is an inherently collisionless kinetic process that the sheet model has been shown to recover~\cite{dawson1962one, dawson1968some}.

To reproduce the expected damping behavior we follow initially a similar procedure to that of Dawson~\cite{dawson1962one}. We produce 100 simulations, each with a duration of $t_{max}=500~\omega_p^{-1}$, of thermal plasmas consisting of $10^4$ sheets for $\lambda_D = 5\,\delta$ and reflecting boundaries. For each simulation time step we compute the mode ``amplitude" $A_m$ and its rate of change $\dot{A}_m$ using the cross-correlation:
\begin{equation}
\eqalign{
    A^t_m = \frac{2}{N} \sum_{i=0}^{N-1} \left(x^t_i - x^t_{eq_i}\right) \ \sin\left( \frac{m\pi}{N}\left(i + \frac{1}{2}\right) \right) \cr
    \dot{A}_m^t = \frac{A_m^t - A_m^{t-1}}{\Delta t}
}
\end{equation}
where the index $i$ indicates the relative ordering of the sheets in the box. 

We then collect trajectories of equal time length every time the mode crosses the region of phase-space $(A_m,\dot{A}_m)$ defined by a ring of radius $R$ and thickness $dR$ ($dR \ll R$). Finally, we rotate the trajectories so that they all start on the same position in phase-space, and compute their average. 

\begin{figure}[tb]
    \centering
    \includegraphics[width=\textwidth]{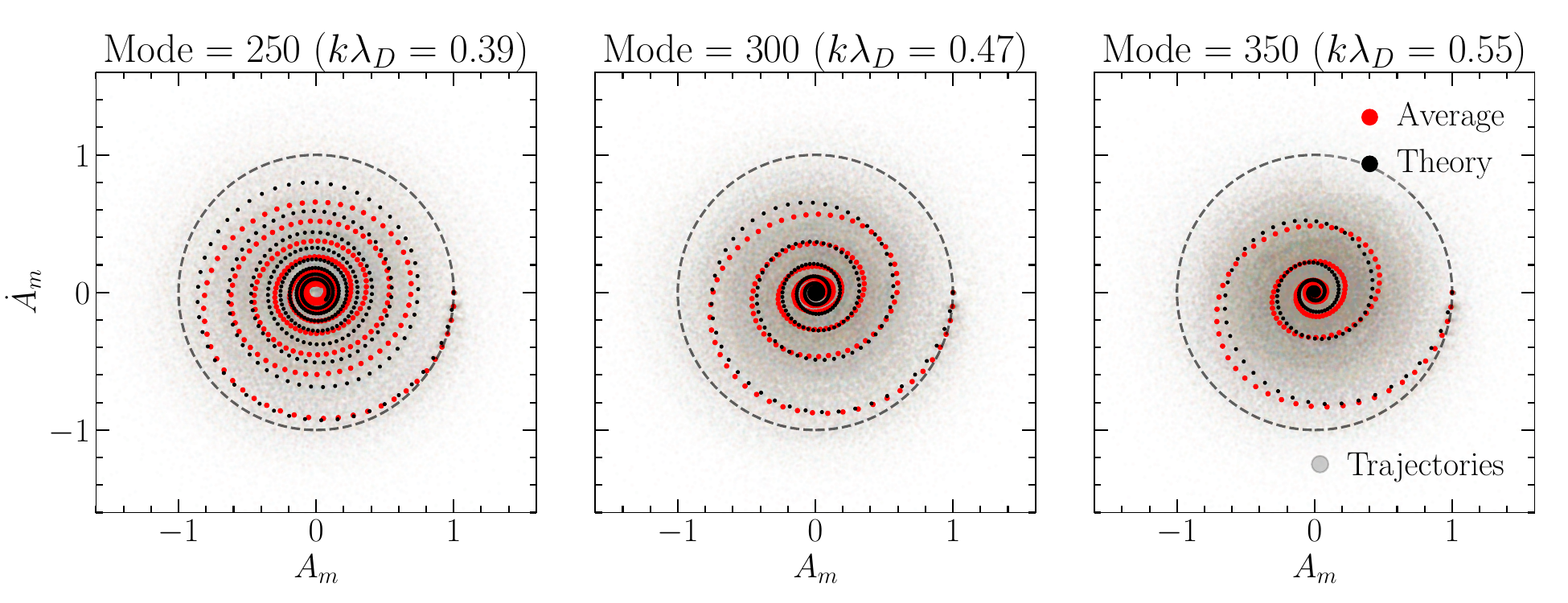}
    \caption{Damping of different modes. The initial mode amplitudes are normalized to $R=0.15\,\delta$. Individual trajectories length is $\Delta t_{traj}=50~\omega_p^{-1}$. The agreement between the theoretical curves and the average mode trajectories demonstrates that the GNS is capable of correctly modeling Landau damping, an inherently kinetic mechanism.}
    \label{fig:landau_damping}
\end{figure}

In \fref{fig:landau_damping} we showcase examples of the results obtained for several modes. It is possible to see that, although for some trajectories the mode is still growing, on average it decreases according to the expected damping time. This demonstrates that the GNS is capable of correctly modeling Landau damping, an inherently kinetic mechanism associated with the collective collisionless dynamics of a plasma.

To further support this claim, we perform a scan over different modes $m\in[180, 448]$ ($k\lambda_D \in [0.28, 0.70]$) and initial mode amplitudes $A^0_m \in [0.08, 0.20]~\delta$ using a resolution of $\Delta m = 2$ and $\Delta A^0_m = 0.005~\delta$ ($135\times 25 = 3375$ data points). For each ($m$, $A^0_m$) pair, we compute the average of the trajectories in phase-space (\fref{fig:landau_damping}). We then proceed to estimate the damping rate, $-\mathrm{Im}(\omega)$, by obtaining the slope of the line that best fits the peaks of $\log|A^t_m|$. The mode angular frequency, $\mathrm{Re}(\omega)$, is estimated as $\pi/\Delta t^{avg}_{peaks}$, where $\Delta t^{avg}_{peaks}$ is the average interval between consecutive peaks of $\log|A^t_m|$ (more details in~\ref{app:landaudamping}).

\begin{figure}[tb]
    \centering
    \includegraphics[width=0.8\textwidth]{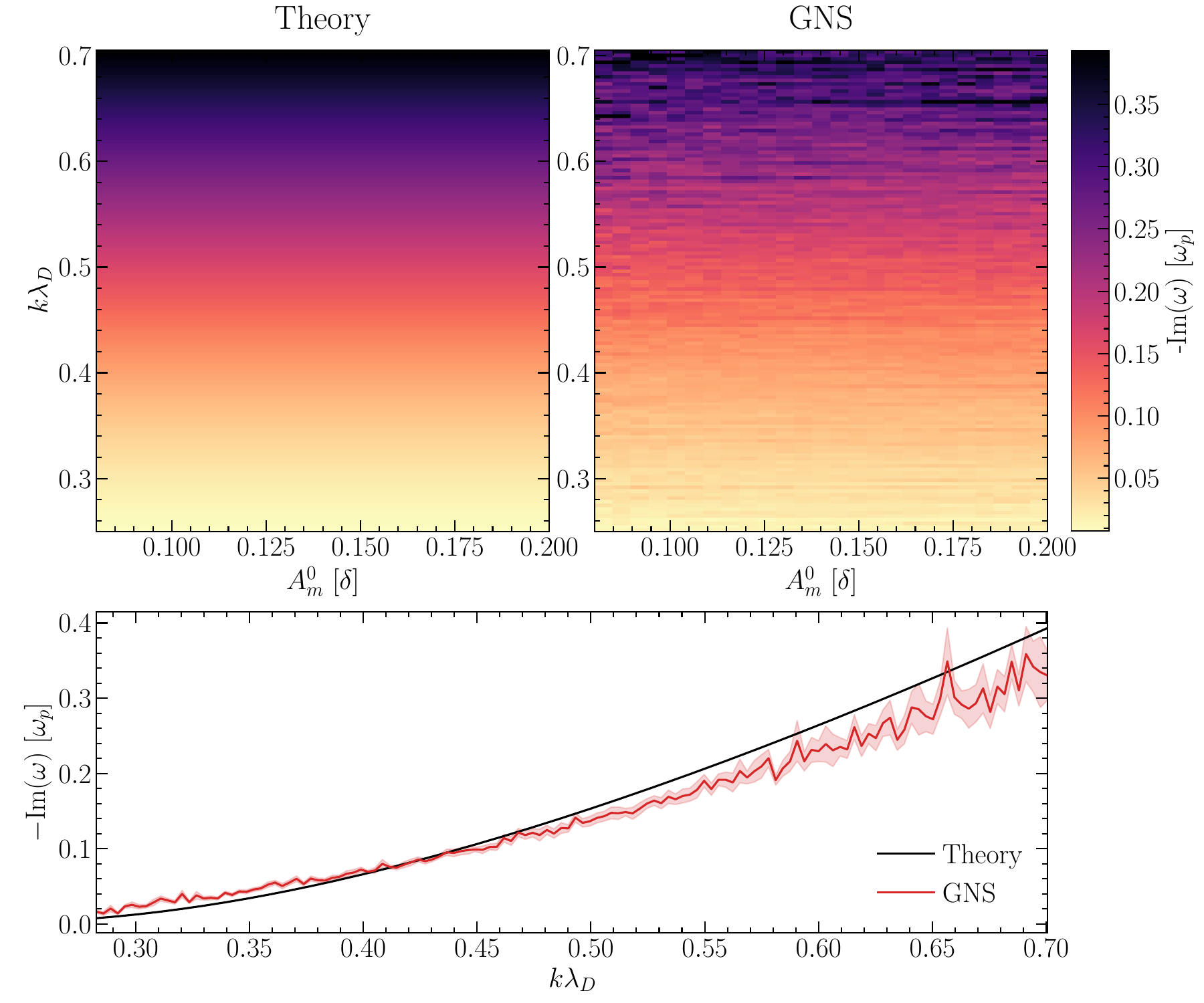}
    \caption{Comparison between the theoretical mode damping rate, $-\mathrm{Im}(\omega)$, and the one measured using the GNS. Results are shown for different mode angular wavenumbers and initial mode amplitudes (no interpolation is performed), as well as the average damping rate across initial mode amplitudes (standard deviation in lighter color). It is observed that the GNS correctly approximates the theoretical damping rate for a broad range of modes.}
    \label{fig:landau_damping_w_im}
\end{figure}

\begin{figure}[tb]
    \centering
    \includegraphics[width=0.8\textwidth]{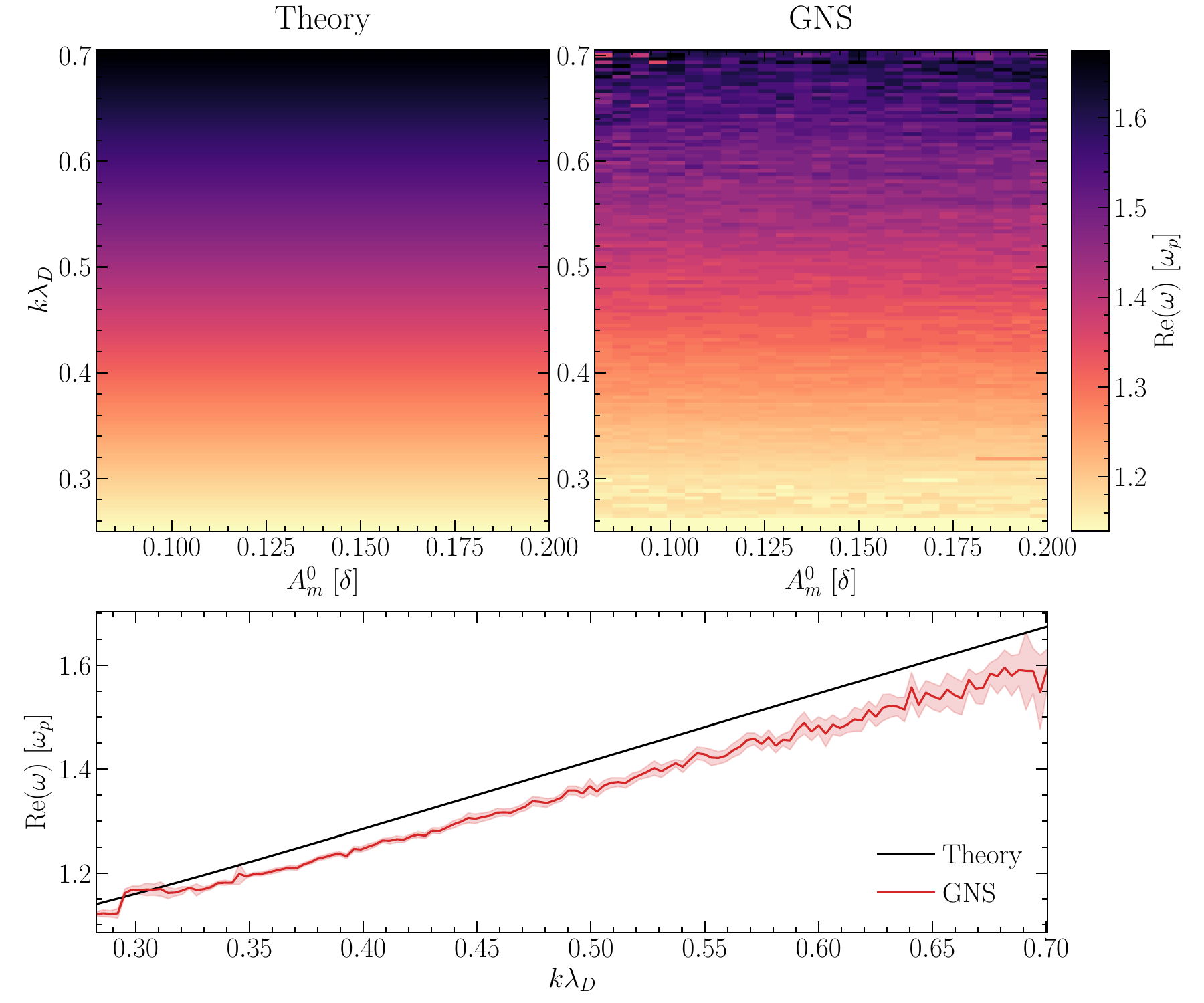}
    \caption{Comparison between the theoretical mode angular frequency, $\mathrm{Re}(\omega)$, and the one measured using the GNS. Similarly to \fref{fig:landau_damping_w_im}, we present results for different mode angular wavenumbers and initial mode amplitudes, as well as the average angular frequency across initial mode amplitudes. Although there is a slight underestimation of the angular frequency, the GNS is capable of approximating the theoretical prediction across a broad range of modes.}
    \label{fig:landau_damping_w_real}
\end{figure}

The Landau damping rates are shown in \fref{fig:landau_damping_w_im}, and the corresponding angular frequency (the real part of $\omega$) in~\fref{fig:landau_damping_w_real}. The average values obtained by the GNS are in good agreement with the theory, and both the damping rate and angular frequency are (as expected) independent of the initial mode amplitude. The higher variations observed for larger values of $k\lambda_D$ are due to having less periods for the estimation of $\omega$ since these correspond to strongly damped modes (for $k\lambda_D > 0.6$ the phase velocity $v_{ph} \leq 2.6~v_{th}$, already in the plasma bulk). 

\subsection{Two-stream instability}
\label{sec:twostream}

As a final example, we show the two-stream instability in the cold beam regime~\cite{chen1984introduction, dawson1962some}. For this scenario, two counter-propagating beams with velocities $\pm v_0$ and no energy spread excite an electrostatic plasma wave that grows exponentially until a significant fraction of particles are trapped inside the electric field, at which point the instability saturates.

From linear theory~\cite{chen1984introduction, dawson1962some}, we expect that for two cold beams with density $n_{beams} = n_0/2$, the fastest growing mode will correspond to $k=\sqrt{3/8} \cdot\omega_p/v_0$ with a corresponding growth rate of $\gamma = \omega_p/\sqrt{8}$. Therefore, to excite mode $m$, with $k_m = m\pi/L = m\pi / N_{sheets} \delta$, we need to set $v_{0_m} = \sqrt{3/8} \cdot \omega_p N_{sheets}\delta / m\pi$. Furthermore, the number of sheets per wavelength of this mode is given by $N_{\lambda_m} = \lambda_m / \delta = 2N_{sheets}/m \cdot$. Note that both $v_{0_m}$ and $N_{\lambda_m}$ are proportional to the number of sheets used. Therefore, to excite a mode whose wavelength must be resolved by a significant amount of sheets, we need to increase $v_0$ proportionally.

For a system of $10^4$ sheets and $m=4$, we obtain $v_{0_m} = 486~ \delta\cdot\omega_p$ and $N_{\lambda_m} = 5\times 10^3$. The chosen velocity is considerably out of the training data range, therefore we expect the energy loss of the GNS to be slightly higher than the one observed for the training scenarios (as shown in~\fref{fig:energy_conservation}). Nonetheless, we observe that the GNS is still able to capture the relevant physics. 

\begin{figure}[t]
    \centering
    \includegraphics[width=0.9\columnwidth]{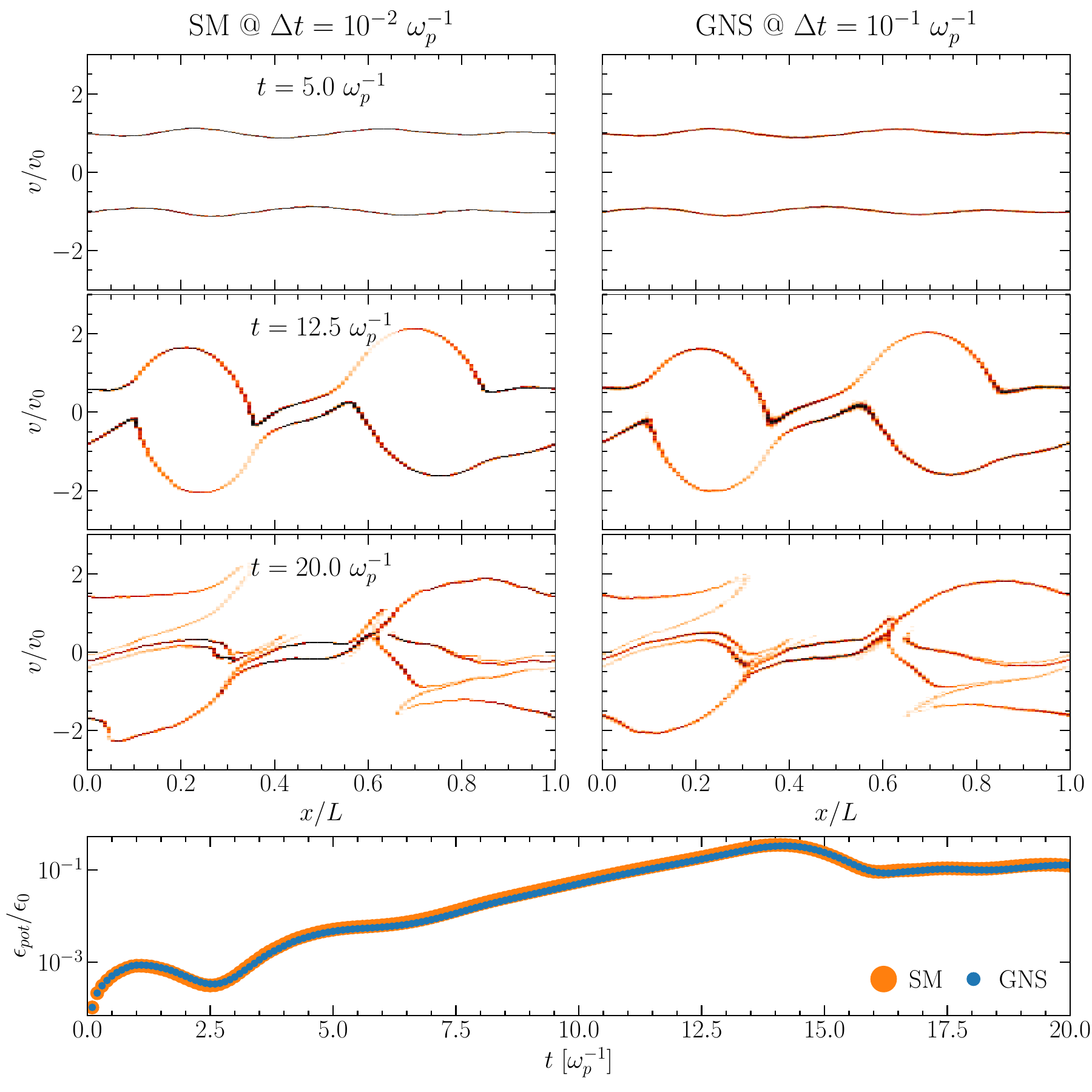}
    \caption{Comparison of the phase-space and potential energy evolution for the sheet model (SM) and the GNS in the two counter-propagating cold beams scenario. The GNS is able to recover the same macrophysics as the high-resolution sheet model, further demonstrating its capabilities to model scenarios significantly outside of its training regime.}
    \label{fig:twostream_sm_vs_gns}
\end{figure}

In \fref{fig:twostream_sm_vs_gns} we provide a comparison between the evolution of the phase-space and the potential energy for the sheet model at $\Delta t = 10^{-2}~\omega_p^{-1}$  and the GNS at lower time resolution $\Delta t = 10^{-1}~\omega_p^{-1}$. The energy variation during the full simulation was $\Delta \epsilon/\epsilon_0 \approx 10^{-6}$ for the sheet model while for the GNS it was $\Delta \epsilon/\epsilon_0 \approx 2\times 10^{-2}$. These values are in accordance with what was measured in~\fref{fig:energy_conservation}.

It is observed that the GNS recovers similar macrophysics when compared to the higher temporal resolution sheet model (which we consider as a good approximation to the ground truth) during the linear phase and up to the saturation time. An extra diffusion in phase-space is observed for the GNS, which is associated with the aforementioned higher energy variation. This is expected since the GNN is not capable of correctly resolving crossings involving more than $2M + 1 = 11$ sheets, and, on average, a sheet moving with $v_{0_m} = 486\,\delta\!\cdot\!\omega_p$ should cross $\sim$49 neighbors in the first timestep (since when the simulation starts all sheets are equally spaced by $\delta$ and, on average, half of the nearest neighbors are counter-propagating with the same absolute velocity). Nonetheless, the overall phase-space structure and growth rates are similar, which provides further support for the generalization capabilities of the model.

Finally, we provide in~\ref{app:twostream} further comparisons for higher order modes (lower $v_{0_m}$) and smaller simulation steps ($\Delta t=10^{-2}~\omega_p^{-1}$), as well as results for the remaining GNNs trained with different seeds (with worse energy conservation capabilities at high $v_{th}$). These results further support the claim that the GNS is consistently able to correctly model the overall dynamics of the instability.

\section{Conclusions}

In this work, we demonstrated that graph neural network-based simulators are capable of fully replacing a one dimensional kinetic plasma physics simulator. By introducing domain knowledge into the graph representation and the overall structure of the simulator, we showed that the GNS is capable of generalizing to a broad range of scenarios and is limited only by its fixed simulation step, training data distribution, graph connectivity, and message passing steps. Furthermore, for larger simulation time steps and crossing frequencies within the training data distribution, we observe that the GNS conserves better the system energy than the synchronous sheet model, while being significantly faster. This happens because the GNS learns an improved (synchronous and parallelized) crossing correction algorithm from higher-temporal resolution simulations (which correctly resolve crossings involving $N>2$ sheets).

In future work, the accuracy of the GNS for higher thermal velocity values can be improved by generating additional ground truth data, ideally using the asynchronous version of the sheet model, which is guaranteed to perfectly resolve crossings. Additionally, although not explored, the developed simulator is fully differentiable, which opens the way to explore gradient-based optimization strategies for the discovery of new physics of interest~\cite{joglekar2022unsupervised}.

It is also important to note that, despite being very simple, the sheet model provides a powerful framework to explore fundamental plasma processes, as demonstrated in this paper, including the relevance of collisional processes. Not only is it a gridless code (thus making fewer approximations than PIC codes) but also provides an exact description of one-dimensional non-relativistic electrostatic plasmas with arbitrary degrees of collisionality.  Furthermore, direct extensions and/or similar models exist, which allow one to study multi-species scenarios~\mbox{\cite{dawson1970electrostatic, gravier2023collision}}, electron-neutral collisions~\mbox{\cite{burger1967elastic}}, and electron-ion 3D scattering events~\mbox{\cite{shanny1967onedimensional}}. Therefore, the demonstration that the GNS can accurately provide a fast surrogate for the sheet model can open new directions for the exploration of the fundamentals of statistical mechanics of plasmas, and further extensions of the framework can be developed to integrate multi-species and stochastic processes.

Additionally, this work suggests that it would be possible to design and train a GNS that accurately models the standard PIC loop. This could be done by extending the developed framework to include grid information in both the graph representation and GNN architecture~\cite{pfaff2020learning}. However,  the standard PIC loop implemented in modern architectures is extremely optimized and does not suffer from the issues that are present in the sheet model scenario (e.g.~time-consuming serial routines). In fact, the PIC loop itself can already be seen from a graph computation perspective: field interpolation and particle push can be seen as message-passing steps from the grid to the particles; current/charge deposition as a message-passing step from the particles to the grid; and the field solver as a message-passing step between neighboring cells. Note that, frequently, most of these operations are linear. Therefore, simply replacing them by a learnable function is counter-productive and should lead to an expected increase in run-time.

The question is then, where might a GNN-based approach introduce some gain compared with traditional PIC?  In our view, two main possibilities arise: a) GNNs could be used to relax the time constraints imposed by the CFL condition~\cite{birdsall2004plasma}; b) GNNs could introduce extra physics that the standard PIC loop fails to model, which usually require additional modules (e.g.~collisions – thus having a model that intrinsically addresses collisions is important on this roadmap). Approaches following the first route should ensure that causality is still verified (similarly to what we have shown for the sheet model), this will result necessarily in increased graph connectivity (e.g.~particles would require information from neighboring grid cells) and will probably require the exploration of multi-scale approaches~\cite{lam2022graphcast, prantl2022guaranteed} or hybrid representations~\cite{wu2022learning} to be competitive. The second route will result in an increased run-time compared to the standard PIC loop but might be competitive against alternative approaches. For example, a standard 1D1V PIC simulation does not correctly capture collisional dynamics. Again, the sheet model, and consequently the GNS framework presented in this work, can do so for arbitrary levels of collisionality, thus providing a valuable source of comparison.

Similarly to what we have done for the sheet model, it would also be important to explore the trade-off between how much is learned by the GNN, and how much prior knowledge about kinetic simulations is embedded into the GNS. As one moves to higher-dimensional and more complex scenarios, this will become even so more important to maximize generability and possibly reduce the run-time. We will further explore these topics in future publications.

\section*{Data availability statement}
The code used for this project is openly available in a git repository: \url{https://github.com/diogodcarvalho/gns-sheet-model}. The training and test data, as well as a set of the final model weights, are available in a zenodo repository: \url{http://doig.org/10.5281/zenodo.10440186}.

\section*{Conflict of interest}

The authors declare no conflict of interest.

\ack{The authors would like to thank P.~J.~Bilbao, B.~Malaca, T.~Grismayer, A.~Joglekar, E.~P.~Alves, and V.~Decyk for helpful discussions and the anonymous referees for their valuable comments which helped improve the quality of this manuscript. This work was supported by the FCT - Fundação para a Ciência e Tecnologia, I.P. under the Project No 2022.02230.PTDC (X-MASER) and PhD Fellowship Grant 2022.13261.BD and has received funding from the European Union's H2020 programme through the project IMPULSE (grant agreement No 871161). The graphics processing units (GPUs) used in this work were donated by NVIDIA Corporation.}

\bibliographystyle{iopart-num}
\section*{References}
\bibliography{references}% Produces the bibliography via BibTeX.

\newpage
\appendix

\section{Sheet model implementation}
\label{app:sheetmodel}

The original publication by Dawson~\cite{dawson1970electrostatic}, detailing the single and multi-species one-dimensional sheet model, is not widely available. Therefore, and for the sake of completion, we provide here a description of our implementation for both the single-species synchronous and asynchronous algorithms. We highlight the differences between our implementation and the original algorithms~\cite{dawson1970electrostatic}, whenever they occur.

We also provide details regarding the modified version of the synchronous algorithm which restricts the number of neighbors checked for crossings. Information regarding the multi-species model, its description and a possible implementation can be found in recent work by Gravier et. al.~\cite{gravier2023collision}.

\subsection{Synchronous Algorithm (SM)}

The main building blocks of the synchronous algorithm, previously shown in~\fref{fig:simulators}, are implemented as follows.

We store 2 sets of 4 arrays, corresponding to the sheet positions $\bi{x}$, equilibrium positions $\bi{x}_{eq}$, velocities $\bi{v}$ and labels $\bi{l}$ at $t$ and $t+\Delta t$. The labels array allows us to track individual sheets, which is necessary to generate training data for the GNS and run additional diagnostics (e.g.~drag on a fast sheet). At initialization, all arrays should be sorted with respect to the initial position of the sheets (which should also result in an array of sorted equilibrium positions).

\subsection*{SM -- Add guards $(t=0)$}

Before starting the simulation, guard sheets are added to the beginning/end of the arrays to model the boundary conditions. For periodic boundaries, these guards represent copies of the sheets closer to the opposite side of the simulation box. For reflecting boundaries, they represent mirrored versions of the sheets closer to the boundary. This is exactly the same approach that is adopted for the GNS.

The number of guard sheets should be large enough that no individual ``real'' sheet crosses all guard sheets in a single time step. For periodic boundaries, it should also be ensured that, in a single time step, not all guard sheets from the left or right wall enter the simulation box. At each iteration of the algorithm we check if these conditions are satisfied. If they are not, the simulation is halted.

\subsection*{SM -- Equation of motion}

When disregarding crossings, each sheet behaves as an independent harmonic oscillator. The update of each sheet position and velocities are then given by the analytical solution of the equation of motion:
\begin{equation}
\eqalign{
    \tilde{x}^{t+\Delta t} &= x^t + \omega_p^{-1}v^t\sin\left(\omega_p\Delta t\right) - \xi^t\left(1 - \cos\left(\omega_p\Delta t\right)\right) \cr
    \tilde{v}^{t+\Delta t} &= v^t\cos\left(\omega_p\Delta t\right) - \omega_p\xi^t\sin\left(\omega_p\Delta t\right)
\label{eq:xv_update}
}
\end{equation}
where $\xi^t = x^t - x_{eq}^t$. If no crossings occur, this is the exact solution for the sheet dynamics,  and the energy of the system is perfectly conserved.

\subsection*{SM -- Resolve crossings}

\begin{figure}[tb]
    \centering
    \includegraphics[width=\textwidth]{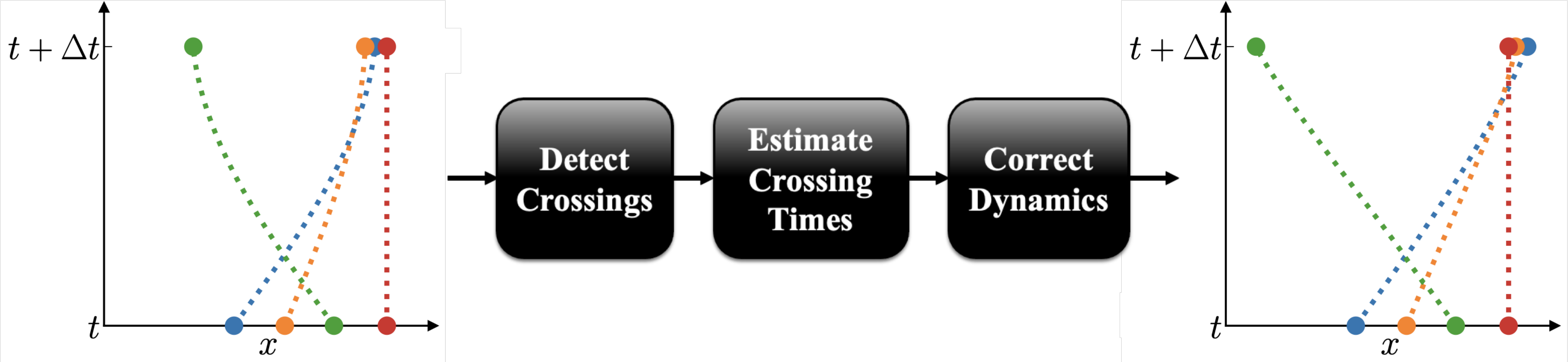}
    \caption{Schematic representation of the crossing correction algorithm (``Handle Crossings'' block in \fref{fig:simulators}). The dotted lines on the left plot represent the independent oscillatory trajectories at higher temporal resolution (not actually computed during an update of the algorithm). On the right plot, they correspond the the high-resolution trajectories when accounting for the set of crossings detected in this iteration (not all crossings that occurred between $t \rightarrow t + \Delta t$ are necessarily detected in this iteration). Each of the three building blocks is illustrated in detail in~\fref{fig:sm_crossing_correction}.}
    \label{fig:sm_crossing_correction_main}
\end{figure}

\begin{figure}[htb]
    \centering    
    \includegraphics[width=0.7\textwidth]{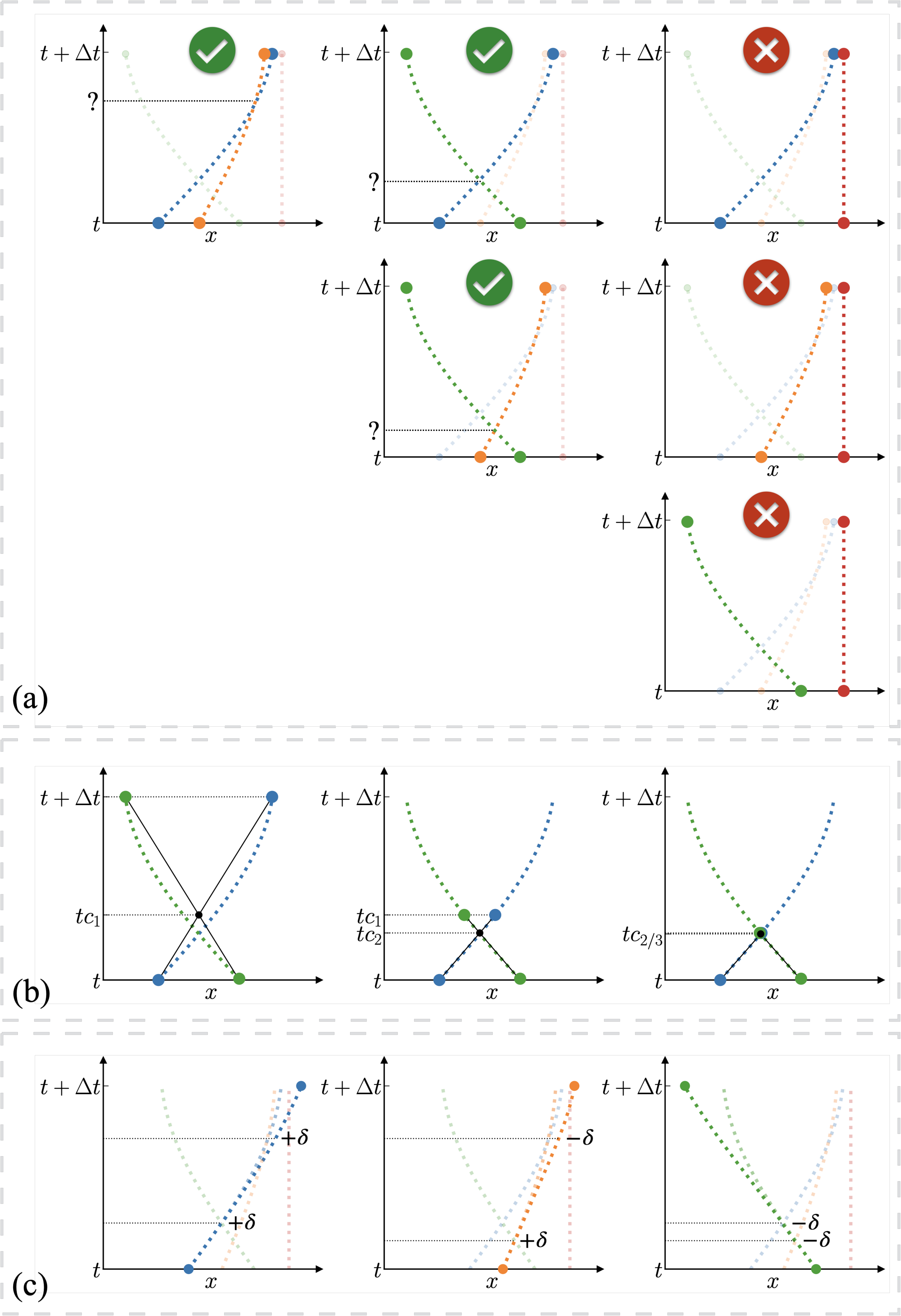}
    \caption{Illustration of the three main building blocks of the crossing correction algorithm shown in~\fref{fig:sm_crossing_correction_main}. For simplicity, we assume $\bi{x}^t$ was initially sorted. (a) First the algorithm detects, from left to right over $\tilde{\bi{x}}^{t+\Delta t}$, if sheets have crossed,i.e.~if \eqref{eq:crossing_detection} is fulfilled; (b) For each detected crossing, the crossing time is estimated using an iterative method \eqref{eq:crossing_time}; (c) Finally, each sheet trajectory is independently corrected by iteratively advancing until its next estimated crossing time, at which point the equilibrium position is updated by $\pm \delta$ (sign related to the direction of the crossing). Note that after these corrections, new crossings might occur. These new crossings will only be resolved in the next iteration of the algorithm ($t + \Delta t \rightarrow t + 2\Delta t$, the estimated $\Delta tc$ will be negative) due to the particularities of the sorting routine (more details in~\fref{fig:sm_sorting}).}
    \label{fig:sm_crossing_correction}
\end{figure}

The algorithm proceeds to correct for crossings which might have happened between $t \rightarrow t + \Delta t$ (or even previous time intervals as it will be later shown). The main steps are illustrated in~\fref{fig:sm_crossing_correction_main} and \fref{fig:sm_crossing_correction}.

To detect crossings, the algorithm checks, from left to right over the position array, for the condition:
\begin{equation}
    \tilde{x}_j^{t+\Delta t} > \tilde{x}_i^{t+\Delta t} \quad, \ \mathrm{for} \quad j > i \ 
    \label{eq:crossing_detection}
\end{equation}
for each $i$-th sheet (note that this sets a constraint on the maximum allowed time-step,i.e.~$\Delta t\leq \pi/2~\omega_p^{-1}$). Since checking all $j > i$ sheets is problematic for larger system sizes, a stopping condition is introduced. 
More precisely, we stop once:
\begin{equation}
    \tilde{x}_j^t - \tilde{x}_i^t > \Delta x^t_{max_i}
\label{eq:sm_stopcriterion}
\end{equation}
where $\Delta x^t_{max_i}$ represents the maximum initial distance that can result in a crossing with the $i$-th sheet. In \cite{dawson1970electrostatic}, Dawson uses the approximation:
\begin{equation}
    \Delta x^t_{max_i} \simeq \left(v^{max}_- + v^t_i\right) \omega_p\Delta t
\end{equation}
where $-v^{max}_- \leq 0$ is the maximum negative velocity across all sheets, whose value is updated every $\omega_p^{-1}$. However, we found that this approximation is not sufficient for non-thermal initial conditions with very large sheet velocities and displacements from equilibrium (e.g.~the two-stream instability simulations shown in Section~\ref{sec:twostream} and \ref{app:twostream}) since it underestimates the real $\Delta x^t_{max_i}$ value. As a consequence, some crossings are not detected, which causes problems when trying to sort sheets after the crossing corrections (i.e. multiple sheets would occupy the same array index).
For this reason, we instead derive a more accurate upper limit from~\eqref{eq:xv_update} that is:
\begin{equation}
    \Delta x^t_{max_i} = \left(v_-^{max} + v_i^t\right)\sin\left(\omega_p\Delta t\right) + \left(\xi_+^{max} - \xi_i^t\right)\left(1-\cos(\omega_p\Delta t)\right)
\end{equation}
which simplifies for the case where $\omega_p\Delta t \ll 1$ to:
\begin{equation}    
    \Delta x^t_{max_i} \simeq \left(v_-^{max} + v_i^t\right)\omega_p\Delta t + \left(\xi_+^{max} - \xi_i^t\right)\frac{(\omega_p\Delta t)^2}{2}
\end{equation}
where $\xi_+^{max} \geq 0$ is the maximum displacement from the equilibrium position across all sheets. Additionally, we update the values of $v_-^{max}$ and $\xi_+^{max}$ at every time step (instead of every $\omega_p^{-1}$) since these values can change significantly for the non-thermal scenario. These changes resolved the aforementioned problems in non-thermal scenarios with fast sheet acceleration and did not impact significantly the run-time.

For each detected crossing, the crossing time $tc = t + \Delta tc$ is estimated using an iterative (recursive) method:
\begin{equation}
    \Delta tc_{k} = \Delta tc_{k-1} \frac{x_j^t - x_i^t}{x_j^t - x_i^t + \tilde{x}_i^{t + \Delta tc_{k-1}} - \tilde{x}_j^{t + \Delta tc_{k-1}}}
    \label{eq:crossing_time}
\end{equation}
where $k$ is the number of iterations performed ($\Delta tc_0 = \Delta t$) and $\tilde{x}^{t+\Delta tc_{k-1}}$ is computed using~\eqref{eq:xv_update}.

Once the full list of crossing times for the $i$-th sheet is compiled, the sheet dynamics in the interval $t \rightarrow t + \Delta t$ are recomputed. This is done by first sorting the list of crossing times. Then, one advances the sheet position and velocity until the first predicted crossing ($t \rightarrow tc^{(1)}$) according to \eqref{eq:xv_update}. At this point, one updates its equilibrium position by $-\delta$ if it was crossed by a sheet from its left, or by $+\delta$ if it was crossed by a sheet from its right.  This process is then repeated to model the dynamics between the subsequent crossings ($tc^{(i)} \rightarrow tc^{(i+1)}$) and the end of the current simulation step ($tc^{(n)} \rightarrow t + \Delta t$).

Although one corrects for multiple crossings in a single simulation step, the crossing time estimation \textit{ignores} the fact that the $i$-th sheet might have crossed other sheets between $t \rightarrow tc^{(i)}$. This means that, if a sheet crosses $n\!>\!1$ other sheets in a single time step, its dynamics, and those of the sheets it crosses, are not correctly modeled (although a good approximation might be obtained). More precisely, the crossing times are overestimated, which results in an extra ``drag'' felt by the sheets (since their equilibrium positions are updated later than they should), leading to an overall energy loss in the system. 

Additionally, the introduced corrections might lead to the appearance of new crossings between $t \rightarrow t + \Delta t$, which were not accounted for (e.g.~crossing of blue and red sheet in~\fref{fig:sm_crossing_correction}). In this case, the final equilibrium positions of the sheets are incorrect,i.e.~we would observe that $\tilde{x}_j^{t+\Delta t} < \tilde{x}_i^{t+\Delta t}$ while $\tilde{x}_{eq_j}^{t+\Delta t} > \tilde{x}_{eq_i}^{t+\Delta t}$. This is not problematic as the algorithm is capable of correcting for these crossings in the next time step due to the particularities of the sorting procedure.

\subsection*{SM -- Sort sheets}

\begin{figure}[tb]
    \centering
    \includegraphics[width=\textwidth]{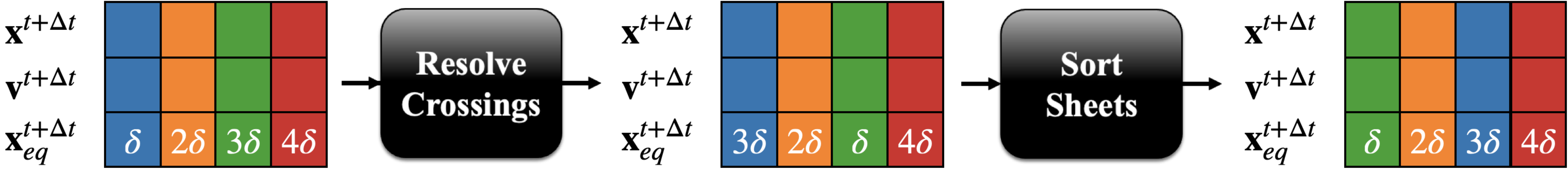}
    \caption{Illustration of changes in the array order after correcting for crossings. Colors represent the labels (array is omitted) and correspond to the sheets in~\fref{fig:sm_crossing_correction_main} and \fref{fig:sm_crossing_correction}. For simplicity, the values indicated in the $\bi{x}_{eq}^{t+\Delta t}$ array represent the equilibrium position $+\delta/2$. Once the corrections for crossings shown in~\fref{fig:sm_crossing_correction} are applied, the final sheet positions, velocities, and equilibrium positions have changed. These arrays are then sorted with respect to the equilibrium positions, and not with respect to the sheet positions. This ensures that in the next time step,  the crossing correction algorithm will be able to correct the crossings that were previously overlooked.}
    \label{fig:sm_sorting}
\end{figure}

After correcting for crossings, a sorting step is applied (see~\fref{fig:sm_sorting}). Each sheet moves $\pm1$ position along the arrays for each detected crossing with a neighbor from its right/left. This is equivalent to sorting all $t+\Delta t$ arrays with respect to $\tilde{\bi{x}}_{eq}^{t+\Delta t}$ (which as explained is not always equivalent to sorting with respect to $\tilde{\bi{x}}^{t+\Delta t}$). The sorting is done in this fashion since the crossing corrections performed during $t\rightarrow t+\Delta t$ might have led to the appearance of new crossings that were not initially accounted for (e.g.~the crossing between blue and red sheets in~\fref{fig:sm_crossing_correction}). In the next iteration of the algorithm ($t+\Delta t \rightarrow t + 2\Delta t$) these crossings will be detected by \eqref{eq:crossing_detection} (which might not have happened if the arrays had been sorted with respect to $\tilde{\bi{x}}^{t+\Delta t}$) and the associated $\Delta tc$ will be negative.

For the case where no correction for crossings between $t \rightarrow t+\Delta t$ is introduced,i.e.~when we bypass the crossing routine, we slightly modify the sorting routine. The position and velocity arrays are sorted with respect to $\tilde{\bi{x}}^{t+\Delta t}$ after the equation of motion update, while no change is applied to the equilibrium position array (i.e. equilibrium positions are re-assigned). 

\subsection*{SM -- Resolve boundary}

Before the next iteration of the algorithm, it is necessary to update the guard sheets to account for crossings (in both boundary conditions) or sheets that might have left the simulation box (when using a periodic boundary).

In the case of reflecting boundaries, we perform two steps: (a) we overwrite the first and last $N_g$ entries of $\tilde{\bi{x}}^{t+\Delta t}$ and $\tilde{\bi{v}}^{t+\Delta t}$ ($N_g$ equals the number of guards) to represent mirrored versions of the current first and last ``real'' sheets (as per the current ordering of the arrays); and (b) we swap the labels of the guard sheets that entered the simulation box with the corresponding ``real'' sheet that left the box.

In the case of periodic boundaries, we perform three steps: (a) we rotate the $\tilde{\bi{x}}^{t+\Delta t}$, $\tilde{\bi{v}}^{t+\Delta t}$, and $\tilde{\bi{x}}_{eq}^{t+\Delta t}$  arrays so that the sheets inside the simulation box at $t+\Delta t$ are centered; (b) we overwrite the first and last $N_g$ entries to represent the equivalent versions of the opposing boundary sheets; and (c) we swap the labels of the guard sheets that entered the simulation box with the corresponding ``real'' sheet that left the box.

This finalizes an iteration of the sheet model algorithm, producing the arrays $\bi{x}^{t+1}$, $\bi{v}^{t+1}$, $\bi{x}_{eq}^{t+1}$ to be used as input for the next iteration.

\subsection*{SM -- Possible parallellization}

No parallelization of the algorithm is mentioned in the original work~\cite{dawson1970electrostatic}. However, it is clear to us that the synchronous sheet model could benefit from a parallelized approach to minimize run-time. This should be straightforward for the equation of motion update, but slight changes might be required to the crossing correction routine to make it thread-safe (in particular the crossing detection routine). Designing and implementing these new parallelized versions is outside the scope of this work.

\subsection{Modified Synchronous Algorithm (MSM)}

To allow for additional comparisons between the sheet model and better understand the limitations of the GNS we implemented a slightly modified version of the algorithm which limits the number of neighbors checked for crossing corrections.

This is achieved by adding an additional stopping criterion for the crossing check~\eqref{eq:crossing_detection},i.e.~we stop if $j\!>\!i\!+\!n$, where $n$ is the maximum number of neighbors checked (to the right). However, this change might lead to different sheets having the same equilibrium position after the crossing corrections, i.e.~we can not apply the sorting mechanism described previously. Therefore we modify the sorting routine to sort the corrected $\tilde{\bi{x}}^{t+\Delta t}$ and $\tilde{\bi{v}}^{t+\Delta t}$ arrays with respect to $\tilde{\bi{x}}^{t+\Delta t}$ (instead of $\tilde{\bi{x}}_{eq}^{t+\Delta t}$) and set $\tilde{\bi{x}}_{eq}^{t+\Delta t}=\bi{x}_{eq}^{t}$ (equivalent to what is done for the GNS). This removes the capability of the algorithm to correct for crossings with neighbors above the established limit in subsequent iterations (leading to poorer energy conservation) and allows for a unique ordering.

This modified version should not be used as an alternative to the original sheet model as it was merely introduced for improved interpretability of the GNS results.

\subsection{Asynchronous Algorithm (ASM)}

\begin{figure}[tb]
    \centering
    \includegraphics[width=\columnwidth]{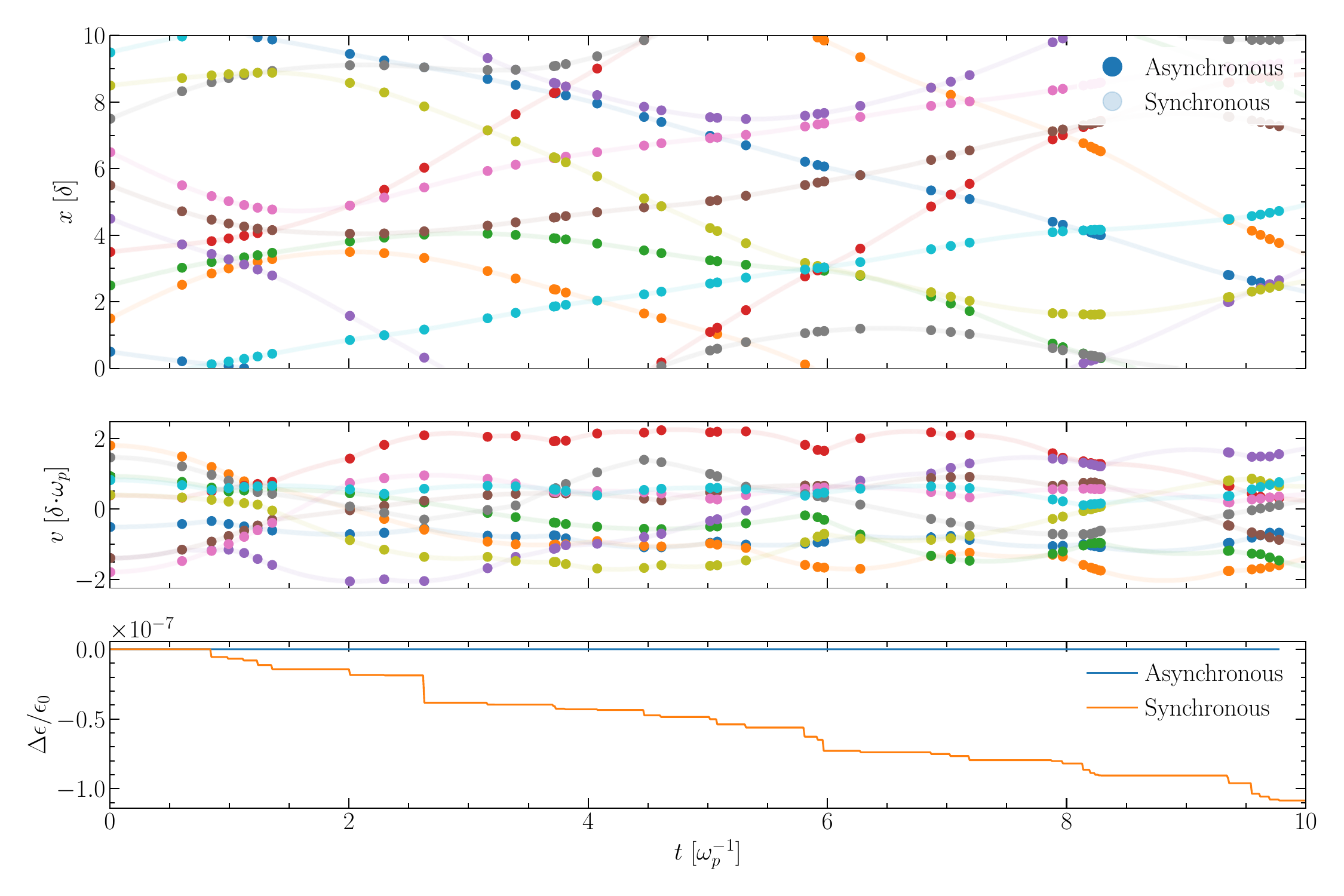}
    \caption{Comparison of asynchronous  \textit{vs.} synchronous sheet model simulation. The synchronous simulation is run with $\Delta t=10^{-2}~\omega_p^{-1}$. Each data point of the asynchronous algorithm corresponds to a timestamp where a sheet crossing occurred.}
    \label{fig:async_vs_sync}
\end{figure}

Unlike the synchronous sheet model, which uses a fixed time step, the asynchronous version of the sheet model algorithm always advances the full system until the next predicted crossing (example in \fref{fig:async_vs_sync}). For this reason, we need only to store one set of arrays $\bi{x}$, $\bi{v}$, $\bi{x_{eq}}$ and $\bi{l}$ which are updated at every crossing.

For this method to work a sorted table containing the crossing times for adjacent sheets needs to be maintained and updated after each crossing. We represent this table as a standard priority queue implemented using a heap (Dawson uses a slightly different mechanism~\cite{dawson1970electrostatic} but with similar end results). This is also the approach followed by Gravier et al.~\cite{gravier2023collision} for the multi-species scenario (refer to this work for illustrations on the heap structure). In our case, each entry of the priority queue corresponds to a list with [$tc_{ij}$, $c$, $\mathrm{tag}_{ij}$] where $tc_{ij}$ is the estimated crossing time between sheets $i$ and $j$ (by definition $j = i+1$), $c$ a unique sequence count (used to break ties for crossings occurring at the exact same time), and $\mathrm{tag}_{ij}$ a tag containing the array indices of the sheets that crossed (not their labels). To find and edit entries in the heap, we initialize an auxiliary dictionary that maps tags to entries.

\subsection*{ASM -- Algorithm iteration}

At initialization, all crossing times between adjacent sheets are computed and the corresponding entries are added to the priority queue. For pairs of sheets that do not cross, no entry is added. 

An iteration of the algorithm then proceeds as follows: (a) we remove from the heap the next crossing,i.e.~the one with the smallest $tc_{ij}$ value; (b) we advance the dynamics until $tc_{ij}$ using \eqref{eq:sheet_motion}; (c) we swap the order of the entries of the sheets that crossed in the $\bi{x}$, $\bi{v}$, and $\bi{l}$ arrays (no change to $\bi{x}_{eq}$); (d) if needed we update the guards; (e) we compute the new crossing time for any pair that involved one of the sheets that crossed, i.e.~($i-1$, $i$), ($i,j$), and ($j$, $j+1$); (f) we flag existing entries in the heap involving these pairs so that they are ignored in later iterations (more specifically, the tags are replaced by a ``removed'' flag); and (g) we add the new crossing entries to the heap and dictionary.

\subsection*{ASM -- Guard sheets}

For the asynchronous model, one needs only to introduce one guard for periodic boundaries (we place it on the left side,i.e.~$i=0$) and two for reflecting boundaries (one on each side,i.e.~$i=0$ and $i=N+1$). The positions, velocities, and equilibrium positions are computed in the same fashion as for the synchronous algorithm. 

These guards will be updated whenever: (a) a crossing between the guard sheet and its neighbor,i.e.~the sheet closer to the respective boundary, occurs; and (b) a crossing involving the ``real'' sheet that the guard sheet represents and any of its neighbors occurs. 

In the case of periodic boundaries, one must also update the ``real'' sheet corresponding to the guard sheet if the latter performs a crossing. Alternatively one can use two guards (as for reflecting boundaries, this is what is present in~\cite{dawson1970electrostatic}) but we found that the chosen approach makes it easier to post-process the sheet trajectories.

\subsection*{ASM -- Analytical solution for the crossing times}

Another difference regarding the synchronous algorithm is that we now use the analytical solution for the crossing times instead of the estimation using the iterative method \eqref{eq:crossing_time}. This was the approach followed by Dawson when implementing the multi-species asynchronous algorithm~\cite{dawson1970electrostatic} and can be similarly applied for the single-species scenario (although the equation to be solved is different). When adopting this strategy, and using a double-precision floating point format, we observe relative energy fluctuations within the range $\Delta \epsilon/\epsilon_0 \in [10^{-15}, 10^{-10}]$ for simulations equivalent to the ones present in the training dataset generated with the synchronous algorithm.

Crossing times between neighboring sheets are computed using the analytical solution to the transcendental equation $x_i(t + \Delta tc) = x_{i+1}(t + \Delta tc)$ with respect to $\Delta tc$. For a solution to exist,i.e.~for neighboring sheets to be able to cross, the following condition has to be satisfied:
\begin{equation}
    \Delta v_i^2 + \left(\Delta x_i - \delta \right)^2  \geq \delta^2
    \label{eq:can_sheets_cross}
\end{equation}
where $\Delta v_i = \omega_p^{-1}(v_{i+1}^t - v_i^t)$, and $\Delta x_i = x_{i+1}^t - x_i^t$.
If this condition is fulfilled, there exist infinite (real) solutions to the crossing time. 

In the case where $\Delta v_i = 0$ and $\Delta x_i = 2\delta$, the solutions are:
\begin{equation}
    \omega_p\Delta tc = \pi + 2n\pi
\label{eq:tc_analytical1}
\end{equation}
where the $2n\pi$ ($n\in\mathbb{Z}$) factor represents solutions equally spaced by a plasma period. For $\Delta v_i \neq 0$ and $\Delta x_i = 2\delta$:
\begin{equation}
    \omega_p\Delta tc = 2\arctan\left(-\frac{\Delta x_i - \delta}{\Delta v_i}\right) + 2n\pi,
\label{eq:tc_analytical2}
\end{equation}
and for all other cases:
\begin{equation}
\omega_p\Delta tc = 2 \arctan\left(\frac{\Delta v_{i}^2 \pm \sqrt{\Delta v_i^2 + \left(\Delta x_i - \delta\right)^2 - \delta^2}}{\left(\Delta x_i- \delta \right)^2  - \delta^2}\right) + 2n\pi.
\label{eq:tc_analytical3}
\end{equation}
In our implementation, we set $n=0$ obtaining, at most, two estimates for $\Delta tc$ (one if $\Delta x_i = 2\delta$,  or two otherwise). Furthermore, we enforce these solutions to be within the interval $]0, 2\pi]$ by computing the modulo of $\omega_p\Delta tc$ with respect to $2\pi$. In the case where there exist two possible solutions within this interval,i.e.~\eqref{eq:tc_analytical3}, we pick the minimum of the two except for the case where the corresponding sheets have just crossed. In this case, we set:
\begin{equation}
    \omega_p\Delta tc = 2 \arctan\left(\mathrm{sgn}(\Delta v_i) \frac{|\Delta v_i| + \sqrt{\Delta v_i^2  + \left(\Delta x_i - \delta\right)^2 - \delta^2}}{\left(\Delta x_i- \delta \right)^2  - \delta^2}\right)
\end{equation}
to avoid the trivial solution $\Delta tc = 0$.  Additionally, to avoid numerical issues for the cases where $\Delta x_i \simeq 2\delta$, we relax the equality conditions of \eqref{eq:tc_analytical1} and \eqref{eq:tc_analytical2} to $|\Delta v_i| < 10^{-5}~\delta$ and $|\Delta x_i - 2 \delta| < 10^{-5}~\delta$.

\section{Collisions vs. crossings}
\label{app:collisions}

It is possible to model the sheet crossings as binary collisions (as previously illustrated in \fref{fig:crossings_vs_collisions}). This was the strategy initially explored in this work, since the Graph Network simulator proposed by Sanchez-Gonzalez et al.~\cite{sanchez2020learning} was introduced for 2D/3D simulations where particle collisions occur. However, the accuracy of the models trained in this fashion was significantly worse. In this subsection, we explain why this happens.

\begin{figure}[htb]
    \centering
    \includegraphics[width=\textwidth]{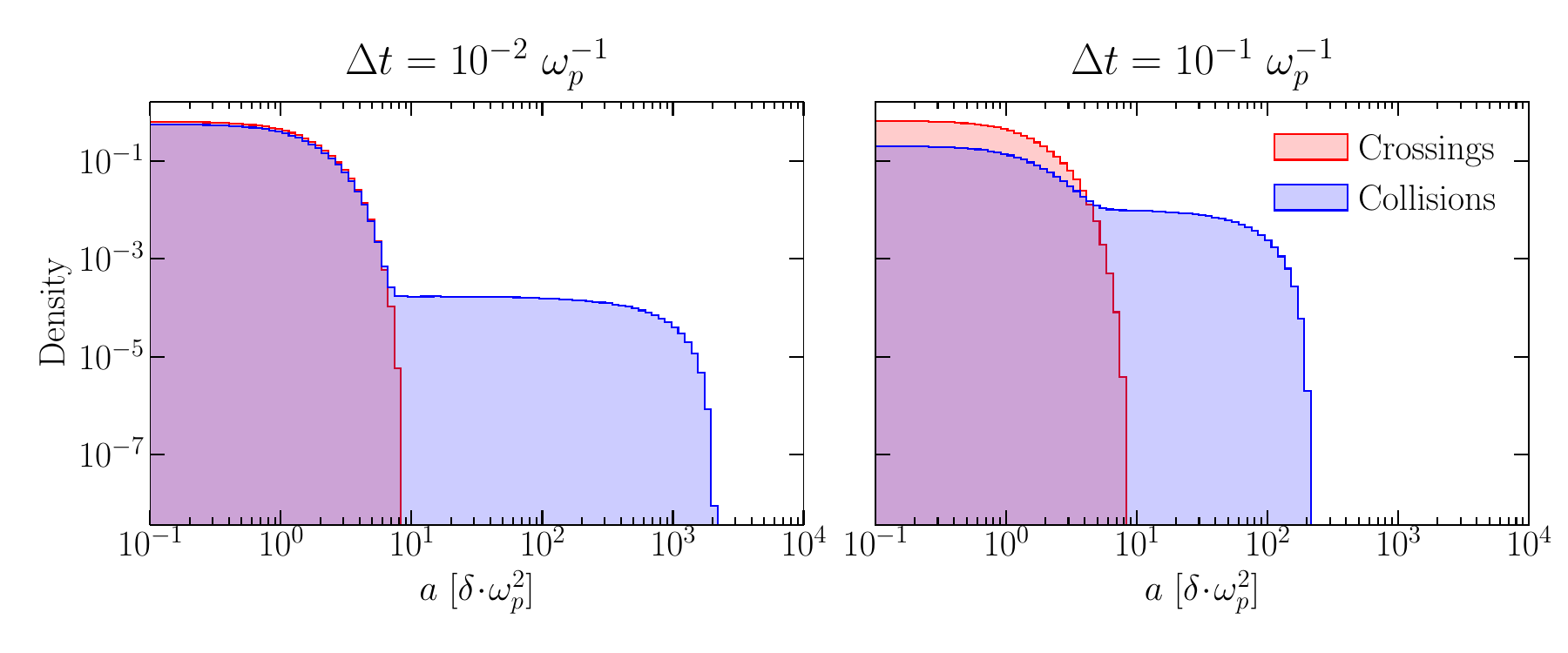}
    \caption{Target acceleration distributions in the training set data when considering collisions or crossings. The range of target accelerations is considerably smaller for the crossing scenario. This makes the training procedure significantly easier.}
    \label{fig:crossings_vs_collisions_targets}
\end{figure}

When trying to predict dynamics considering binary collisions, the range of target accelerations changes significantly (see \fref{fig:crossings_vs_collisions_targets}) since at the moment of collision, sheets feel a force that is orders of magnitude larger than the one felt during their oscillatory motion. This broadening makes it harder for the network to accurately model the full dynamic range of accelerations (even if we normalize the targets to unit variance).

Additionally, in the time step where sheets collide, their respective finite difference velocities significantly decrease, and information is lost about the initial momentum of the sheets (especially for larger $v$ and small $\Delta t$). Therefore, to allow for correct momentum transfer,  it is necessary to include the previous $C$ step velocities in the node representation $\textbf{n}^t_i = \left[\xi^t_i, v^t_i, ..., v^{t-C+1}\right]$ as in Sanchez-Gonzalez \textit{et al.}~\cite{sanchez2020learning}. This introduces a new hyperparameter that requires tuning and whose value depends on the frequency of collisions.

For larger collision frequencies, it then becomes hard to correctly model momentum transfer since the model needs to predict a larger dynamics range of accelerations, and to take into account dynamics across multiple time steps and larger neighboring regions. In our case, this resulted in considerably larger neural network architectures, whose accuracy was still significantly worse than models trained to predict crossing dynamics. 

\begin{figure}[tb]
    \centering
    \includegraphics[width=\textwidth]{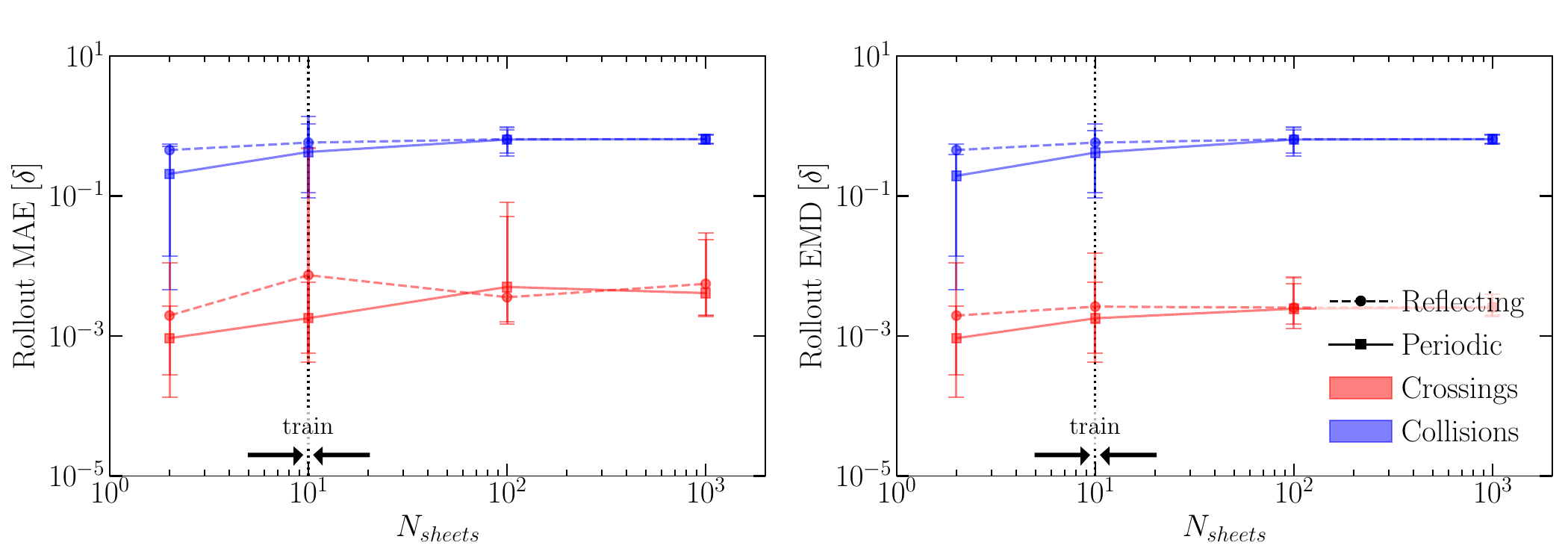}
    \caption{Accuracy of equivalent models trained to predict crossing or collisional dynamics. Both models use $\Delta t = 10^{-1}~\omega_p^{-1}$. The rollout error of the GNS model trained for crossing dynamics is significantly lower. This is the main reason why we opted to model sheet dynamics considering crossings in the main body of the paper.}
    \label{fig:benchmark_crossings_vs_collisions}
\end{figure}

To illustrate this, we showcase in \fref{fig:benchmark_crossings_vs_collisions} a comparison between the rollout accuracy difference presented in the main body of the paper for $\Delta t = 10^{-1}~\omega_p^{-1}$, and an equivalent one trained for collisional dynamics using $C=5$ (same neural network architecture and number of message passing steps). It is clear that the accuracy of the simulator degrades significantly when attempting to predict collisions. The significant difference in performance does not disappear when introducing noise during training, a strategy proposed in Sanchez-Gonzalez et al.~\cite{sanchez2020learning} to improve rollout accuracy, or by normalizing the targets to unit variance (latter already included in the model shown in~\fref{fig:crossings_vs_collisions}).

It is possible that for larger simulation steps, the problems stated above become less pronounced since the effect of collisions is smoothed out over larger time steps (leading to a smaller dynamic range of target accelerations as observed in \fref{fig:crossings_vs_collisions_targets}, when moving from $\Delta t=10^{-2}~\omega_p^{-1}$ to  $\Delta t=10^{-1}~\omega_p^{-1}$). However, in the 1D sheet model scenario, our results indicate that modeling sheet interactions as crossings is the best and most natural option in terms of model/graph size and accuracy, while also allowing the tracking of particular sheets (useful, for example, for the fast sheet diagnostic).

Finally, it is important to highlight that when using reflecting boundaries the sheet trajectories contain ``collisions'' with the wall even when considering crossing dynamics. This is the reason why we first train our models on periodic boundaries and only at test time demonstrate that it is possible to simulate reflecting boundaries (and not the opposite). It would be possible to bypass this issue, while maintaining the same graph structure if, for example, we introduced an extra preprocessing step during training that masks out time steps for which collisions with the wall happen (or correct the target accelerations). However, this removes any interaction with the wall from the training data (defeating the purpose of having reflecting boundaries in the first place) and would require us to store extra information during the generation of the training simulations.

\section{Parameter scans}
\label{app:parameter_scans}

The main goal of this work was to understand if we can build a surrogate simulator capable of recovering kinetic plasma processes with acceptable run-time. We did not perform extensive hyperparameter tuning and instead opted for a coarse scan. For the models showcased in the main body of the text, we prioritized simplicity in favor of smaller gains in performance. Additionally, we decided to use the same architecture for both training simulation steps, except for the number of message-passing steps, as this parameter is related to the crossing frequency.

\begin{figure}[htb]
    \centering
    \includegraphics[width=\columnwidth]{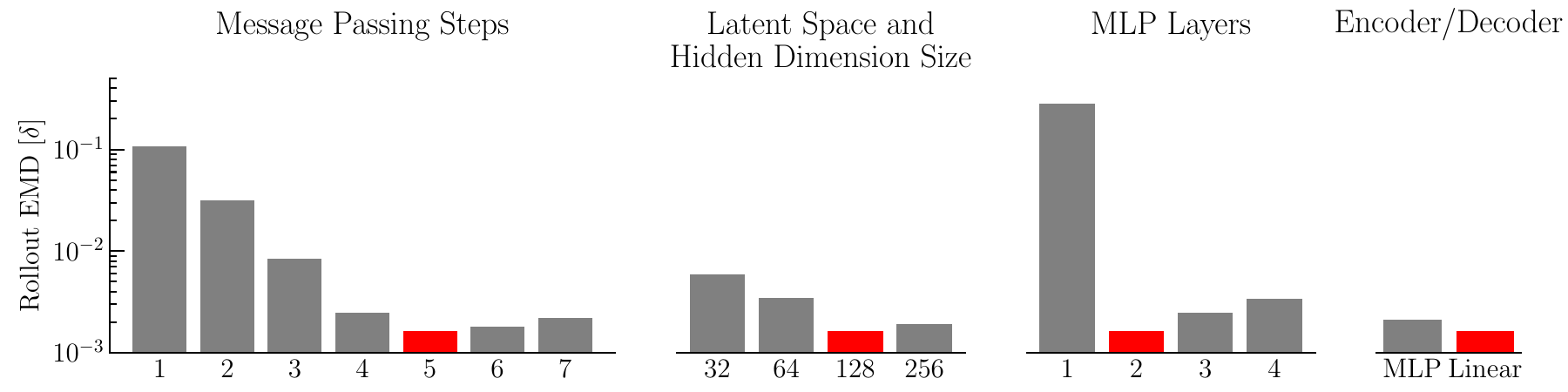}
    \caption{Parameter scan results for GNS trained with $\Delta t = 10^{-1}~\omega_p^{-1}$. The rollout EMD values were computed on the validation set. The final chosen hyperparameter values are highlighted in red. The most important parameter to tune is the number of message-passing steps, whose optimal value depends on the expected sheet crossing frequency.}
    \label{fig:param_scan}
\end{figure}

The rollout accuracy on the validation set for models trained with different hyper-parameters for $\Delta t = 10^{-1}~\omega_p^{-1}$ is shown in~\fref{fig:param_scan}. The most impactful parameter is the number of message-passing steps. Increasing the number of layers and hidden dimensions size of the MLP networks does not affect performance significantly and might lead the model to overfit the training data earlier in the training process. Similarly, introducing extra complexity by using MLPs for the encoder/decoder instead of linear transformation does not result in significant performance changes.

We also checked if the introduction of residual connections between message-passing blocks would improve performance for larger numbers of message-passing steps, but it did not provide significant improvements.

\section{Equivariant \textit{vs} non-equivariant architectures}
\label{app:equivariantgnn}

The Graph Network block presented in Section~\ref{sec:gnn} can be altered to use solely the sent edges in the node update mechanism, i.e.:
\begin{equation}
\mathbf{v}_{i}^{m+1} = \phi^{v}\left(\overline{\mathbf{e}}^{m+1}_{r_i}, \mathbf{v}^{m}_i\right)
\end{equation}
which corresponds to the update mechanism presented in the original work by Battaglia~{et al.}~\cite{battaglia2018relational}.

Additionally, one can make the full architecture (1D) rotationally equivariant by modifying the architecture as follows. The encoder should now be given by:
\begin{equation}
\eqalign{
    \mathbf{v}_i = \varepsilon^v\left(\mathbf{n}_{i} \cdot \mathrm{sgn}\left(\xi_i\right)\right) \cr
    \mathbf{e}_{ij} = \varepsilon^e\left(\mathbf{r}_{ij} \cdot \mathrm{sgn}\left(\xi_i\right)\right)
}
\end{equation}
where $\mathrm{sgn}(\cdot)$ represents the sign function. In this case the inputs become rotationally invariant, and, therefore, the latent representations $\mathbf{v}_i$ and $\mathbf{e}_{ij}$ are also guaranteed to be invariant. The multiplication by $\mathrm{sgn}\left(\xi_i\right)$ can be interpreted as a change of coordinate system (or a projection since in 1D the $\mathrm{sgn}(x) \equiv x/||x||$) to a frame where the $i$th sheet is on the right-hand side of its equilibrium position (i.e. $\xi > 0$). For the case where $\xi_i = 0$ (which would result in $\mathrm{sgn(0)}=0$), we use instead the velocity (i.e. $\mathrm{sgn}(v_i)$). If the latter is also zero, we set the value to 1. These extra steps are required to ensure that the architecture maintains its equivariance properties.

Similarly, the processor is also made rotationally invariant with a slight modification:
\begin{equation}
\eqalign{
    \mathbf{e}_{ij}^{m+1} & = \phi^e\left(\mathbf{e}^{m}_{ij}, \mathbf{v}^m_i, \mathrm{sgn}\left(\xi_i\xi_j\right) \cdot \mathbf{v}^m_j\right) \cr 
    \overline{\mathbf{e}}_{r_i}^{m+1} &= \sum_{j \in \mathcal{N}(i)}  \mathbf{e}^{m+1}_{ij} \cr
    \mathbf{v}_{i}^{m+1} &= \phi^{v}\left(\overline{\mathbf{e}}^{m+1}_{r_i}, \mathbf{v}^{m}_i\right) \label{eq:gn_equiv}
}
\end{equation}
since $\mathbf{e}^0_{ij}$, $\mathbf{v}^0_{i,j}$, and $\mathrm{sgn}\left(\xi_i\xi_j\right)$ are invariant to this transformation, making all subsequent updates invariant as well. In this case, the multiplication of the sender node latent vector $\mathbf{v}^m_j$ by $\mathrm{sgn}\left(\xi_i\xi_j\right)$ is required to break the symmetry introduced by the encoder (mirrored sheets around the equilibrium position have the same latent representation) and can be interpreted as a transformation to the frame of the receiver latent node $\mathbf{v}^m_i$ (i.e. where $\xi_i > 0$).

Finally, the modified decoder enforces equivariance as:
\begin{equation}
    \mathbf{y}_i = \delta^v\left(\mathbf{v}_i^M\right) \cdot \mathrm{sgn}\left(\xi_i\right)
\end{equation}
since $\mathbf{v}_i^{M}$ is guaranteed to be invariant under reflections.

\begin{figure}[t]
    \centering
    \includegraphics[width=0.5\columnwidth]{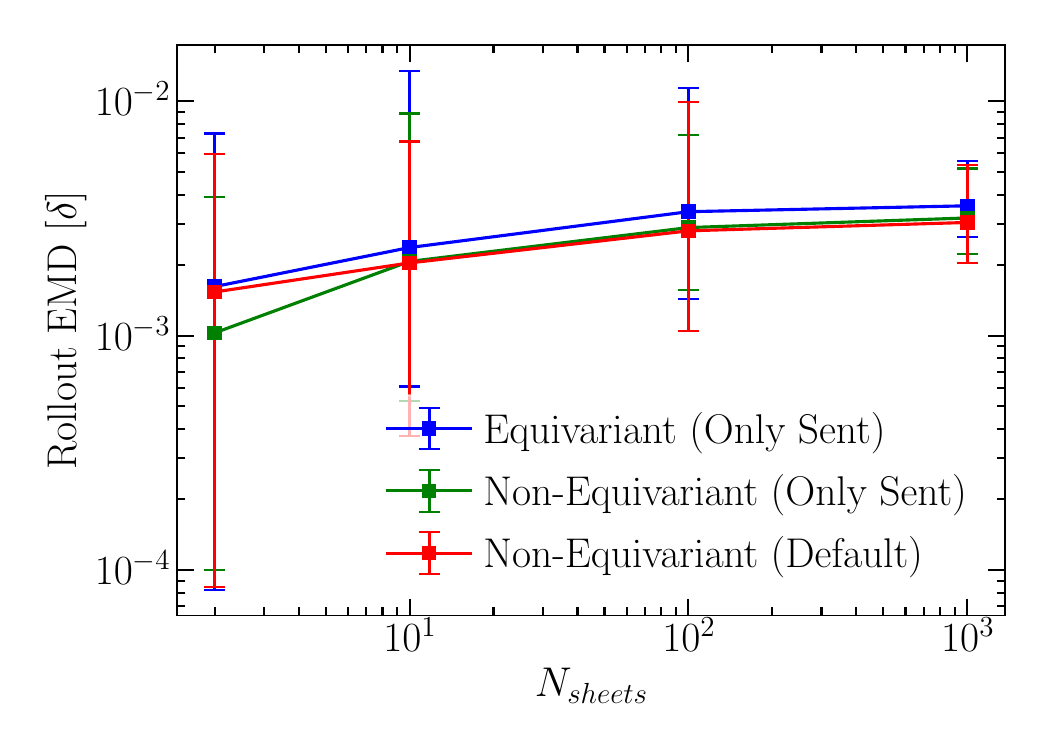}
    \caption{Comparison of rollout accuracy of different architectures. We use the test sets described in Section~\ref{sec:trajectoryprediction} that consider periodic boundary conditions. Mean values for each architecture are computed over 5 equivalent models (different random seeds), and error bars represent minimum/maximum values observed over all models. All models are trained considering a time step $\Delta t=10^{-1}~\omega_p^{-1}$ using the same hyperparameters described in Section~\ref{sec:gnn} and the training procedure described in Section~\ref{sec:training}. Non-Equivarient models are trained with data augmentation along both axes ($x$ and $t$) while equivariant models are only provided with data augmented along the $t$-axis. No significant differences in rollout accuracy are observed across the different architectures.} \label{fig:benchmark_equivariant_rollout}
\end{figure}

\begin{figure}[t]
    \centering
    \includegraphics[width=\columnwidth]{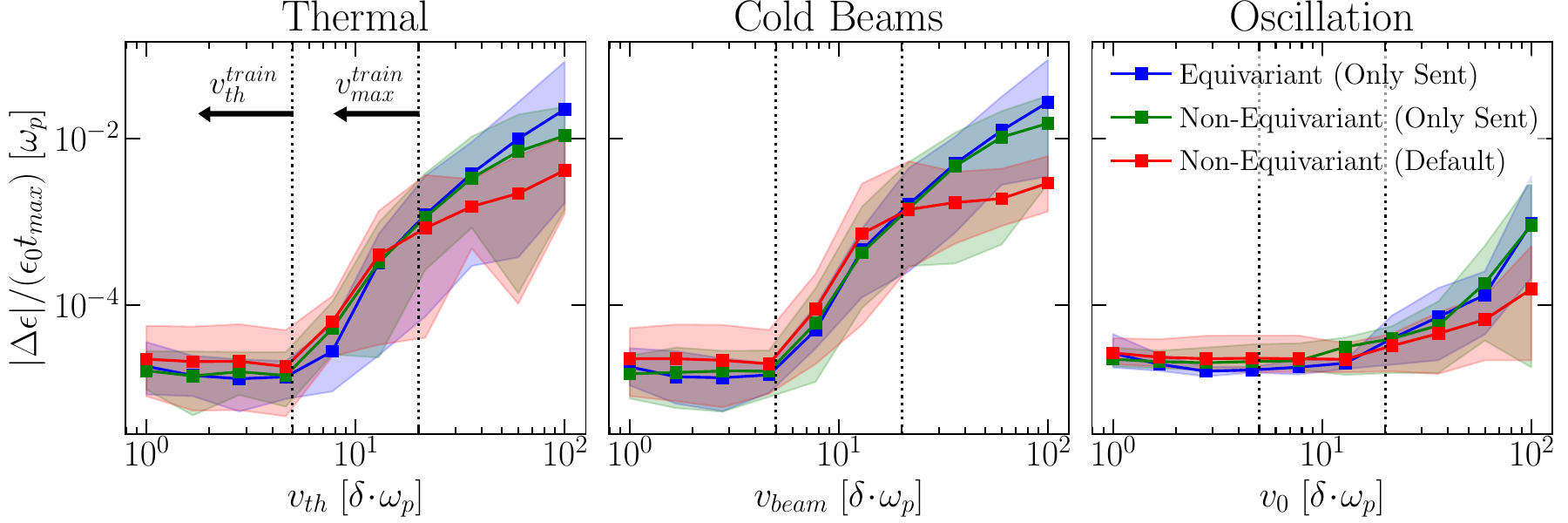}
    \caption{Comparison of energy conservation capabilities of different architectures for different initial conditions. We use the same energy conservation diagnostic described in Section~\ref{sec:energy_conservation}. All sheets are initialized at their equilibrium position and their initial velocities are sampled from a normal distribution (thermal), chosen randomly from $\pm v_{beam}$ (cold beams), or all equal to $v_0$ (oscillation). Mean values are computed over 5 equivalent models (different random seeds). All models are trained considering a time step $\Delta t=10^{-1}~\omega_p^{-1}$ using the same hyperparameters described in Section~\ref{sec:gnn} and the training procedure described in Section~\ref{sec:training}. Non-Equivariant models are trained with data augmentation along both axes ($x$ and $t$) while equivariant models are only provided with data augmented along the $t$-axis.} \label{fig:benchmark_equivariant_energy_all}
\end{figure}

\begin{figure}[t]
    \centering
    \includegraphics[width=.5\columnwidth]{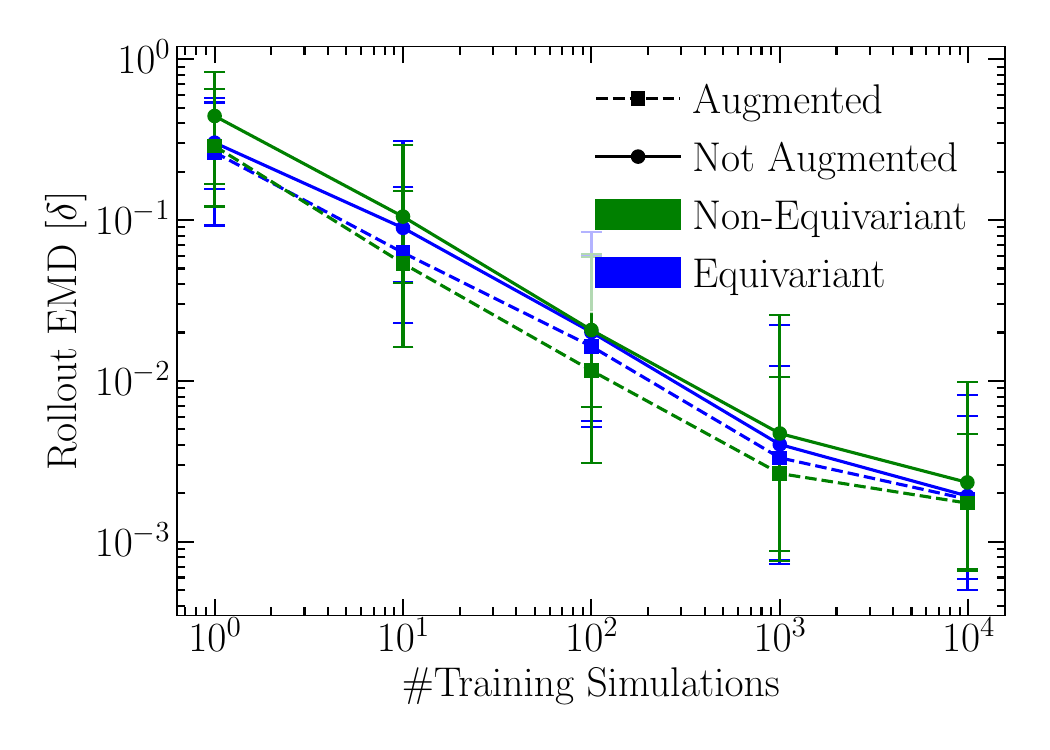}
    \caption{Rollout accuracy as a function of the number of training simulations. We consider architectures that only use the sent edges for the latent node update. A single model is trained for each dataset size. Training is slightly modified compared to previous results. We use a larger number of maximum gradient updates ($2\times10^6$) and early stopping with a patience of 100 epochs. Augmented datasets include reflections along both axes ($x$ and $t$). Overall, it is observed that there are no significant differences in performance between the architectures.} \label{fig:benchmark_equiv_nsimulations}
\end{figure}

Comparisons between the rollout accuracy and energy conservation capabilities of the architecture used in the main body of the paper (``Non-Equivariant Default''), and the variants that only use the sent edge information (``Non-Equivariant Only Sent'' and ``Equivariant Only Sent'') are presented in \fref{fig:benchmark_equivariant_rollout}, \fref{fig:benchmark_equivariant_energy_all}, and \fref{fig:benchmark_equiv_nsimulations}.

It is observed that the rollout accuracy is similar across all architectures (\fref{fig:benchmark_equivariant_rollout}). Similarly, the models demonstrate equivalent energy conservation within the training data thermal velocity range (\fref{fig:benchmark_equivariant_energy_all}). However, as one further increases the thermal velocity of the plasma, the non-equivariant architectures demonstrate better energy conservation capabilities, in particular the one used in the main body of the paper. In terms of data requirements, we do not observe relevant performance gains for a varying number of training simulations (\fref{fig:benchmark_equiv_nsimulations}). There is a slight improvement by using the equivariant architecture in the case where no data augmentation is used. This effect disappears once data augmentation is used at train time.

There can be several reasons for the lack of performance improvement when considering the equivariant architecture. Firstly, we are dealing with a relatively simple one-dimensional system, and the symmetry that we are trying to enforce might be easily learned by the network. This can explain why we do not observe significant performance improvements in the low data regime when compared to other works which learn to model charged-particle dynamics in 3D using equivariant architectures~\cite{ brandstetter2022geometric, satorras2021n}. Further support for this thesis is the fact that other experiments that included augmented data solely along the $x$-axis did not cause significant changes in the performance of the non-equivariant architecture (only using both $x$ and $t$ augmentation produced improvements).

Secondly, it can be argued that the way we enforce the equivariance is not ideal, since we are constraining significantly the latent node representations with the introduction of the multiplication term $\textrm{sgn}(\xi_i\xi_j)$ in the edge update. This was the option we found to work better. Furthermore, the performance of the equivariant and the equivalent non-equivariant architecture (the one which uses solely the sent edges for the node update) is basically the same across all metrics (rollout accuracy, energy conservation, etc.) except for considerably large thermal velocities ($v_{th} \gg v^{train}_{th}$). Therefore, we conclude that the imposed constraints are not affecting significantly the network learning capabilities, and the lack of improvements is mostly due to the simplicity of the setup and symmetry.

Finally, note that we do not propose an equivariant architecture that uses the sent edges for the node update mechanism. This is due to the fact that, in the proposed equivariant scheme, the sent edges are computed in the frame of the \textit{receiving} node. Therefore, one would have to add an operation that transforms them back to the sender node frame, or duplicate the amount of operations in the network in order to store and update the (latent) edges in both frames. 

At this point, it is not clear why using the sent edges in the node update of the non-equivariant architecture improves the energy conservation capabilities at larger thermal velocities. We do not observe any relevant difference in the validation loss between the different non-equivariant architectures, which leads us to believe that this effect is not due simply to a larger number of degrees of freedom of the network which allows for a better fit. Instead, the network seems to learn a structurally different update mechanism, which is more resilient to higher crossing rates.

Lastly, it is important to mention that \textit{all} architectures are capable of reproducing the kinetic plasma process presented in Section~\ref{sec:plasma_processes} with the exception of the Two-Stream Instability. The latter is only recovered using the non-equivariant architecture presented in the main body of the paper, due to the aforementioned improved energy conservation capabilities for larger sheet velocities.

\section{Boundary crossings re-injection}
\label{app:boundary_crossings}

For reflecting boundary conditions, the new positions, velocities, and equilibrium positions of the sheets are updated according to:
\begin{equation}
   \eqalign{
    x_i^{t+1} &=  
        \begin{cases}
            -\tilde{x}_i^{t+1} &,\textrm{ if} \ \tilde{x}_i^{t+1} < 0 \\
            \tilde{x}_i^{t+1} - L &,\textrm{ if} \ \tilde{x}_i^{t+1} \geq L \\
            \tilde{x}_i^{t+1} &,\textrm{ elsewhere}
        \end{cases} \cr
    v_i^{t+1} &=
        \begin{cases}
            -\tilde{v}_i^{t+1} &,\textrm{ if} \ \tilde{x}_i^{t+1} < 0 \ \lor \ \tilde{x}_i^{t+1} \geq L \\
            \tilde{v}_i^{t+1} &,\textrm{ elsewhere}
        \end{cases} \cr
    \bi{x}_{eq}^{t+1} &= \bi{x}_{eq}^{t}
    }
\label{eq:reflecting_boundary}
\end{equation}
where $\tilde{x}$, $\tilde{v}$ represent the positions and velocities after the ODE integration step. 

For periodic boundary conditions, the update rule is:
\begin{equation}
   \eqalign{
   x_i^{t+1} &=  
        \begin{cases}
            L + \tilde{x}_i^{t+1} &,\textrm{ if} \ \tilde{x}_i^{t+1} < 0 \\
            \tilde{x}_i^{t+1} - L &,\textrm{ if} \ \tilde{x}_i^{t+1} \geq L \\
            \tilde{x}_i^{t+1} &,\textrm{ elsewhere}
        \end{cases}\cr
    v_i^{t+1} &= \tilde{v}_i^{t+1} \cr
    \bi{x}_{eq}^{t+1} &= \bi{x}_{eq}^{t} + (n_{left} - n_{right}) \delta 
    }
\label{eq:periodic_boundary}
\end{equation}
where $n_{left}$, $n_{right}$ represent the number of particle that crossed the left and right boundary. For both boundary conditions, $\bi{x}^{t+1}$ is sorted after the update (and $\bi{v}^{t+1}$ accordingly) so that its indices match the equilibrium positions array $\bi{x}^{t+1}_{eq}$ (i.e.~the correct equilibrium position is attributed).

\section{Impact of training dataset size and data augmentation}
\label{app:dataset_size}

\begin{figure}[t]
    \centering
    \includegraphics[width=0.55\textwidth]{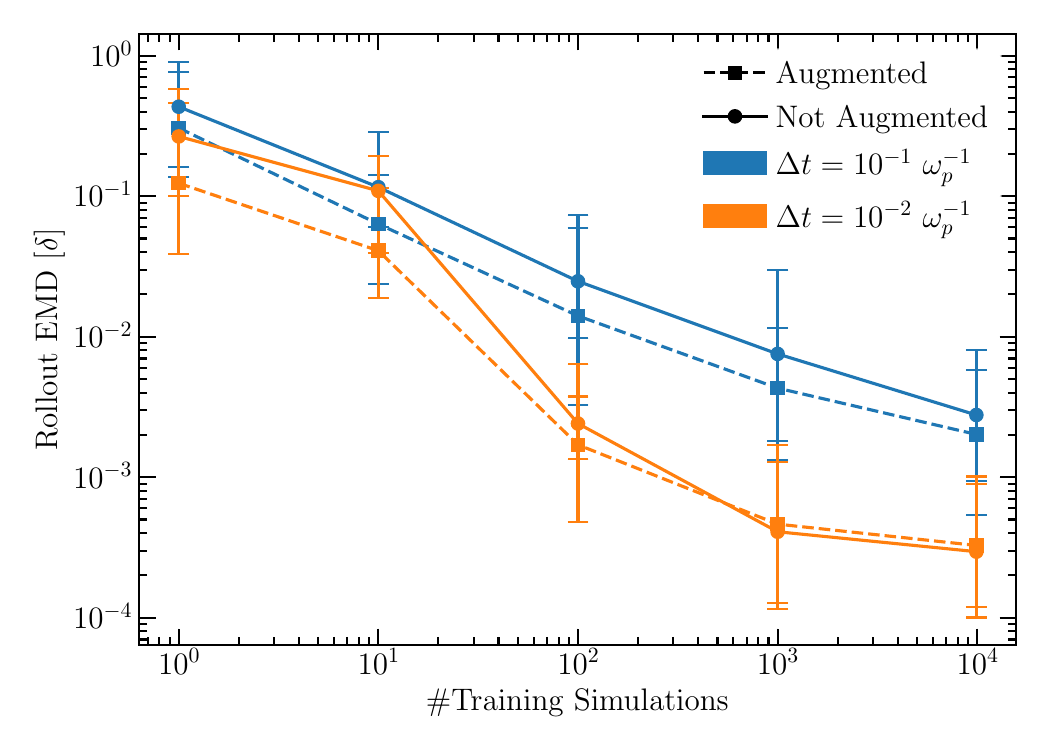}
    \caption{Model accuracy as a function of the number of training simulations. The test set contained  100 simulations of 10 sheets moving on a periodic box (same as in \fref{fig:benchmark_rollout}). It is observed that rollout EMD follows approximately a power law for $\Delta t=10^{-1}~\omega_p^{-1}$ meaning that further improvements can be achieved by increasing the dataset size. For $\Delta t=10^{-2}~\omega_p^{-1}$ the performance has already plateaued. Overall, the proposed data augmentation strategy is proven to be beneficial.}
    \label{fig:benchmark_nsimulations}
\end{figure}

In \fref{fig:benchmark_nsimulations}, we show the accuracy of equivalent models to those presented in the main body of the paper, when trained on datasets of different sizes. For $\Delta t = 10^{-2}~\omega_p^{-1}$ the accuracy plateaus before the maximum amount of training simulations used. The same does not happen for a $\Delta t = 10^{-1}~\omega_p^{-1}$, which exhibits a consistent power law scaling, meaning that increasing the dataset size could still lead to performance gains. We believe the difference in behavior occurs not only because more training time steps are available per simulation for smaller $\Delta t$, but also because it is easier to model the dynamics in smaller time steps since fewer crossings per simulation step occur.

Additionally, the results demonstrate the impact of using data augmentation, especially at smaller training dataset sizes. Further investigation revealed that this improvement in performance is mainly due to the time reversal augmentation since introducing the data reflected along the $x$-axis did not produce significant differences. We investigated if the data requirements could be further lowered by enforcing the symmetries expressed by the data augmentation procedure in the network architecture. However, due to the finite difference nature of the velocities, solely the spatial rotational symmetry could be enforced. No significant improvements were observed for a GNN architecture that enforced the latter, reinforcing the notion that this symmetry is easily learned/approximated by the network. More details are provided in~\ref{app:equivariantgnn}.

\section{Model benchmark -- Extra}
\label{app:benchmark}

\subsection*{Performance across different seeds}
\label{app:seeds}

\begin{figure}[htb]
    \centering
    \includegraphics[width=\columnwidth]{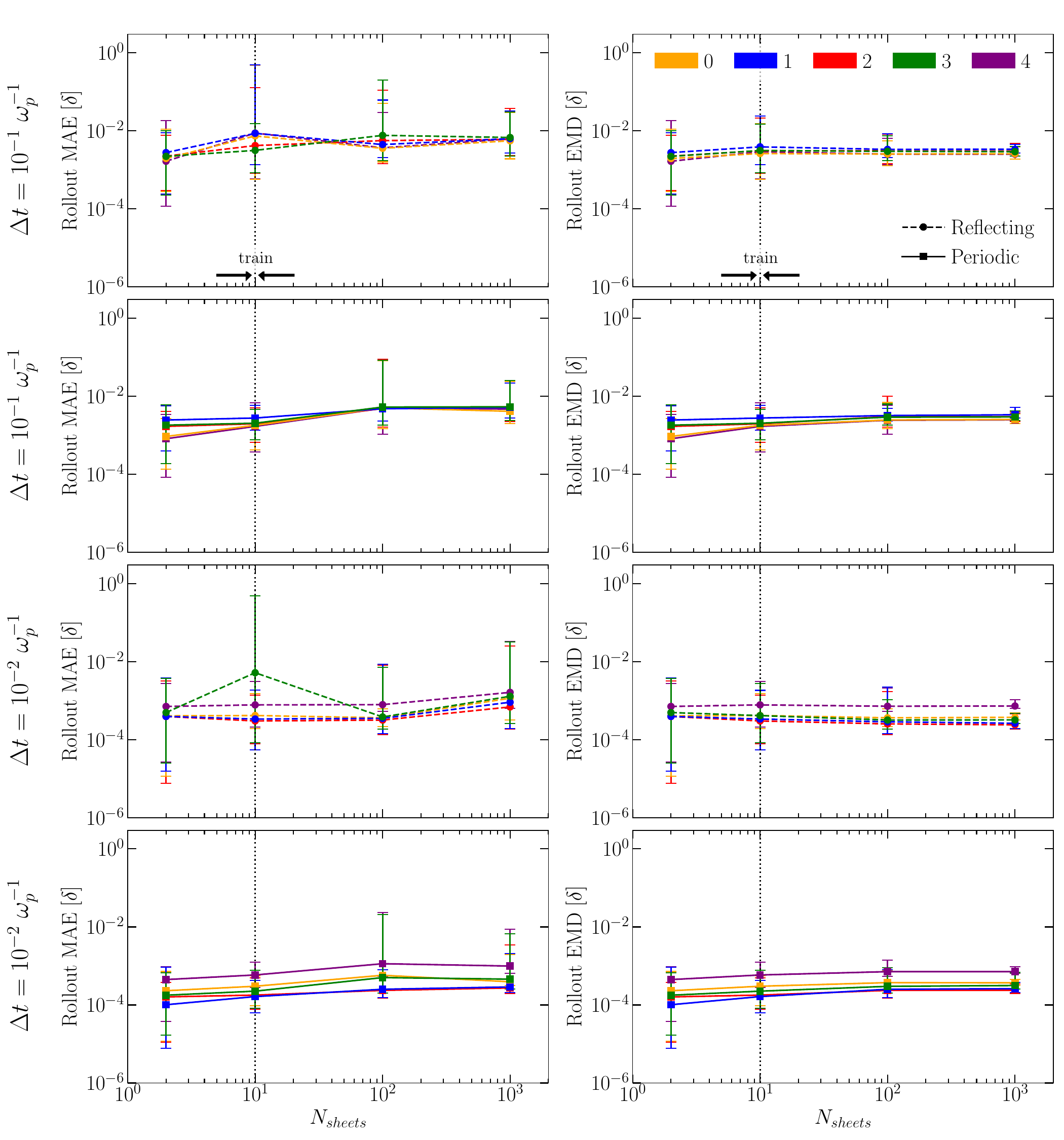}
    \caption{Rollout error metrics for all models used in the benchmark tests presented in Section~\ref{sec:benchmark} (5 models for each $\Delta t$, trained in a similar fashion using different random seeds). Mean values for each model are computed by averaging over sheets, time steps, and simulations. The error bars represent the minimum and maximum rollout error achieved for the corresponding set of test simulations.  We observe that equivalent models perform similarly, with the exception of model \#4 for $\Delta t=10^{-2} \ \omega_p^{-1}$. This difference in performance was already expected after completing the training procedure since the validation loss was approximately two times that of equivalent models (i.e.~achieved worse single-step acceleration prediction).}
    \label{fig:benchmark_rollout_all_seeds}
\end{figure}

\begin{figure}[htb]
    \centering
    \includegraphics[width=\columnwidth]{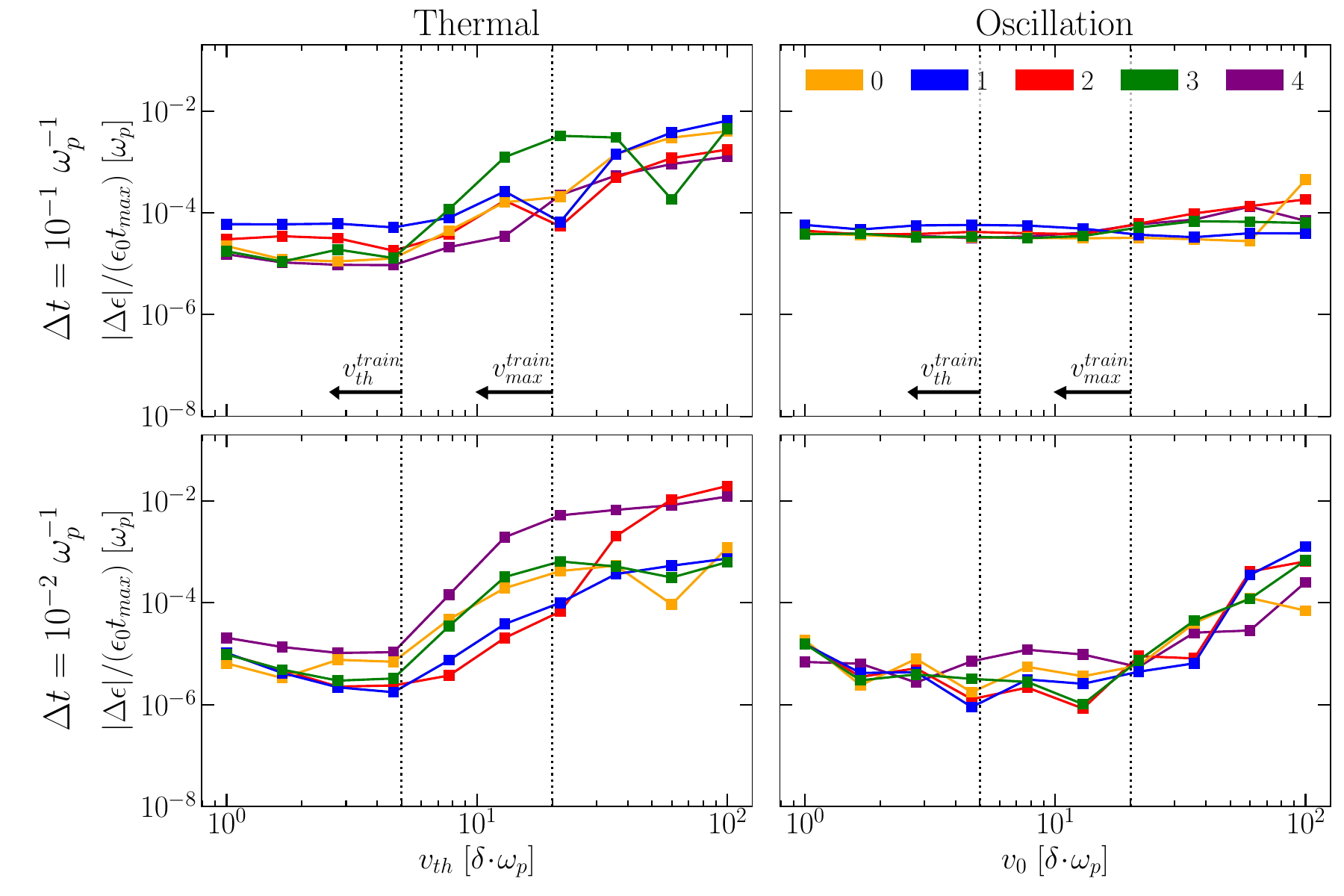}
    \caption{Energy conservation metrics for all models used in the benchmark tests presented in Section~\ref{sec:benchmark} (5 models for each $\Delta t$, trained in a similar fashion using different random seeds). Similarly to the rollout error results presented in~\fref{fig:benchmark_rollout_all_seeds}, the energy conservation does not vary significantly with the random seed. The exception is once again model \#4 for $\Delta t = 10^{-2}~\omega_p^{-1}$, which causes a significant increase in the average energy variation in~\fref{fig:energy_conservation} (see ``Thermal'').}
    \label{fig:benchmark_energy_all_seeds}
\end{figure}

In \fref{fig:benchmark_rollout_all_seeds} and \fref{fig:benchmark_energy_all_seeds} we show the same metrics presented in Section~\ref{sec:benchmark} but now presented for all the individual GNN models trained (instead of the average value and min/max range over equivalent models). The results clearly identify the existence of a consistently worse performing model for $\Delta t = 10^{-2}~\omega_p^{-1}$, both in terms of rollout accuracy and energy conservation. We attribute this difference in performance to a worse single-step prediction capability, which was already clear in its training performance ($2\times$ larger validation loss than equivalent models). 

\begin{figure}[htb]
    \centering
    \includegraphics[width=\columnwidth]{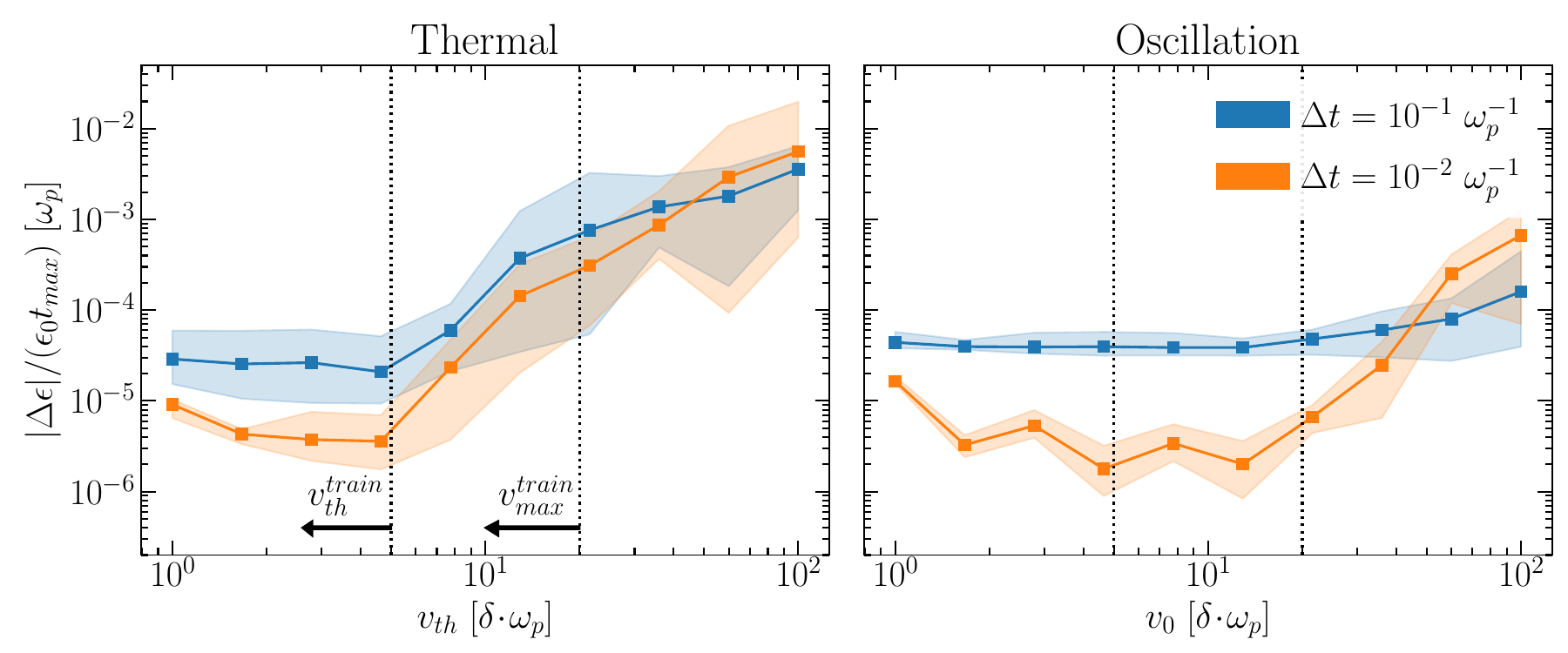}
    \caption{Same diagnostic presented in~\fref{fig:energy_conservation} without considering the contribution of model $\#4$ for $\Delta t=10^{-2}~\omega_p^{-1}$, which was shown to converge to a significantly worse validation loss, rollout accuracy, and energy conservation capabilities. The behavior of the GNS curves becomes more consistent across the different time resolutions.}
    \label{fig:benchmark_energy_no4}
\end{figure}

\begin{figure}[htb]
    \centering
    \includegraphics[width=\columnwidth]{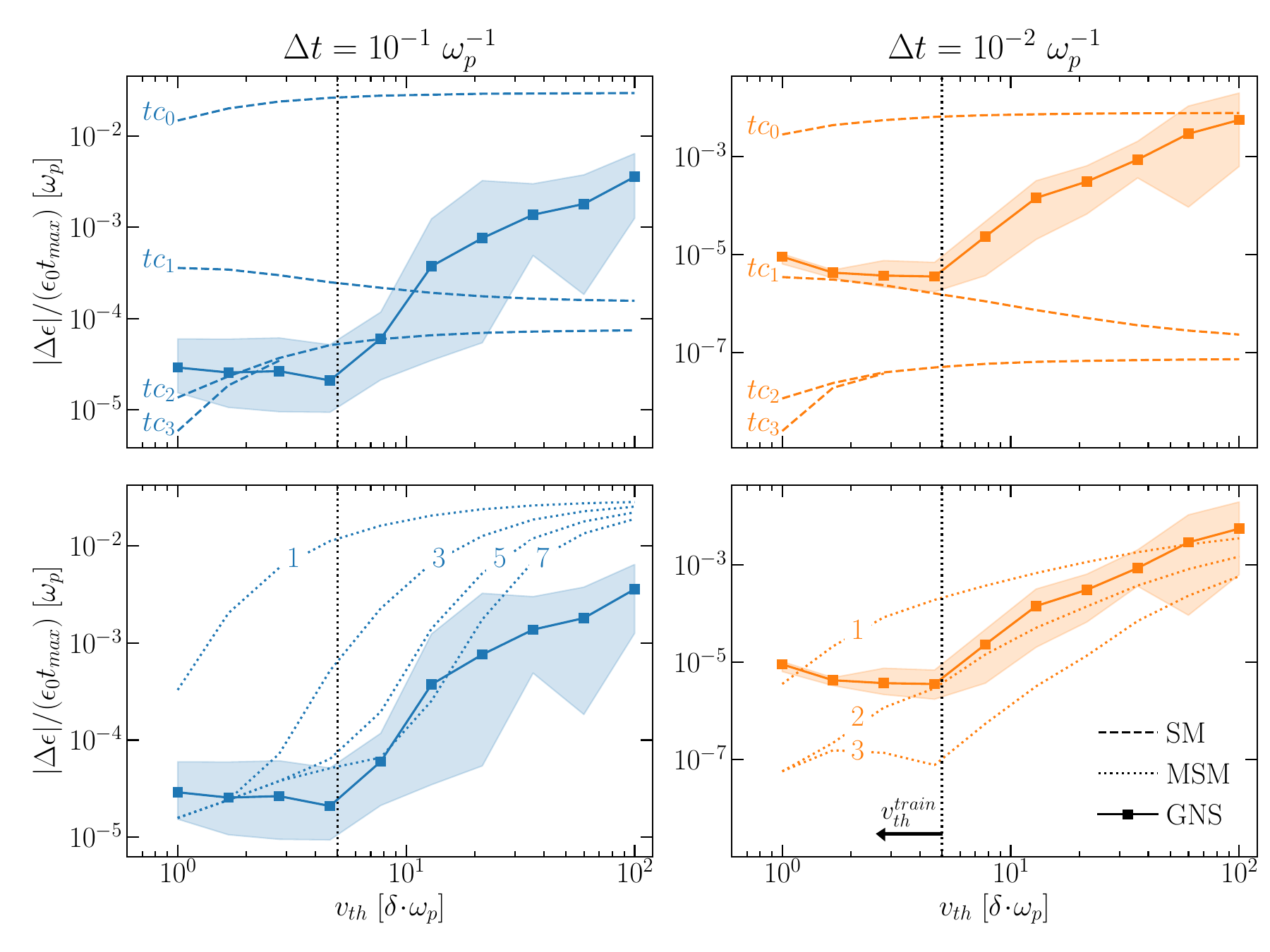}
    \caption{Same diagnostic presented in~\fref{fig:energy_conservation_comparison} without considering the contribution of model $\#4$ for $\Delta t=10^{-2}~\omega_p^{-1}$. The behavior of the GNS at $\Delta t = 10^{-2}~\omega_p^{-1}$ becomes overall comparable to the MSM which resolves crossings up to the 2nd neighbor.}
\label{fig:benchmark_energy_comparison_no4}
\end{figure}

When removing this model from the set considered for the energy conservation diagnostic (\fref{fig:benchmark_energy_no4}), the behavior for different time resolutions ($\Delta t =10^{-1}~\omega_p^{-1}$ and $\Delta t =10^{-1}~\omega_p^{-2}$) is significantly more consistent than what was observed in~\fref{fig:energy_conservation}. Furthermore, the average GNS at $\Delta t=10^{-2}~\omega_p^{-1}$ becomes comparable to the MSM which resolves crossings up the 2nd closest neighbor (\fref{fig:benchmark_energy_comparison_no4}) instead of the solely the closest neighbor (\fref{fig:energy_conservation_comparison}).

\clearpage

\subsection*{Energy conservation diagnostic}
\label{app:energy_conservation}

\begin{figure}[htb]
    \centering
    \includegraphics[width=\columnwidth]{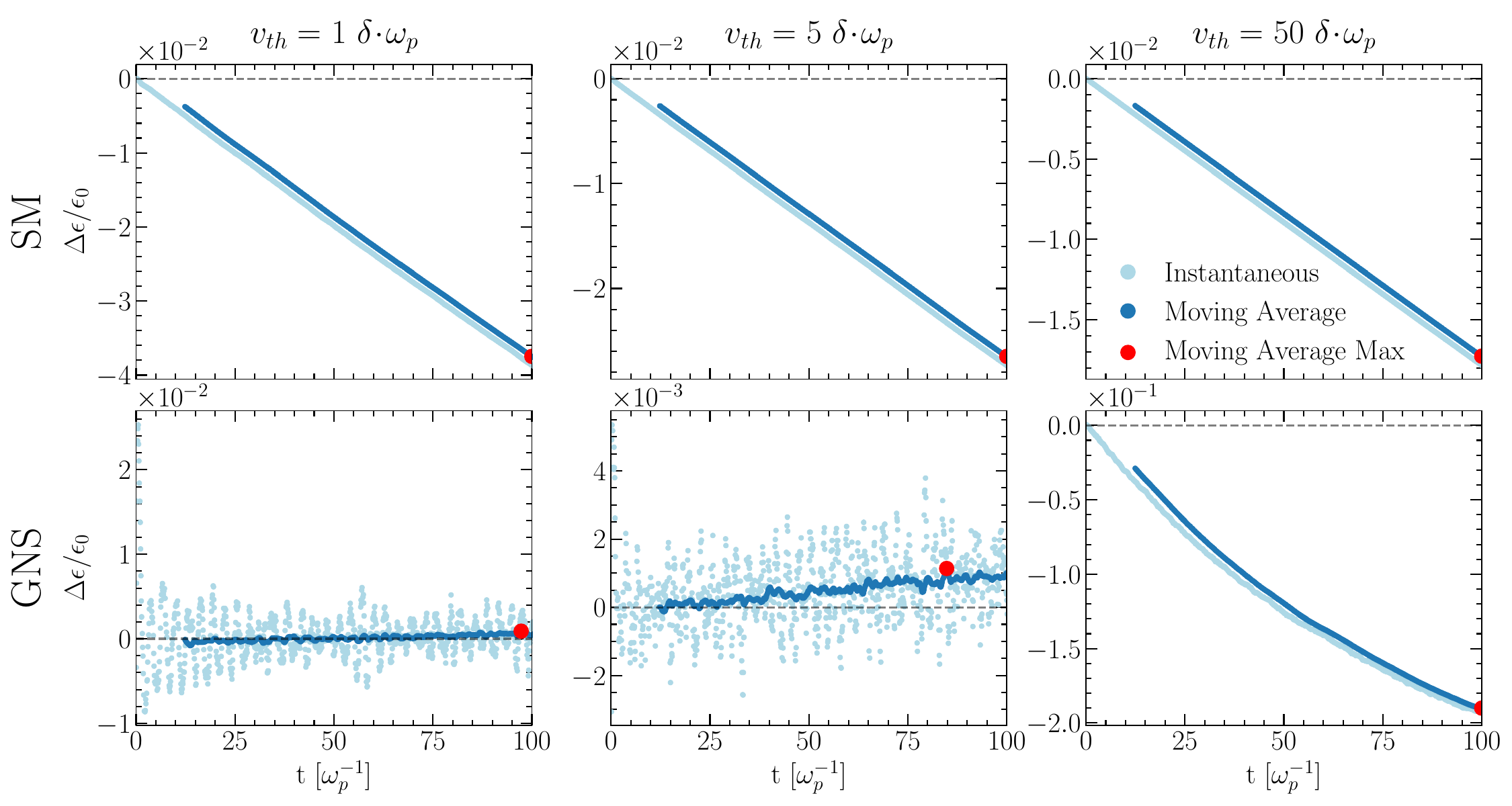}
    \caption{Illustration of the energy conservation diagnostic calculation. All systems consist of $10^3$ sheets inside a periodic box. Both the sheet model and the GNS use a time step of $\Delta t = 10^{-1}~\omega_p^{-1}$. The results illustrate why it is important to use the moving average of the energy over a plasma oscillation ($\Delta t = 2\pi~\omega_p^{-1}$) for a fair comparison between algorithms (since the GNS uses finite difference velocities). Additionally, it shows the reason for not considering the first plasma oscillation, since the finite difference velocities introduce a large oscillation in the estimation of the energy before crossings occur (i.e.~initial time-steps).} \label{fig:energy_conservation_ex}
\end{figure}

In \fref{fig:energy_conservation_ex} we provide examples of the observed energy variation during a simulation rollout for the sheet model and the GNS. The moving average ($\Delta t = 2 \pi~\omega_p^{-1}$) as well as its maximum value (what we consider for the results in \fref{fig:energy_conservation}) are highlighted. The results illustrate the importance of using the moving average for a fair comparison between algorithms since the usage of finite difference velocities for the GNS introduces an oscillation in the energy calculation (period equal to half a plasma oscillation). It also showcases why we do not use the first plasma period, since before crossings occur (i.e.~first time-steps) the initial ``energy increase'' introduced by the finite difference velocities is not compensated in the remaining of the plasma period (which would lead to an incorrect estimation of the energy using the moving average strategy). 

It is important to highlight that the initial energy of the system, which is considered for both the sheet model algorithm and the GNS, is computed using the initial instantaneous velocities  (i.e.~the input used for the sheet model simulation) which are different from the initial finite difference velocities used for the GNS. We compute the initial finite difference velocities by performing a simulation backward in time until $t=-\Delta t_{GNS}$ using the sheet model at a very high temporal resolution ($\Delta t_{SM} = 10^{-4}~\omega_p^{-1}$). This ensures that the initial conditions of the different simulators are equivalent.

\subsection*{Impact of crossings present in the training datasets}

\begin{figure}[htb]
    \centering
    \includegraphics[width=\columnwidth]{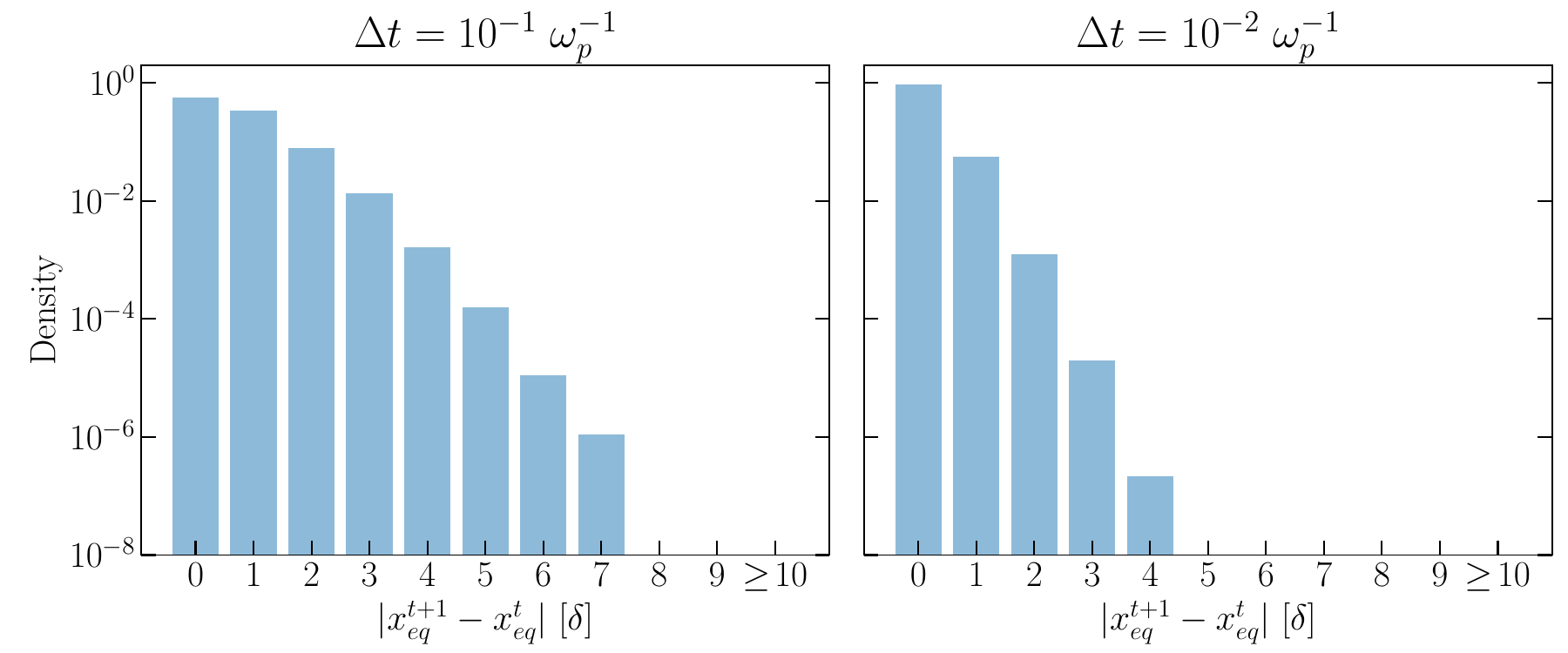}
    \caption{Distribution of sheet rank changes at different temporal resolutions for the test set consisting of 100 simulations of $10^3$ sheets moving over a reflecting box (used for rollout accuracy diagnostic in~\fref{fig:benchmark_rollout}). The training distribution is expected to follow a similar behavior.  There is a clear cut-off in the maximum rank change, which decreases with an increased temporal resolution, and a clear imbalance in the percentage of higher-order crossings. This will limit the capability of the GNS to model crossings involving a larger number of sheets.}
    \label{fig:dataset_rankchange}
\end{figure}

In~\fref{fig:dataset_rankchange} we show the distribution of relative sheet rank change ($x_{eq}^{t+1} - x_{eq}^t$) across all time steps of the test set of $100$ simulations containing $10^3$ sheets on reflecting boundary conditions (previously used for the rollout accuracy measurement in \fref{fig:benchmark_rollout}). The rank change is a proxy (lower limit) on the number of sheets involved in a crossing, thus allowing us to understand what is the percentage of data points that include crossings with the $n$th neighbor. The distribution in the test data shown should be equivalent to the one present in the training data, which we do not show since for periodic boundary sheets can cross through the boundary, leading to higher values of $x_{eq}^{t+1} - x_{eq}^t$ which are not meaningful for understanding the number of sheets involved in the crossings.

What we observe is that there is a cut-off in the rank changes present for different time resolutions, which decreases for higher temporal resolutions. For $\Delta t=10^{-2}~\omega_p^{-1}$ the majority of time-steps either no crossing occurs ($>90\%$) or no sheet moves $\pm2$ positions to the left/right ($>99.999\%$). We believe this is the main reason why the GNS fails to outperform the MSM which checks for crossings up to the third neighbor at this time resolution,i.e.~it overfits the oscillatory motions and it is not provided with enough training samples for higher-order crossings.

\clearpage

\section{Landau damping -- Extra}
\label{app:landaudamping}

\begin{figure}[htb]
    \centering
    \includegraphics[width=0.8\textwidth]{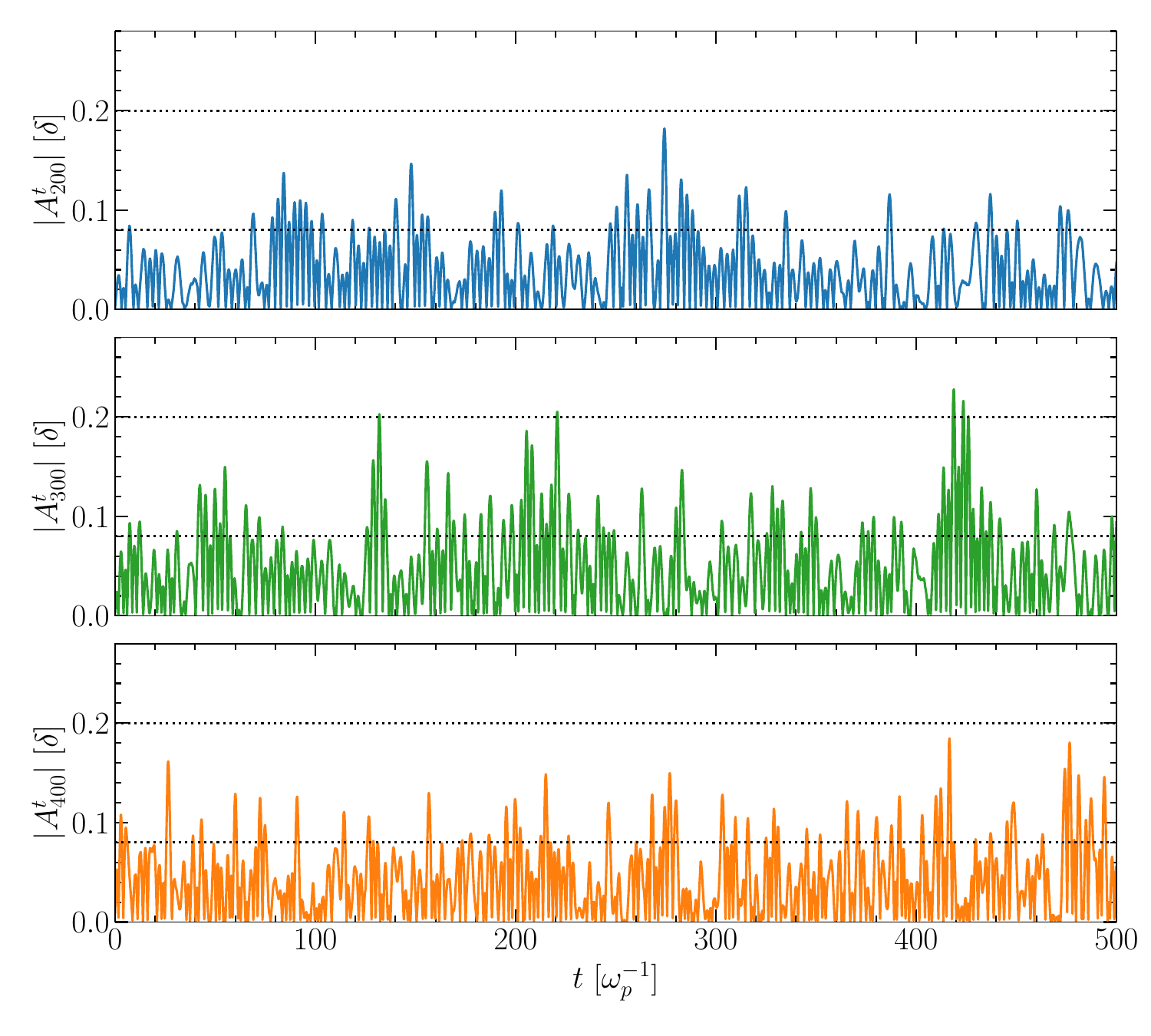}
    \caption{Examples of mode amplitude evolution for one of the simulations used in Section~\ref{sec:landau_damping}. The black dashed lines delimit the region used in the Landau Damping diagnostic results presented in~\fref{fig:landau_damping_w_im} and~\fref{fig:landau_damping_w_real}. This region is chosen such that we start tracking the mode evolution above the typical ``noise'' level while guaranteeing that enough statistics exist to compute a meaningful average trajectory.}
    \label{fig:ld_mode_evolution_ex}
\end{figure}

The range used for the initial mode amplitudes $A_m^0 \in [0.08, 0.2]~\delta$ in Section~\ref{sec:landau_damping} was chosen so that: a) the minimum initial amplitude was above the usual ``noise'' level, i.e.~the values at which on average most nodes stop decaying; b) the maximum value was not too high to guarantee we had enough statistics (number of trajectories) to obtain a good estimate of the average trajectory across all modes. To further illustrate the reasoning behind the chosen range we provide in~\fref{fig:ld_mode_evolution_ex} the amplitude of several modes over time for one of the simulations produced in Section~\ref{sec:landau_damping}.

The range of the modes used, $k\lambda_D \in [0.28, 0.7]$, was selected such that enough particles exist to resolve the distribution function near the mode phase velocity (for $k\lambda_D=0.28$, $v_{ph} \simeq 4 v_{th}$ meaning only $\simeq 100$ particles have $v > v_{ph}$) and that the mode does not decay too fast (for $k\lambda_D = 0.7$, $-\mathrm{Im}(\omega)^{-1} \simeq 2.5~\omega_p^{-1}$). It is therefore expected that, closer to these limits, the dynamics are not as well approximated.

To automatically compute the decay rate and the angular frequency we proceed as follows. We first obtain the average trajectory of the mode in the ($A_m$, $\dot{A}_m$) phase-space as explained in Section~\ref{sec:landau_damping}. The length of this trajectory is chosen so that it corresponds to the maximum of: a) the time it takes for the mode to decay to $1/5$-th of its initial amplitude (using the theoretical estimate of the damping rate); b) 3 mode periods (using the theoretical estimate of the angular frequency). The first condition is the preferred one for slowly damped modes, and the second for fastly damped modes.

\begin{figure}[htb]
    \centering
    \includegraphics[width=0.8\textwidth]{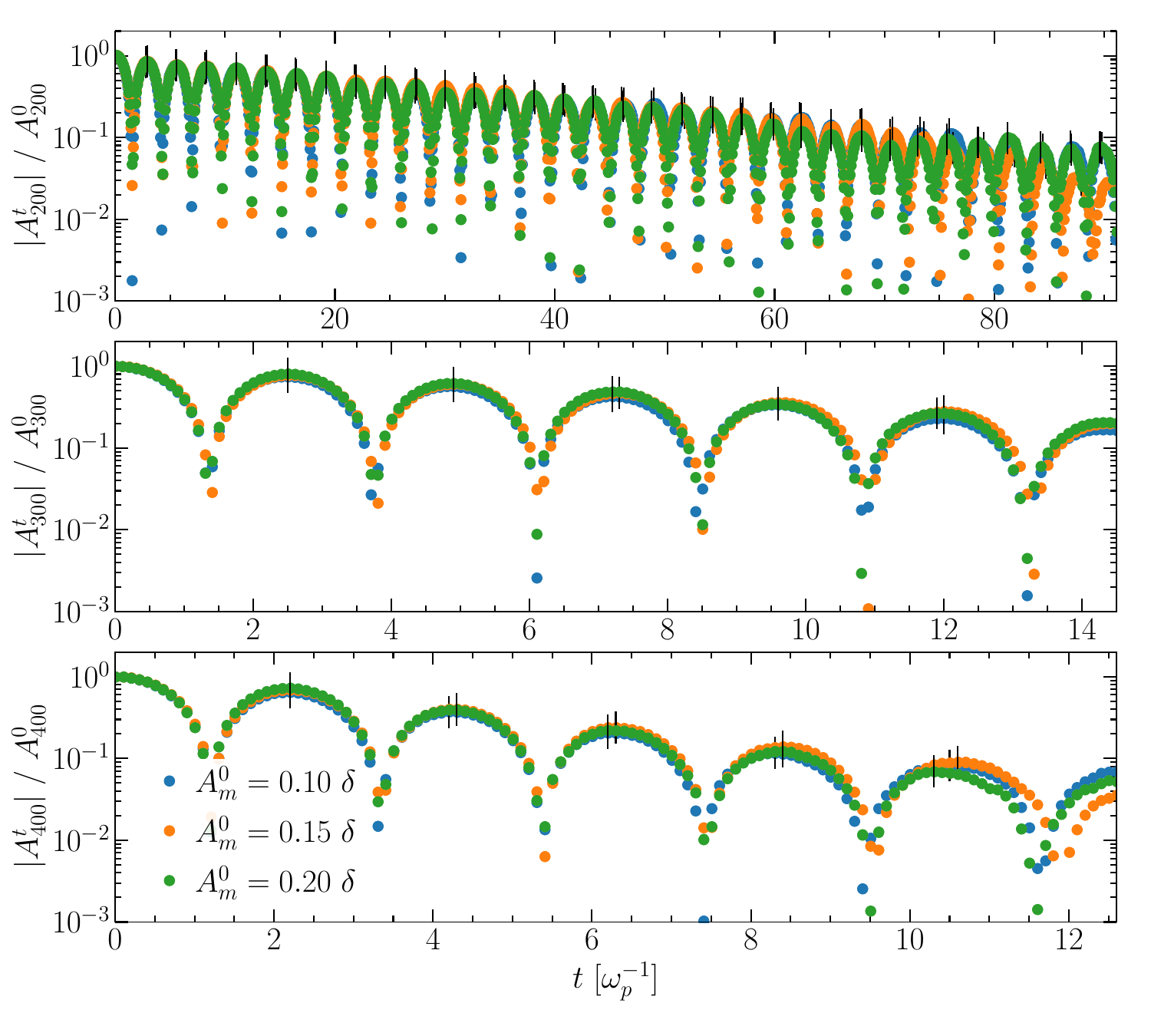}
    \caption{Illustration of the Landau Damping diagnostic used to retrieve the damping rate and angular frequency for \fref{fig:landau_damping_w_im} and \fref{fig:landau_damping_w_real}. For each mode $m$ and initial mode amplitude $A^0_m$ we compute the average mode amplitude over time. The maximum for each half period (shown by a black line) is then automatically calculated. The mode damping rate and the angular frequency are extracted based on these maxima.}
    \label{fig:ld_w_estimation_ex}
\end{figure}

After obtaining the average trajectory, we split it into equal regions of size $\Delta t = \pi/\omega_{t}$ where $\omega_{t}$ is the theoretical angular frequency of the mode. For each time interval, we retrieve the maximum value of $|A_m^t|$ and the corresponding time. This process is illustrated in \fref{fig:ld_w_estimation_ex}. The decay rate is then obtained from the slope of the $\log|A_m^t|$ maxima curve, and the angular frequency from the average distance between consecutive maxima. For highly damped modes (larger $m$, or equivalently larger $k\lambda_D$) this estimate is noisier (e.g.~$m=400$ in \fref{fig:ld_w_estimation_ex}), resulting in the larger variations presented in \fref{fig:landau_damping_w_im} and \fref{fig:landau_damping_w_real}.

\clearpage

\section{Two-stream instability -- Extra}
\label{app:twostream}

\subsection*{Different modes}

\begin{figure}[htb]
    \centering
    \includegraphics[width=0.9\columnwidth]{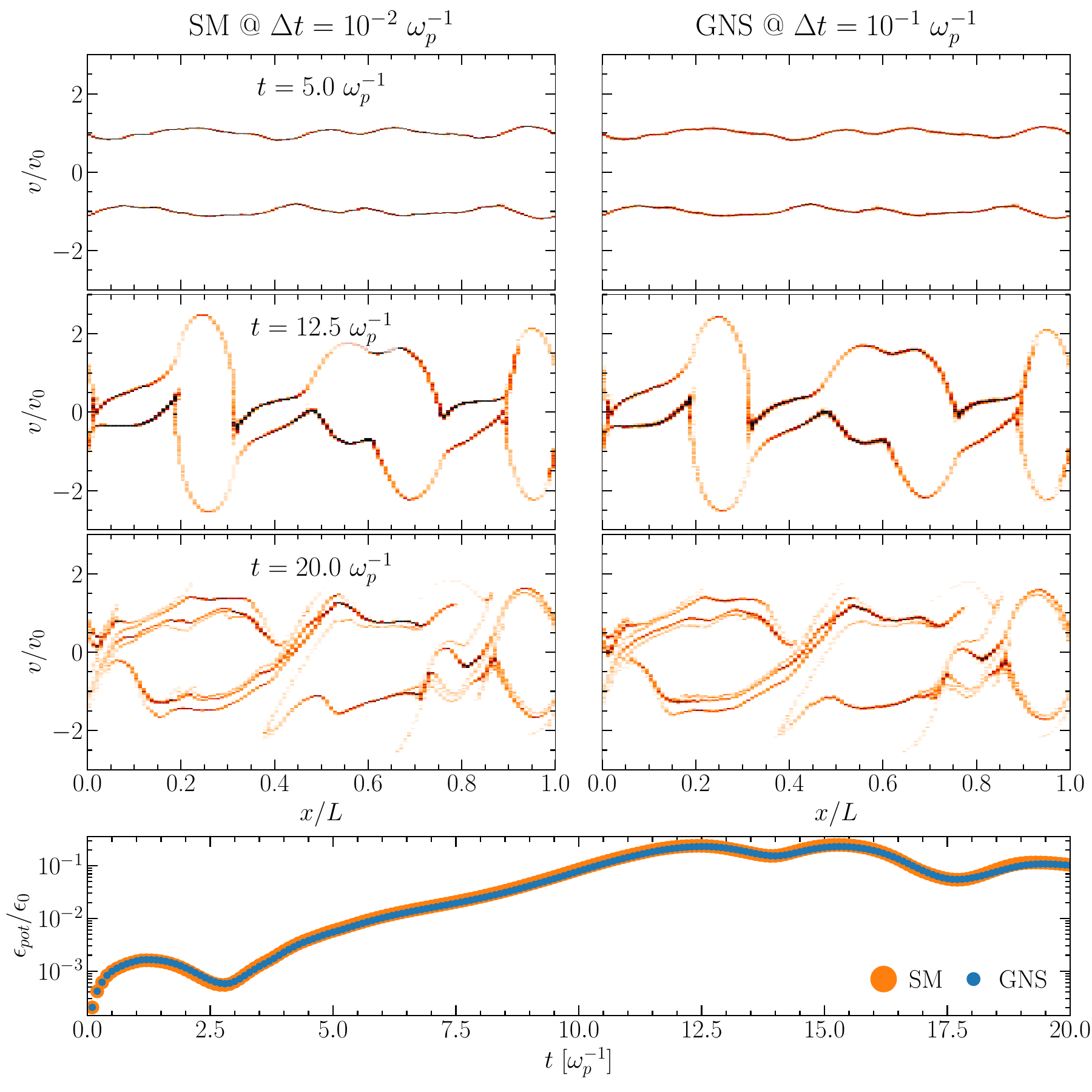}
    \caption{Two stream instability simulation which excites the $m=8$ mode ($v_0 = 244~\delta\!\cdot\!\omega_p$). We use the same model and setup as described in Section~\ref{sec:twostream}. The GNS is  once again capable of recovering the same dynamics as the higher temporal resolution sheet model.}
    \label{fig:ts_m8}
\end{figure}

\begin{figure}[htb]
    \centering
    \includegraphics[width=0.9\columnwidth]{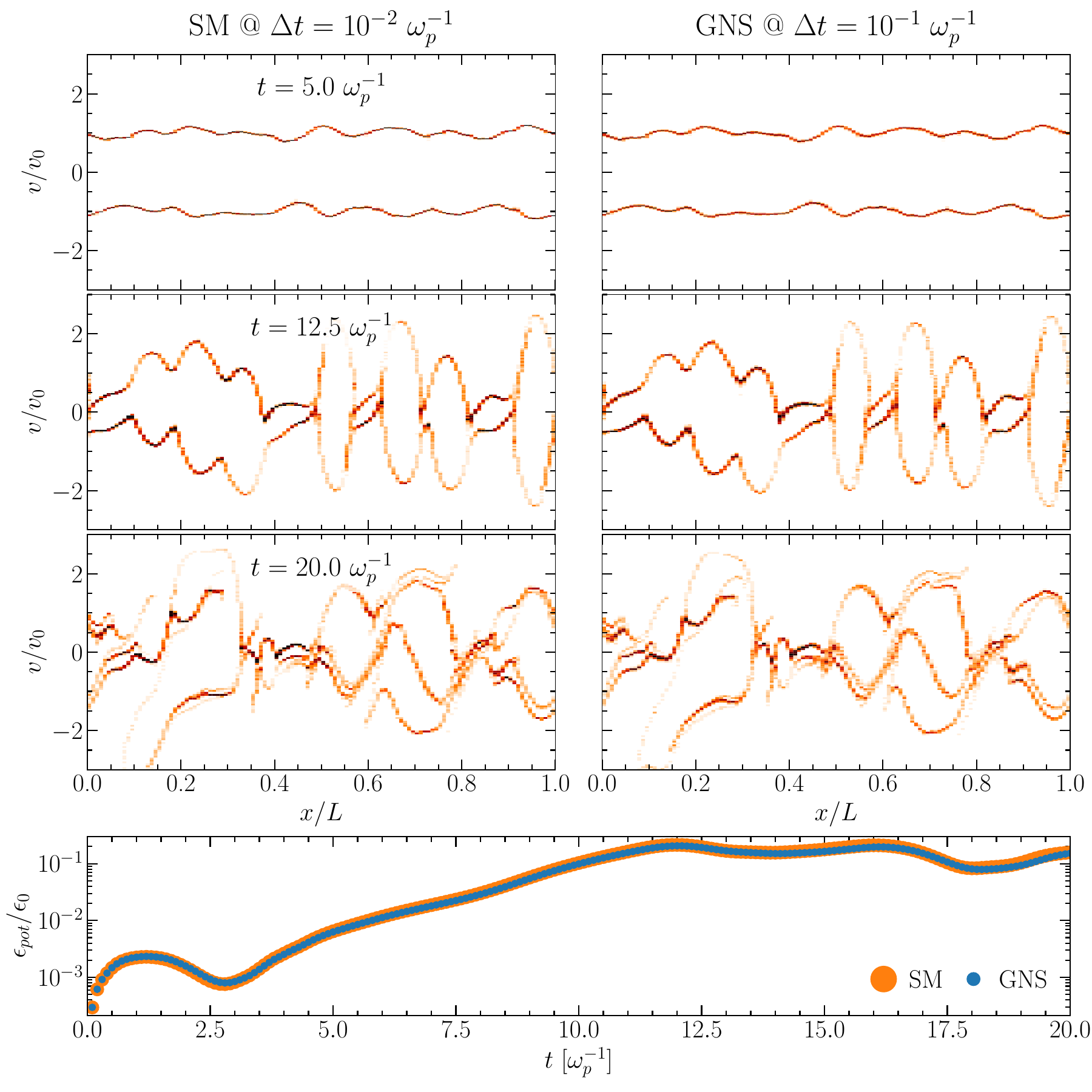}
    \caption{Two stream instability simulation which excites the $m=12$ mode ($v_0 = 162~\delta\!\cdot\!\omega_p$). We use the same model and setup as described in Section~\ref{sec:twostream} and \fref{fig:ts_m8}. Results provide further demonstration of the GNS capability to recover the instability dynamics.}
    \label{fig:ts_m12}
\end{figure}

In \fref{fig:ts_m8} and \fref{fig:ts_m12} we showcase a similar setup as the one shown in Section~\ref{sec:twostream} with the difference that the main modes excited are now $m=8$ ($v_0 = 244~\delta\!\cdot\!\omega_p^{-1}$) and $m=12$ ($v_0 = 162~\delta\!\cdot\!\omega_p^{-1}$). We observe once again that the GNS is capable of modeling the same macrophysics as the sheet model at a considerably higher time resolution. The relative energy variation of the system (for $m=8$ and $m=12$) is similar to the one presented for $m=4$ in Section~\ref{sec:twostream} ($\Delta\epsilon_{\mathrm{SM}}/\epsilon_0 \approx 10^{-6}$, and $\Delta\epsilon_{\mathrm{GNS}}/\epsilon_0  \approx 2\times10^{-2}$).

\clearpage

\subsection*{Different model seeds}

\begin{figure}[htb]
    \centering
    \includegraphics[width=0.9\columnwidth]{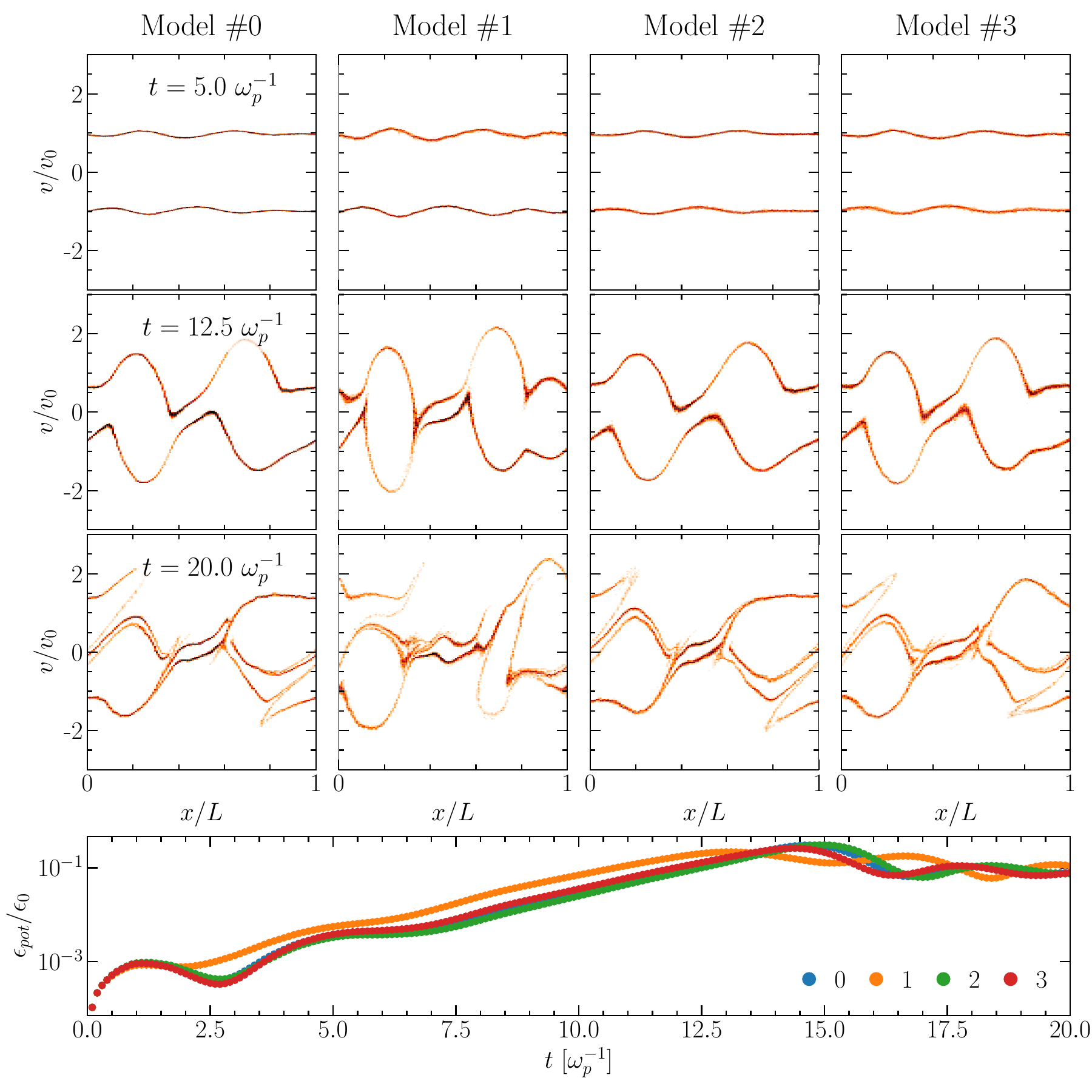}
    \caption{Two-stream instability results ($m=4$) for GNNs initialized with different random seeds. We recover similar dynamics to those presented in the main body of the paper (\fref{fig:twostream_sm_vs_gns}) while using models with worse energy conservation capabilities.}
    \label{fig:ts_m4_seeds}
\end{figure}

In~\fref{fig:ts_m4_seeds} we reproduce the two-stream simulation from Section~\ref{sec:twostream} now using equivalent GNNs trained with different random seeds. As previously mentioned, these models showcase (overall) worse energy conservation capabilities for out-of-training distribution velocities (\fref{fig:benchmark_energy_all_seeds}). In particular, for the two-stream instability simulation presented in~\fref{fig:ts_m4_seeds}, we observe relative energy variations between $10-16\%$ in comparison with the $2\%$ reported in Section~\ref{sec:twostream}. Nonetheless, the macrophysics obtained for these models is consistent with the results shown in the main body of the paper (\fref{fig:twostream_sm_vs_gns}). The only exception is Model $\#1$, whose phase space looks considerably different (the potential energy growth rate still matches that of other models).

\clearpage

\subsection*{Different time step}

\begin{figure}[htb]
    \centering
    \includegraphics[width=0.9\columnwidth]{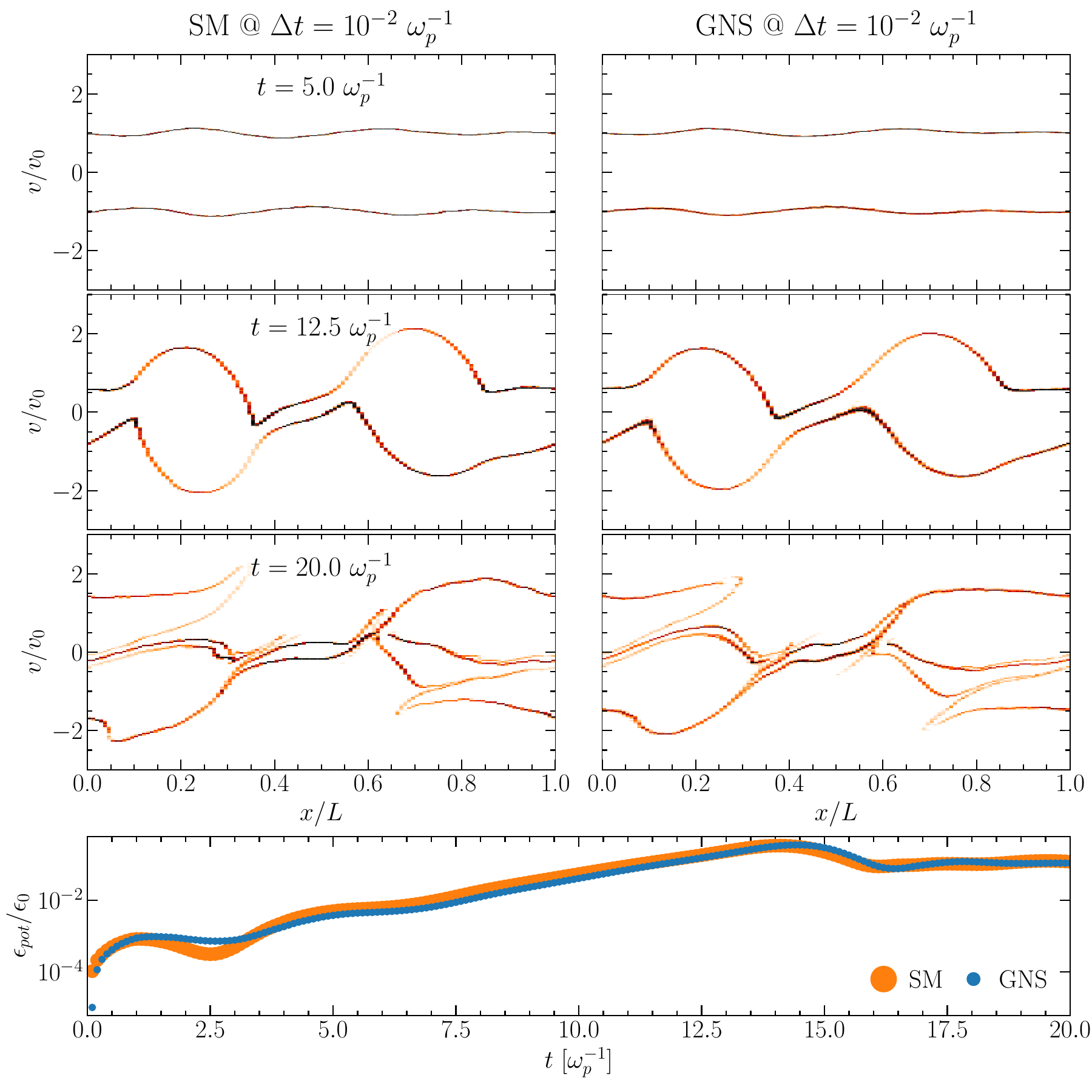}
    \caption{Comparison of the phase space and potential energy evolution of the two-stream scenario (with $m=4$) using a smaller time step ($\Delta t = 10^{-2}~\omega_p^{-1}$). We observe similar results to those obtained for the GNS at lower time resolutions (\fref{fig:twostream_sm_vs_gns}).}
    \label{fig:twostream_sm_vs_gns_001}
\end{figure}

In \fref{fig:twostream_sm_vs_gns_001}, we provide an additional example of the two-stream instability test performed in Section~\ref{sec:twostream}, now performed for the GNS at $\Delta t = 10^{-2}~\omega_p^{-1}$. The relative energy variation is approximately $\Delta\epsilon/\epsilon_0 \approx 2\times10^{-2}$ (Model $\#0$). Once again, the GNS is able to correctly recover the dynamics. The reasons why we do not observe improved energy conservation capabilities for the GNS at $\Delta t = 10^{-2}~\omega_p^{-1}$ when compared to models trained at $\Delta t = 10^{-1}~\omega_p^{-1}$ have already been discussed in Section~\ref{sec:energy_conservation}.

\begin{figure}[htb]
    \centering
    \includegraphics[width=0.9\columnwidth]{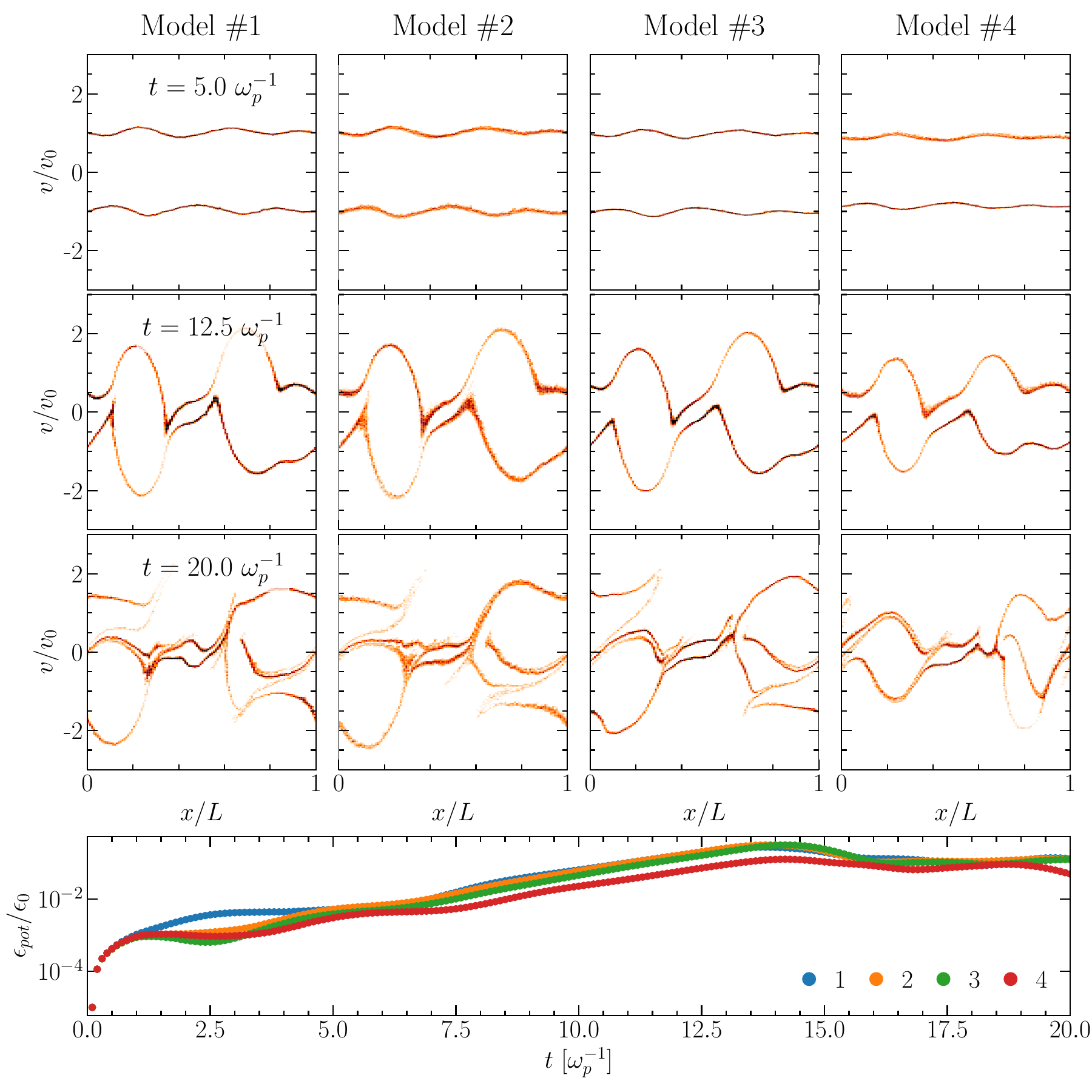}
    \caption{Two-stream instability ($m=4$) results obtained for the remaining GNN models trained with a time step $\Delta t = 10^{-2}~\omega_p^{-1}$. Apart from model $\#4$ (whose problems were already identified in~\ref{app:benchmark}), all others retrieve similar dynamics than those observed in~\fref{fig:twostream_sm_vs_gns_001}.}
    \label{fig:ts_m4_001_seeds}
\end{figure}

The results obtained across GNN models trained with different seeds for this higher temporal resolution are also consistent (\fref{fig:ts_m4_001_seeds}), with the exception of model $\#4$ ($51\%$ energy loss compared to $2-7\%$ for the remaining seeds) whose problems have already been identified and discussed in~\ref{app:seeds}.

\end{document}